\documentclass[superscriptaddress,reprint, nofootinbib,
 amsmath,amssymb,aps]{revtex4-1}

\usepackage{graphicx}
\usepackage{gensymb}
\usepackage{xcolor}
\usepackage{subcaption}
\usepackage{graphicx}% Include figure files
\usepackage{dcolumn}% Align table columns on decimal point
\usepackage{bm}% bold math
\usepackage{float}
\usepackage{placeins}
\usepackage[%
  colorlinks=true,
  urlcolor=blue,
  linkcolor=blue,
  citecolor=blue
]{hyperref}

\begin{document}

%\preprint{APS/123-QED}

\title{Superconducting Incommensurate Host-Guest Phases in Compressed Elemental Sulfur}

\author{Jack Whaley-Baldwin}
\email{jajw4@cam.ac.uk}
\affiliation{TCM Group, Cavendish Laboratory, University of Cambridge}

\author{Michael Hutcheon}
\affiliation{TCM Group, Cavendish Laboratory, University of Cambridge}

\author{Chris J. Pickard}
\affiliation{Advanced Institute
for Materials Research, Tohoku University, Sendai, Japan}

\date{\today}

\begin{abstract}
We use Density Functional Theory (DFT) and structure searching methods to show that the ground state of elemental sulfur is a Barium-IVa type incommensurate host-guest (HG) phase between $386$-$679$ GPa, becoming the first group-VI element predicted to possess such a structure. Within the HG phase, sulfur undergoes a series of transitions in which adjacent guest chains are not aligned but rather offset by different amounts, which can be described as a rearrangement of the stacking order of the chains. We show that these chain rearrangements are intimately coupled to modulations of the host and guest atoms which prove crucial to stabilising the HG structure. Unlike the high-pressure HG phases of other elements, sulfur does not exhibit interstitial charge localisation, and instead features strongly localised `voids' that are depleted of electronic charge. Prior to adopting the HG structure, we predict that sulfur possesses an orthorhombic structure of $Fdd2$ symmetry. We calculate the superconducting critical temperatures of these newly discovered phases, and show that $T_c$ is expected to peak between $24.8$ and $28.2$ K at $271$ GPa.
\end{abstract}

\maketitle

\section{\label{intro}Introduction}

Sulfur (S) is an element renowned for its structural diversity, possessing a considerable array of metastable allotropes at ambient pressure alone \cite{sulfur_allotropes,Meyer_sulfur}. This complexity persists at high pressures, where sulfur adopts elaborate catenated spiral structures above $1.5$ GPa \cite{chain_structures_sulfur} and an incommensurately modulated structure at $83$ GPa \cite{Degtyareva_and_Hemley_PRB,Luo_et_al}, which is driven by a charge density wave (CDW) instability.
\par
Self-hosting incommensurate host-guest (HG) structures, which comprise two distinct elemental sublattices embedded within one another in such a way that the ratio of their lattice lengths along a certain crystal axis is an irrational number, are well-known within the literature \cite{HG_Barium,HG_alkali_metals,Rb-IV,HG_Aluminium,HG_group_V,HG_Bismuth_SC,Haussermann_2002,HG_Strontium,HG_scandium,HG_Antimony,HG_Calcium}. The reasoning behind the stability of such host-guest phases is often explained through so-called Fermi surface effects, sometimes referred to as the Jones mechanism \cite{Jones_mechanism}, in which lattice distortions introduce Fourier components into the potential that lower the energy of Fermi level states.
\par
The physical properties of these elemental host-guest structures are of great interest and vary significantly between elements. Several examples are superconducting, and often exhibit high critical temperatures owing to favourable low-frequency \textit{quasi}-acoustic phonon modes \cite{HG_Bismuth_SC}, with a record $T_c$ of $29$ K among all elements (excluding hydrogen) being calcium in just such a phase (Ca-VII at $210$ GPa \cite{SC_elements_review,Calcium_experimental,chain_ordered_calcium}). Conversely, other realisations of elemental HG structures feature large pseudogaps in their electronic densities of states at the Fermi level, and form poorly-conducting electrides with significant amounts of electronic charge localised in interstitial regions \cite{HG_alkali_metals,HG_Aluminium}.
\par
X-ray diffraction studies on S up to $165$ GPa \cite{Degtyareva_and_Hemley_PRB,Luo_et_al} and $250$ GPa \cite{highest_experimental_pressure_sulfur,drozdov_hydrogen_sulfide_experiment} have determined that S transforms from the incommensurately-modulated CDW phase into the trigonal $\beta$-Po structure of $R\bar{3}m$ symmetry at ${\sim} 150$ GPa. Several first-principles structure searching investigations \cite{USPEX,Whaley_Baldwin,sulfur_no_SC} have predicted that the $R\bar{3}m$ structure remains stable up to ${\sim} 500$ GPa, where S then adopts a body-centered cubic (bcc) structure.
\par
The correct determination of the crystal structure of S at high pressures is of interest to experimentalists, in order to identify its presence as a decomposition product in high pressure hydrogen sulfide experiments on H$_2$S and H$_3$S and to resolve the resulting diffraction pattern \cite{hydrogen_sulfide_Goncharov,novel_sulfur_hydrides}. Knowledge of the true ground state is also important from a structure searching standpoint, so that the enthalpy of candidate H$_2$S / H$_3$S structures can be measured against decomposition into H and S \cite{hydrogen_sulfide_Cui,dissociation_H2S}.
\par
The high-pressure superconducting properties of pure S have long attracted attention, and at one time it held the highest experimentally verified $T_c$ in an element \cite{record_Tc_sulfur}. Further, the fact that the incommensurately modulated phase of S is superconducting with such a high $T_c$ is unusual, as the Fermi-level gaps opened up by CDW formation often work to suppress superconductivity \cite{CDW_sulfur}. As well as H$_2$S and H$_3$S, sulfur is a key constituent of several other high-$T_c$ compounds, including a recently claimed room-temperature superconductor \cite{RT_supercond}.
\par
In this work, we investigate the phase diagram of S up to $700$ GPa and report that under sufficient compression, S adopts a superconducting incommensurate HG structure similar to that of Barium in the IVa phase \cite{HG_Barium}, in addition to a previously unreported orthorhombic phase. We show that the guest structure in S undergoes several changes with increasing pressure, and link this to modulations of the host and guest atoms. We further study the electronic structure and charge density distribution of the HG phase, and compare the latter to other high-pressure HG structures.
\newpage

\section{\label{comp_details}Computational Details}

We used the \textit{ab-initio} Random Structure Searching (AIRSS) package \cite{airss_PRL,airss_JPCM} for our structure searching, with the plane-wave DFT code \verb|CASTEP| \cite{castep} performing the underlying electronic structure calculations. We generated a sulfur pseudopotential specifically designed for high-pressure work using the OTFG code in \verb|CASTEP| (see Supplemental Material \cite{supp}). Due to small enthalpy differences in certain parts of the phase diagram, we used an $850$ eV plane-wave cutoff and a \textbf{k}-point spacing of $2\pi \times 0.01$ \AA$^{-1}$, sufficient for absolute enthalpy convergence to $\pm 0.2$ meV. Relative enthalpies between structures were converged to an even tighter tolerance. Geometry optimisations were carried out using the L-BFGS algorithm and structures were relaxed until the forces on each atom were $5 \times 10^{-4}$ eV / \AA$^{-1}$ or smaller, and the residual stresses $1 \times 10^{-2}$ GPa or smaller.
\par
We performed structure searches at $300$, $400$, $500$ and $750$ GPa, using randomly generated structures containing between $1$ and $15$ atoms and possessing between $1$ and $48$ symmetry operations. We then performed searches at these same pressures with $16$, $21$, $24$ and $32$ atoms. Finally, we performed much smaller searches in peturbed supercells of our various host-guest approximants, which involved up to $288$ atoms. In total, we considered around $7500$ fully relaxed structures.

\section{\label{results}Results and Discussion}
\subsection{\label{relative_enths}Relative Enthalpies}

\begin{figure}[h!]
    \hspace{-0.22cm}
    \includegraphics[width=0.49\textwidth]{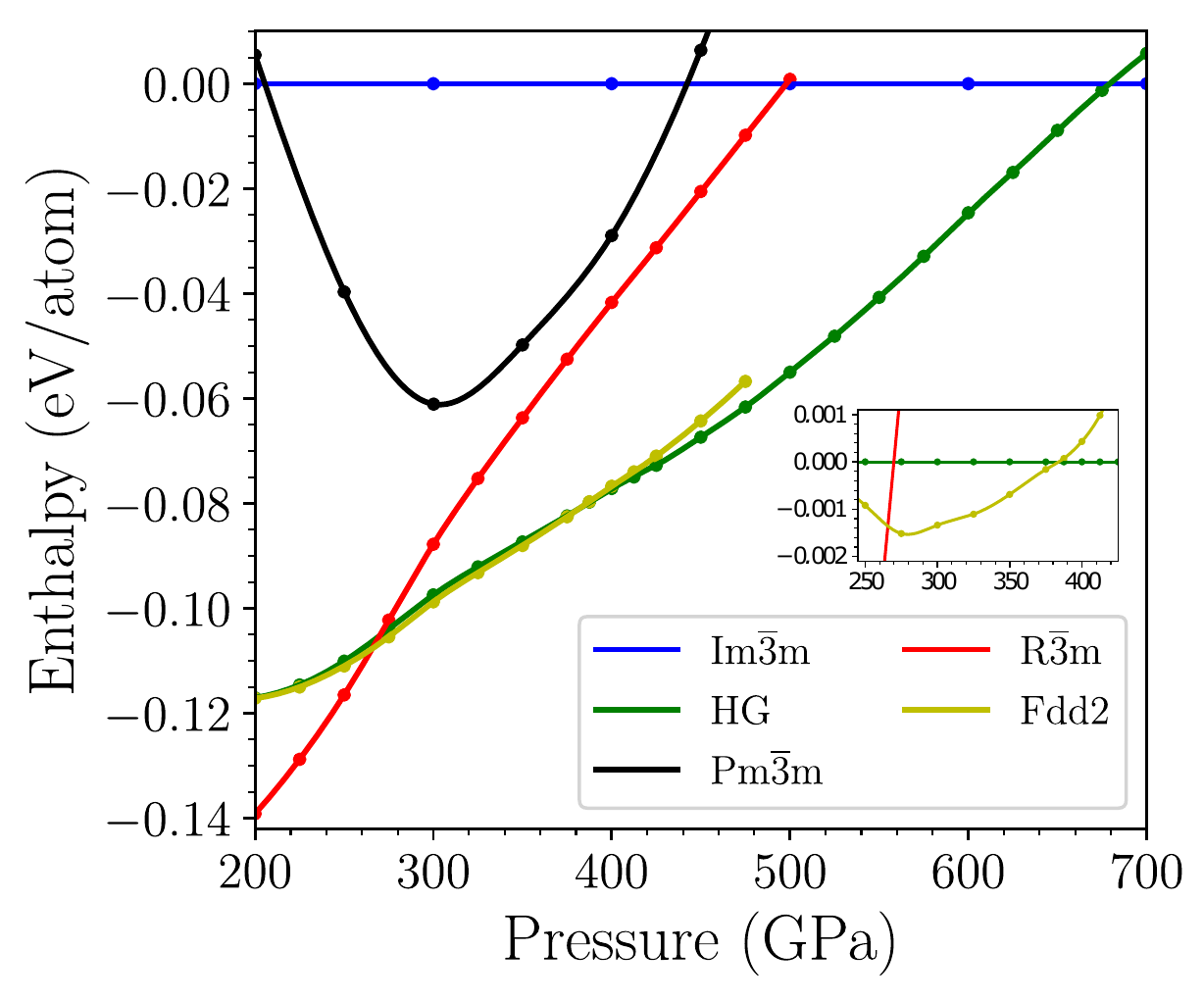}
    
    \vspace{0.25cm}
    
    \includegraphics[width=0.22\textwidth]{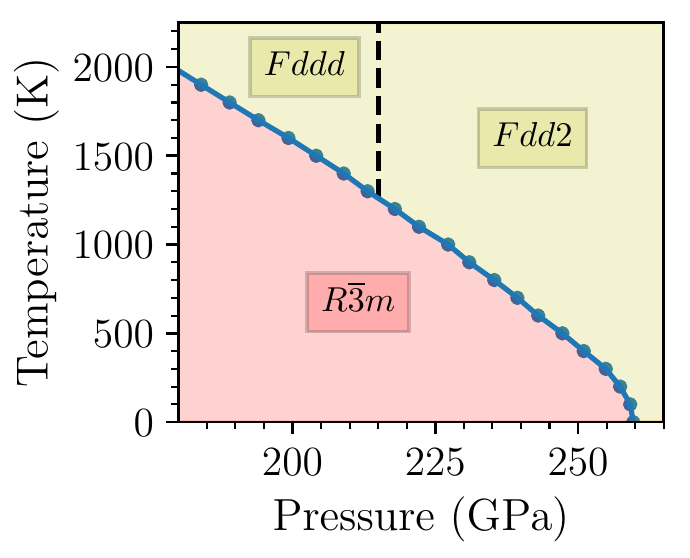}
    \hspace{0.35cm}
    \includegraphics[width=0.22\textwidth]{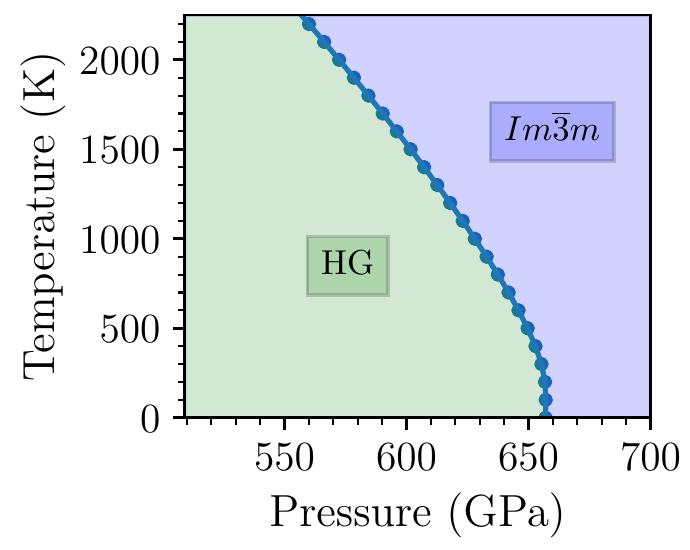}
    
    \caption{\textbf{Top:} Static-lattice phase diagram of S calculated with the PBE functional. `HG' denotes the host-guest phases, and for clarity, they have been plotted as a single curve. The inset shows the small enthalpy differences between the $Fdd2$ structure and HG phase. The simple-cubic $Pm\bar{3}m$ structure considered in previous studies \cite{Whaley_Baldwin,USPEX,sulfur_no_SC,Rudin,sulfur_and_Se_Rudin} is shown for comparison. \textbf{Bottom:} Phase diagrams with vibrational effects included (at the level of the harmonic approximation) in the vicinity of the $R\bar{3}m \rightarrow Fdd2$ transition (left) and HG $\rightarrow Im\bar{3}m$ transition (right). The black dashed line on the left diagram marks the boundary of the second order $Fdd2 \rightarrow Fddd$ phase transition.}
    
    \label{phase_diagram}
\end{figure}

\begin{figure*}
    \centering
    \begin{subfigure}[b]{0.3\textwidth}
        \hspace{-1cm}\includegraphics[width=0.2\textwidth]{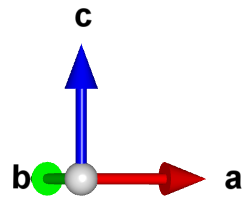}
        \includegraphics[width=0.75\textwidth]{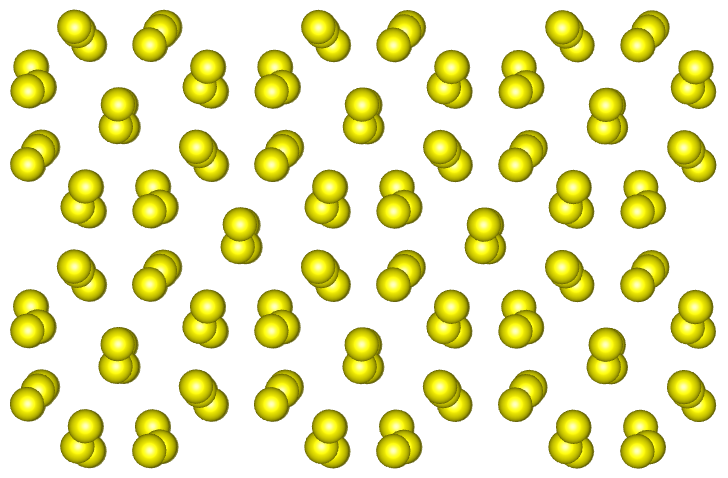} % 0.85
        \caption{Looking down the pseudo-guest chain axes. From top-to-bottom, the pseudo-guest chains are arranged in an AB stacking.\newline\newline}
        \end{subfigure}
        \hspace{0.5cm}
        \begin{subfigure}[b]{0.3\textwidth}
        \hspace{-1cm}\includegraphics[width=0.2\textwidth]{yellow_axes.png}
        \includegraphics[width=0.75\textwidth]{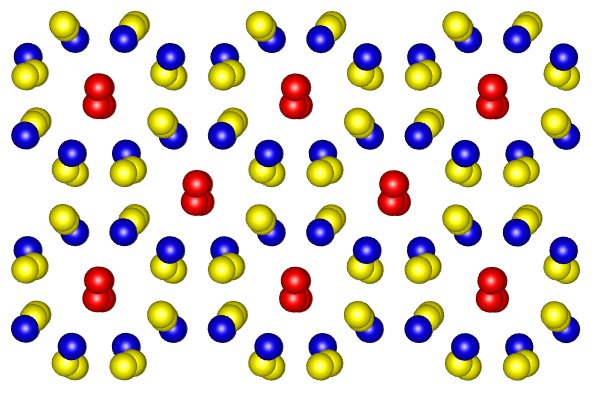}
        \caption{The same structure, but with the two different types of pseudo-guest chains artificially coloured red and blue, shown looking down the axis of the red guests. \newline}
        \end{subfigure}
        \hspace{0.5cm}
        \begin{subfigure}[b]{0.3\textwidth}
        \hspace{-0.8cm}\includegraphics[width=0.164\textwidth]{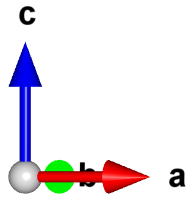}
        \includegraphics[width=0.75\textwidth]{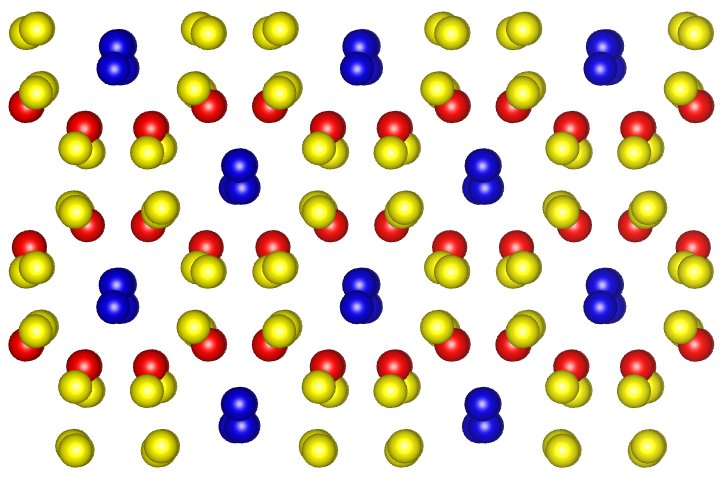}
        \caption{The same as the image on the immediate left, but rotated $45 \degree$ anticlockwise around an axis running top-to-bottom of the page (i.e. perpendicular to the chains). The result is an equivalent structure.}
        \end{subfigure}
        
    \caption{The commensurate $Fdd2$ structure, viewed down the `pseudo-guest' chains. The relation of this view to the lattice vectors of the conventional (64-atom) orthorhombic cell is shown in Fig. \ref{Fdd2_birds_eye_view}.}
    \label{Fdd2_detail}
    \vspace{0.25cm}
\end{figure*}

\begin{figure}[h!]
    \hspace{-0.70cm}
    \includegraphics[width=0.075\textwidth]{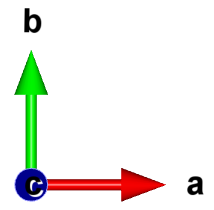}
    \hspace{+0.15cm}
    \includegraphics[width=0.30\textwidth]{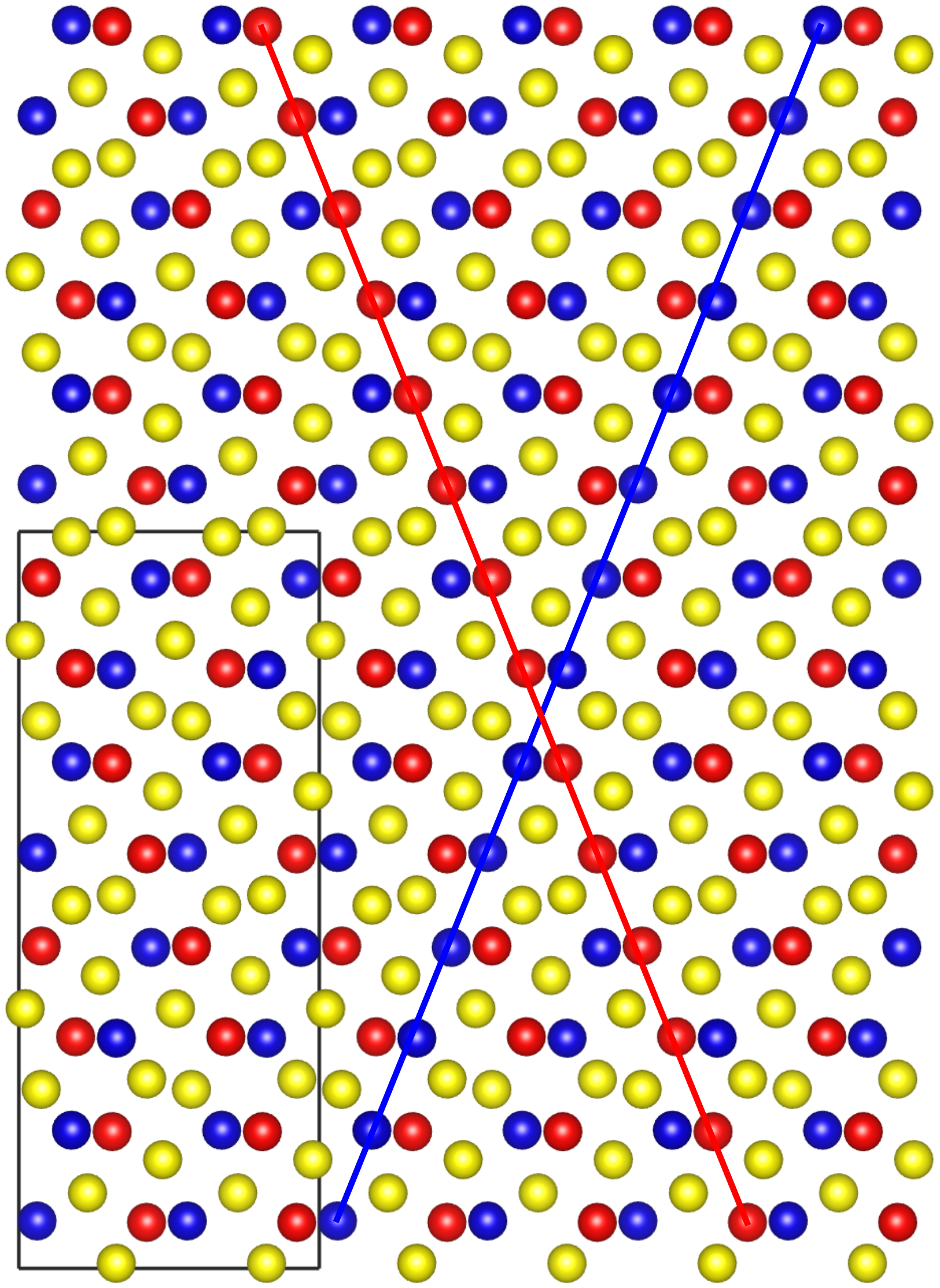}
    
    \vspace{0.325cm}
    
    \hspace{-0.70cm}
    \includegraphics[width=0.075\textwidth]{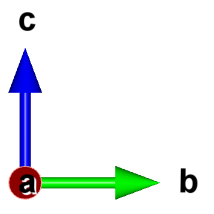}
    \hspace{+0.15cm}
    \includegraphics[width=0.32\textwidth]{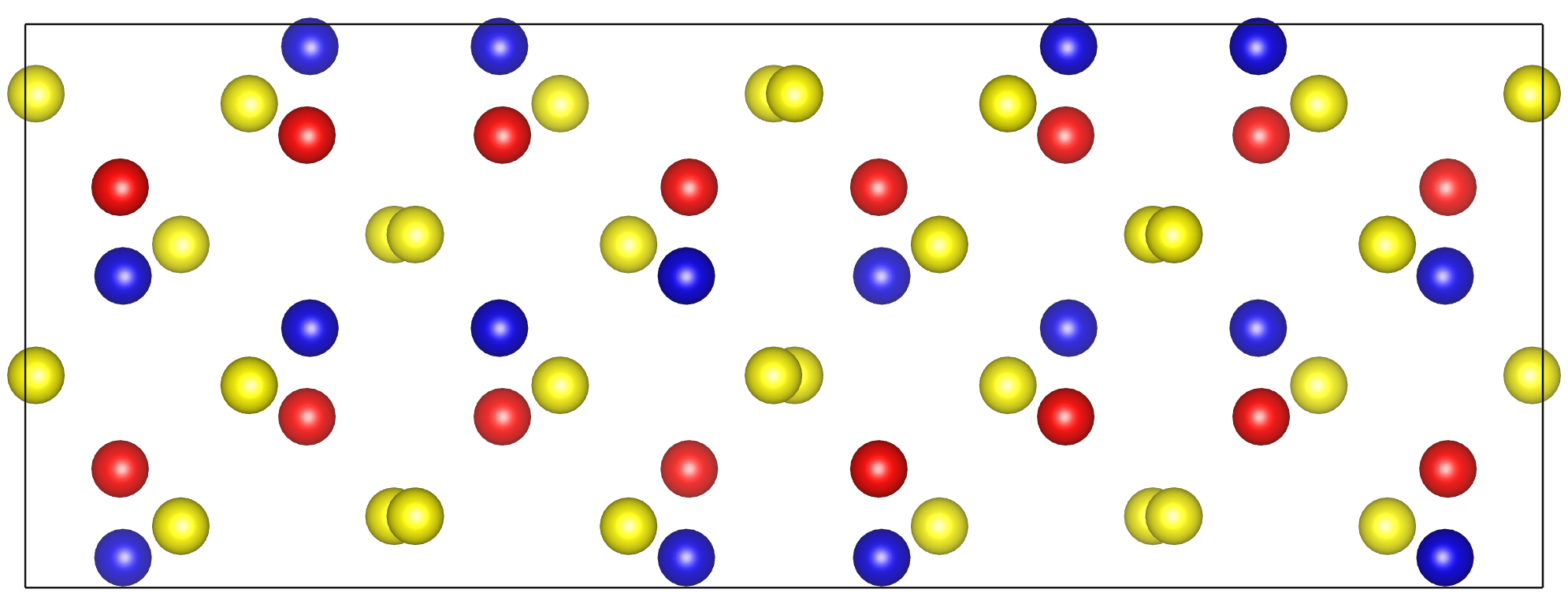}
    
    \vspace{0.325cm}
    
    \hspace{-0.70cm}
    \includegraphics[width=0.075\textwidth]{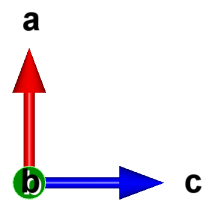}
    \hspace{+0.15cm}
    \includegraphics[width=0.125\textwidth]{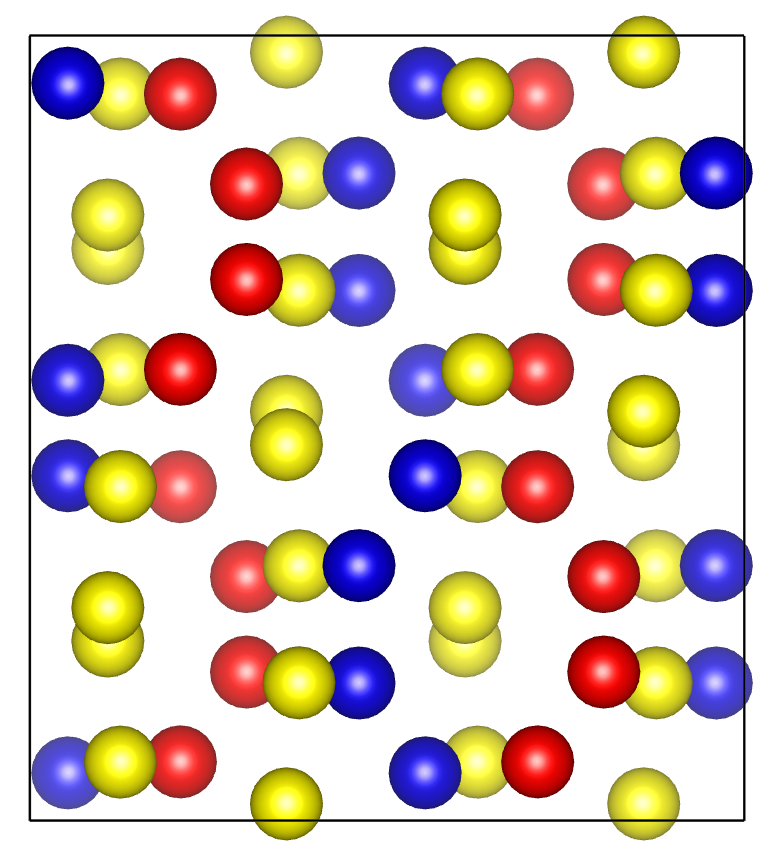}
    
    \caption{The $Fdd2$ structure, in the same colour scheme as Fig. \ref{Fdd2_detail}. Two example chains (like those shown in Fig. \ref{Fdd2_detail}) have been highlighted with red and blue lines (top figure only). The conventional (64-atom) orthorhombic unit cell is indicated in black.}
    \label{Fdd2_birds_eye_view}
\end{figure}

Fig. \ref{phase_diagram} shows our phase diagram of S calculated using the PBE functional. In addition to PBE, table \ref{transitions_table} shows the (static-lattice) transition pressures according to the LDA and PBEsol; the order of phase transitions is unaffected by the choice of functional. Using PBE, we determine that at $266$ GPa, S transitions from the $R\bar{3}m$ ($\beta$-Po) phase to a structure of $Fdd2$ symmetry with $16$ ($64$) atoms in its primitive (conventional) cell, which visually resembles a strongly distorted version of the Ba-IVa HG phase.

\addtolength{\tabcolsep}{+1.25pt} % increase column spacing
\begin{table}[!htbp]
\begin{tabular}{c c c c}
\hline\hline
Functional & $R\Bar{3}m \rightarrow$ $Fdd2$ & $Fdd2 \rightarrow$ HG & HG $\rightarrow Im\bar{3}m$ \\ \hline
LDA & $249$  & $378$ & $700$ \\ %\hline
PBE & $266$  & $386$ & $679$ \\ %\hline
PBEsol & $245$  & $366$ & $644$ \\ %\hline
\hline\hline
\end{tabular}
\caption{Static-lattice transition pressures in GPa according to LDA, PBE and PBEsol.}
\label{transitions_table}
\end{table}
\addtolength{\tabcolsep}{-1.25pt} % restore original column spacing 

This $Fdd2$ structure is detailed in Figs. \ref{Fdd2_detail} and \ref{Fdd2_birds_eye_view}. Whilst it clearly resembles the Ba-IVa HG structure, it has the property of being self-similar when rotated $45 \degree$ about an axis perpendicular to to the pseudo-guest chains. We use the term pseudo-guest, as this property implies that there is no longer a unique chain axis, and therefore it cannot be equivalent to the `standard' Ba-IVa type HG structure (although, as we will show, S does eventually adopt the Ba-IVa HG structure at higher pressures). The equivalence through a $45 \degree$ rotation means that there are two nonintersecting pseudo-guest chains, running in different directions in offset planes, as detailed in Fig. \ref{Fdd2_birds_eye_view}. Each pseudo-guest chain forms part of the host structure when looking down the other chain's axis (and vice versa), and thus it is no longer possible to decompose the structure into two distinct host and guest lattices (see Figs. \ref{Fdd2_detail}(b) and \ref{Fdd2_detail}(c)). Whilst the pseudo-guest chains in the same layer are in alignment, those in the layer below are shifted by half of the intra-chain atomic separation, thus giving the chains an AB ordering when viewed from the side.
\par
We tested the stability of this structure for a few other elements that possess a high-pressure HG phase (such as Al and As), but in each case it was never the lowest-enthalpy phase. The `two-chain' motif is reminiscent of the structure of the ternary incommensurate compound Hg$_{3-\delta}$AsF$_6$ \cite{incommensurate_mercury_compound_axe,incommensurate_mercury_compound_buiting,incommensurate_mercury_compound_degroot}, in which chains of mercury atoms, running at $90 \degree$ to one another, sit within an AsF$_6$ framework. However, unlike this compound, and the incommensurate Ba-IVa phase, our calculations suggest that the $Fdd2$ structure is in fact a commensurate crystal and is not \textit{quasi}-periodic. To demonstrate that this is the case, we first show that at higher pressures, S \textit{does} exhibit a truly incommensurate Ba-IVa type HG structure, and then use these findings to show that the $Fdd2$ phase cannot be incommensurate.
\par
The Ba-IVa HG structure, which occurs (with minor variations) in multiple elements across the periodic table \cite{HG_Barium,HG_alkali_metals,Rb-IV,HG_Aluminium,HG_group_V,HG_Bismuth_SC,Haussermann_2002,HG_Strontium,HG_scandium,HG_Antimony,HG_Calcium}, comprises a tetragonal `host' structure of $I4/mcm$ symmetry with $8$ atoms in the (conventional) unit cell, and one of two possible coexisting `guest' structures; either a $C$-face centered tetragonal structure, or an $M$-face centered monoclinic structure, with atoms at $(0,0,0)$ and $(1/2,1/2,0)$ in both cases. The host and guest structures are incommensurate along their $c$-axes with an ideal ratio $\gamma_0$ that is remarkably close to $4/3$. In some cases, such as Bi-III and Sb-II \cite{modulated_Bi_and_Sb,Kartoon_Bismuth} or Ca-VII \cite{chain_ordered_calcium}, the positions of both the host and guest atoms in the HG structure are distorted/modulated relative to their `ideal' HG positions, giving both the channels and chains a `wiggly' or worm-like appearance.
\par
We find that above $386$ GPa, a Ba-IVa type HG approximant of $C2/c$ symmetry ($\gamma=4/3$), with $16$ (32) atoms in its primitive (conventional) cell, becomes lower in enthalpy than the $Fdd2$ structure. Similar to the $Fdd2$ structure, the atoms in the $C2/c$ approximant are distorted from their ideal positions; however, a unique chain axis for the guest atoms now exists, and thus the $C2/c$ approximant is of the `standard' Ba-IVa type. Below $386$ GPa, the $C2/c$ approximant is dynamically unstable, and it develops an imaginary optical $\Gamma$-point phonon mode that leads to the $Fdd2$ structure. The structural and enthalpic differences between the $C2/c$ approximant and the $Fdd2$ structure are small, and over the stability range of the $Fdd2$, they differ in enthalpy by at most $1.5$ meV per atom (see inset in Fig. \ref{phase_diagram}).

\begin{figure}[h!]
    \hspace{-1.5cm}
    \includegraphics[width=0.495\textwidth]{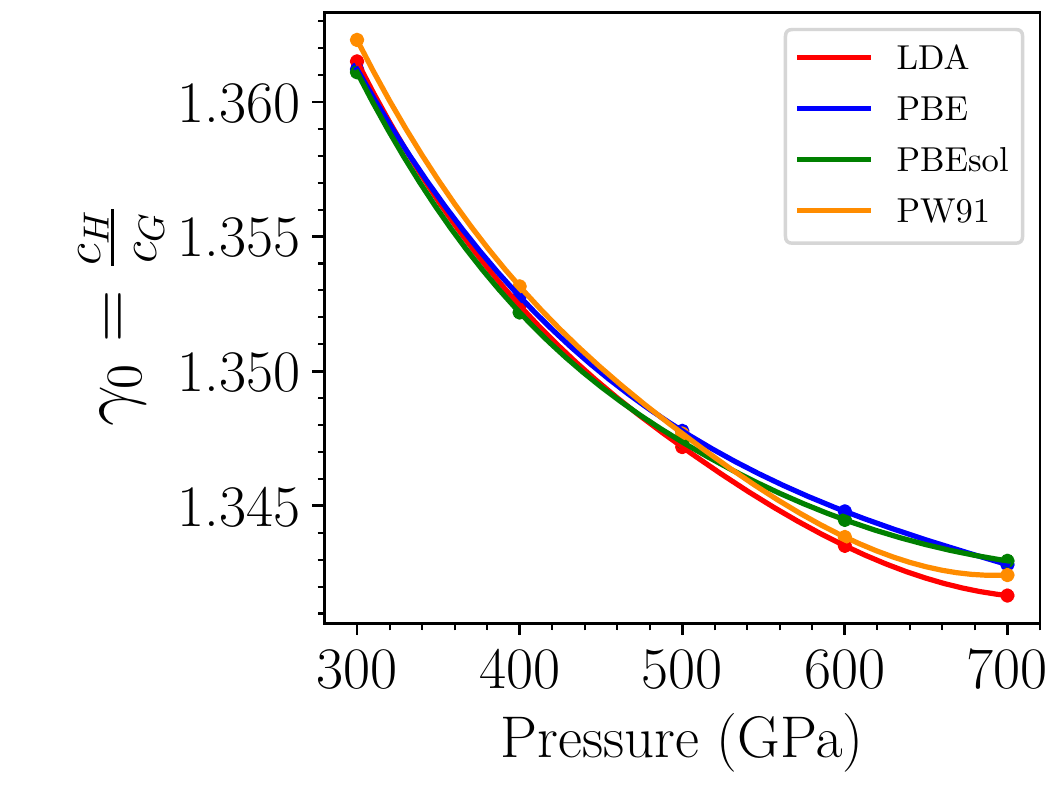}
    \caption{Dependence of the ideal incommensurate ratio with pressure. Data was taken in $100$ GPa intervals and fitted to a cubic spline.}
    \label{ratio_with_pressure}
\end{figure}

In order to calculate the ideal incommensurate host-guest ratio $\gamma_0$ at which the enthalpy is a minimum, we fitted the enthalpy of various commensurate approximants to a quadratic curve (see Supplemental Fig. 3 \cite{supp}), from which the ideal incommensurate ratio can be read off as the minimum. Fig. \ref{ratio_with_pressure} shows that $\gamma_0$ slowly decreases monotonically with increasing pressure, and that the choice of exchange-correlation functional does not affect the results significantly. This slow monotonic decrease in $\gamma_0$ with pressure is also seen in the HG phase of sodium \cite{HG_alkali_metals}, but is opposite to the cases of aluminium \cite{HG_Aluminium} and bismuth \cite{Kartoon_Bismuth}. In potassium and rubidium $\gamma_0$ initially decreases with pressure, before this behaviour reverses and $\gamma_0$ instead increases with pressure \cite{HG_alkali_metals}.

The procedure used to calculate $\gamma_0(p)$ allows us to propose that the $Fdd2$ structure is \textit{not} a commensurate approximant to an incommensurate phase, despite its visual similarity to the Ba-IVa structure; of all the approximants used to construct our $H(\gamma)$ quadratic fits ($9$ in total), only the $16$-atom approximant collapsed into the two-chain pattern at low pressures; approximants with other $\gamma$ values simply remained in the `single chain' Ba-IVa type structure. This suggests that the $Fdd2$ structure does not belong to a family of HG approximants, but rather is a $16$-atom structure in its own right. Additionally, we re-emphasise the fact that since one cannot decompose the $Fdd2$ into a unique host and a unique guest structure, it is not possible to even define a $\gamma$ value for the $Fdd2$.
\par
Relative to the more densely packed bcc ($Im\bar{3}m$) and trigonal ($R\bar{3}m$) phases, the HG structures are stabilised by a relatively large reduction in the total electronic energy. At 500 GPa, despite being ${\sim} 0.719\%$ larger in volume, the HG phase has a total electronic energy $0.190$ eV per atom lower than the bcc structure, which is more than enough to offset the increased $pV$ term and make it a favourable structure. The idea that S becomes more open-packed at these pressures had previously been considered by Rudin \textit{et. al.} \cite{Rudin,sulfur_and_Se_Rudin} who proposed an $R\bar{3}m \rightarrow$ simple-cubic ($Pm\bar{3}m$) transition, but neither that work, nor two previous high-pressure structure searching studies \cite{USPEX,Whaley_Baldwin}, located the $Fdd2$ and HG phases. We further note the significance of our findings on the construction of convex hulls for sulfur compounds (e.g. H$_2$S and H$_3$S) at these pressures; the HG phase of S is lower in enthalpy than the previously-assumed ground state ($R\bar{3}m$) by 11 meV per atom at 300 GPa, and up to 57 meV at 500 GPa.
\par
The inclusion of nuclear vibrational effects does not change the phase transition sequence, and at low temperatures, vibrational Zero-Point Energies (ZPEs) cause only small deviations from the static-lattice pressure boundaries (see Fig. \ref{phase_diagram}). For example, at the $R\bar{3}m \rightarrow Fdd2$ transition, we find the ZPE of the $R\bar{3}m$ phase to be  $1.4$ meV per atom higher than that of the $Fdd2$ structure, which lowers the transition pressure by ${\sim}4$ GPa from the static-lattice value in Table \ref{transitions_table}. The ZPEs of the $Fdd2$ structure and the $C2/c$ approximant are identical to within our convergence limits. At the HG $\rightarrow Im\bar{3}m$ transition, we find the HG ZPE to be higher than that of the $Im\bar{3}m$ phase by $4.8$ meV per atom, and the transition pressure reduces by ${\sim}22$ GPa compared to the value in Table \ref{transitions_table}. At higher temperatures ($\gtrsim 300$ K), vibrational effects reduce the transition pressures of both the $R\bar{3}m \rightarrow Fdd2$ and HG $\rightarrow Im\bar{3}m$ transitions. The lowering of the transition pressure for the $R\bar{3}m \rightarrow Fdd2$ transition is particularly interesting for two reasons. Firstly, it provides an experimental route to access the $Fdd2$ structure at slightly lower pressures, somewhat reducing the difficulty of any diamond anvil experiments. Secondly, it allows for observation of a new structure of $Fddd$ symmetry, also with 16 (64) atoms in its primitive (conventional) unit cell. This $Fddd$ structure is derived from the $Fdd2$ structure, which continuously develops an imaginary $\Gamma$-point phonon mode below $215$ GPa, pushing it into the $Fddd$ symmetry - the resulting transition between the phases is thus second order. We note that this $Fddd$ structure is distinct from the well-known ambient-pressure orthorhombic ground state of sulfur (also of $Fddd$ symmetry) which comprises S8 rings. Our calculations show that at temperatures above $\sim 1260$ K, the $Fddd$ phase should become the ground state at $215$ GPa (again, see Fig. \ref{phase_diagram}). We note that a recent publication \cite{sulfur_melting_curve} on the high-pressure melting curve of S achieved temperatures of $\sim 1700$ K at $51$ GPa using a laser-heated diamond anvil cell (LHDAC), and extrapolation of their data using a Simons-Glatzel fit \cite{simons-glatzel} suggests that the melting temperature of S should be at least $\sim 3850$ K at $200$ GPa.

\subsection{\label{chain_ordering}Chain Ordering}

\begin{figure}[h!]
    
    \begin{subfigure}[t]{0.2\textwidth}
        \includegraphics[width=\linewidth]{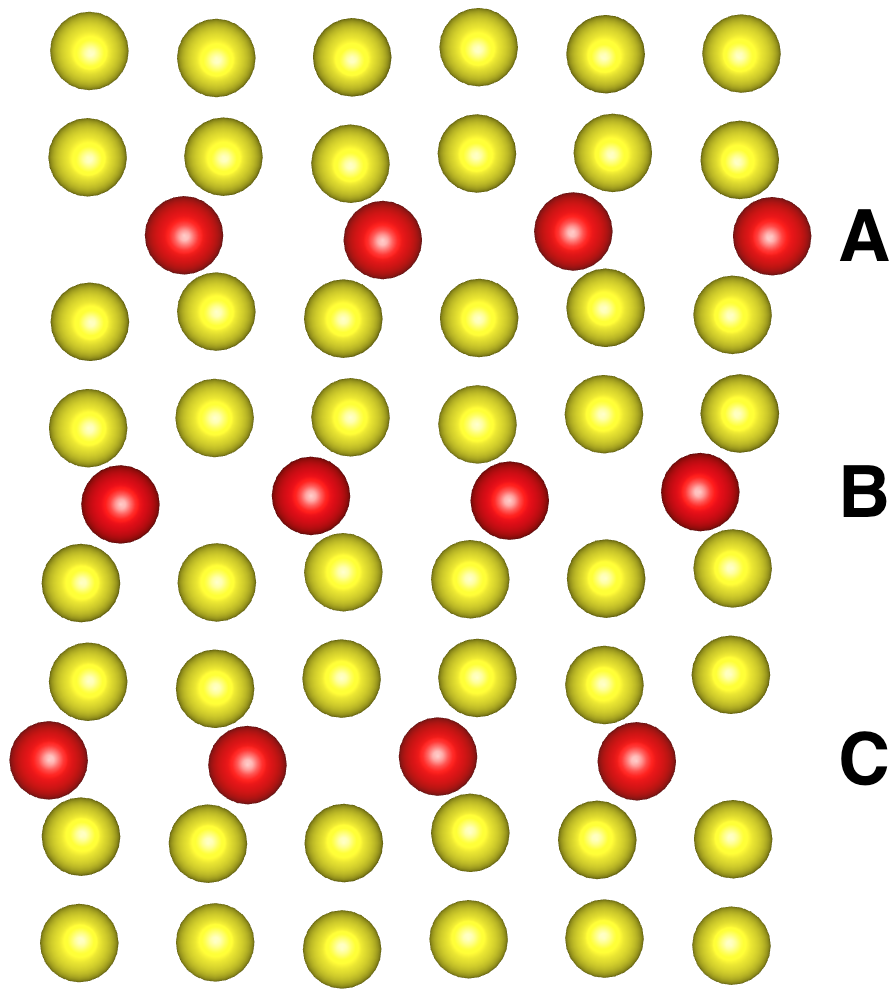}
    \end{subfigure}
    \hfill
    \begin{subfigure}[t]{0.21\textwidth}
        \hspace{-1.1cm}
        \includegraphics[width=\linewidth]{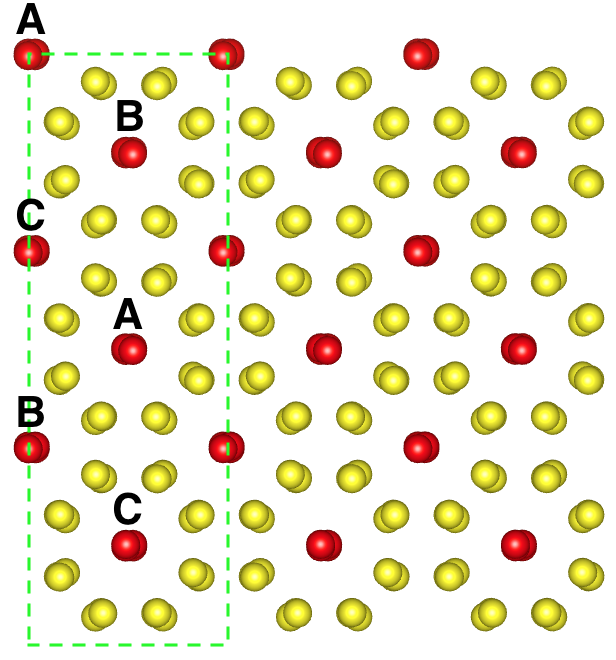}
    \end{subfigure}
    
    \vspace{0.521cm}
    
    \begin{subfigure}[t]{0.2\textwidth}
        \includegraphics[width=\linewidth]{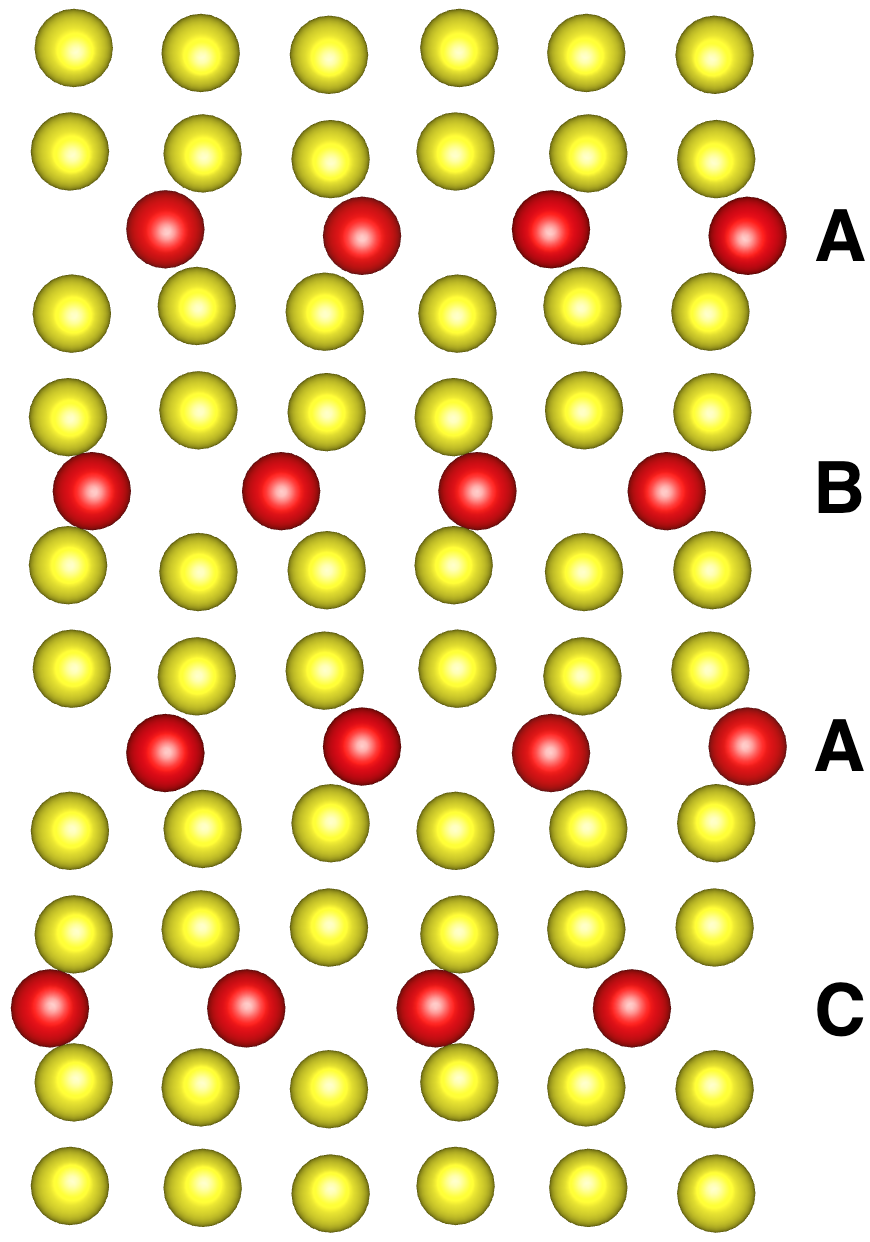}
    \end{subfigure}
    \hfill
    \begin{subfigure}[t]{0.21\textwidth}
        \hspace{-1.1cm}
        \includegraphics[width=\linewidth]{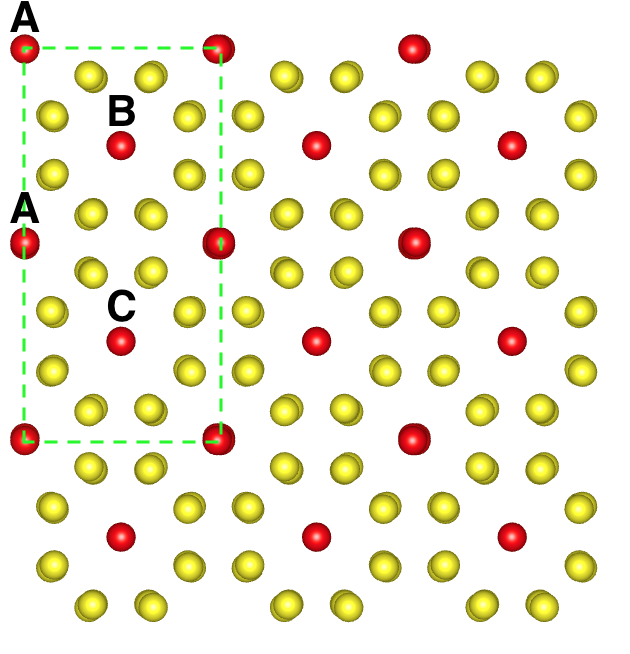}
    \end{subfigure}
    
    \vspace{0.521cm}
    
    \begin{subfigure}[t]{0.2\textwidth}
        \includegraphics[width=\linewidth]{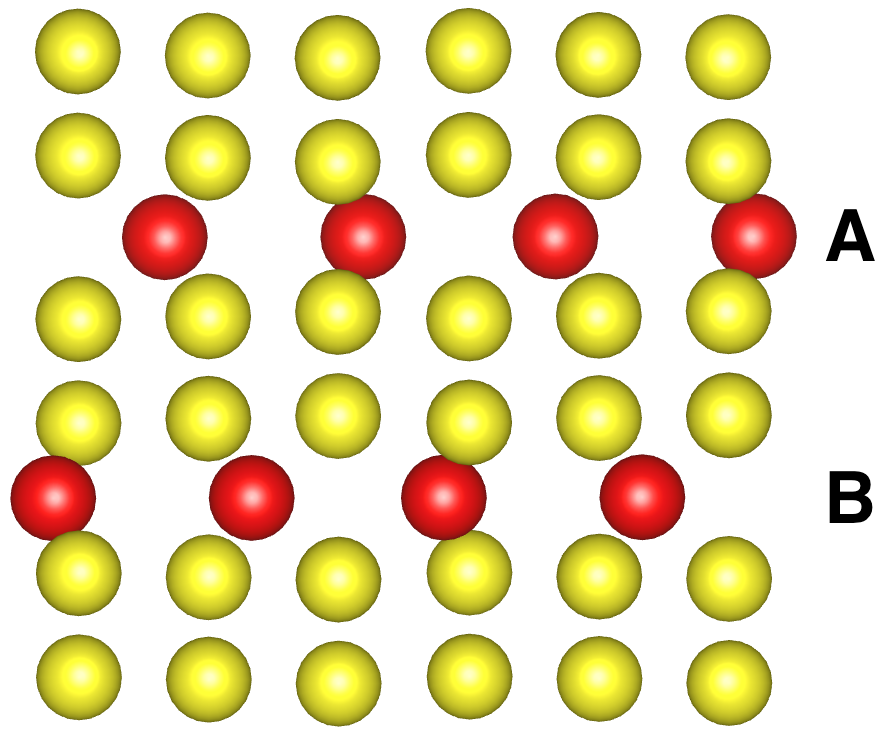}
    \end{subfigure}
    \hfill
    \begin{subfigure}[t]{0.22\textwidth}
        \hspace{-1.05cm}
        \includegraphics[width=\linewidth]{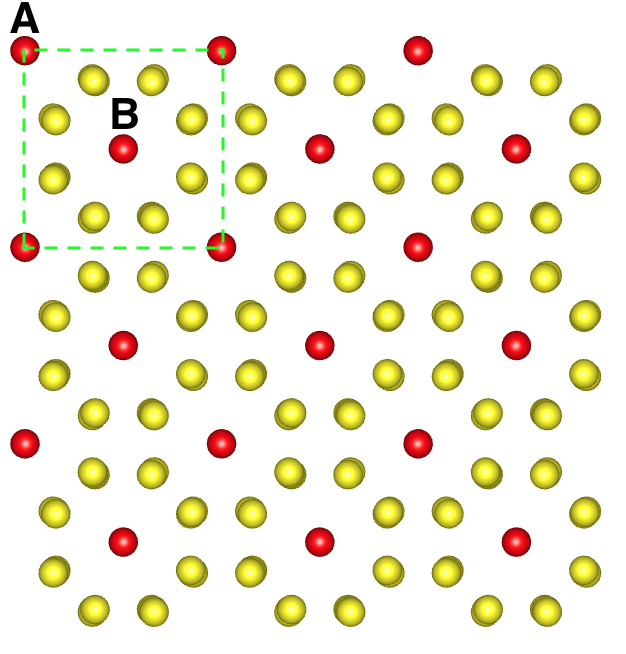}
    \end{subfigure}
    
    \vspace{0.31cm}
    
    \hspace{-0.5cm}\includegraphics[width=0.082\textwidth]{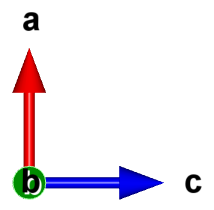}
    \hspace{+4.5cm}\includegraphics[width=0.0794\textwidth]{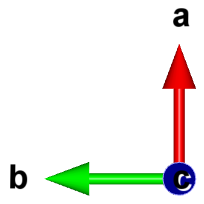}
    
    \vspace{0.298cm}
    
    \hspace{-0.6cm}\includegraphics[width=0.475\textwidth]{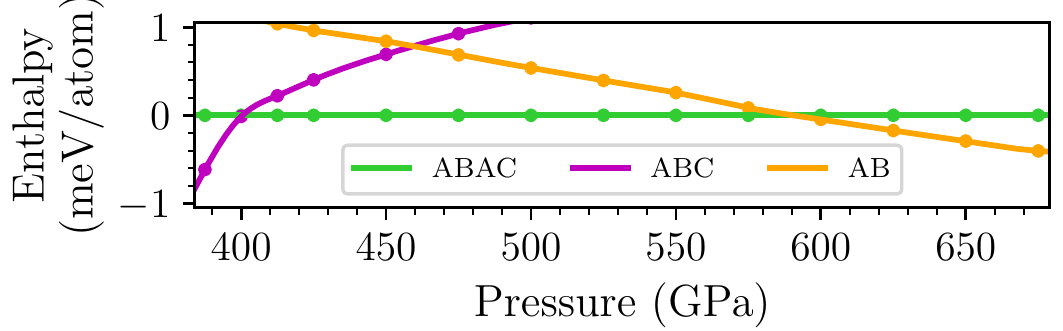}
    
    \caption{\textbf{Top:} ABC $(0,\frac{1}{3},\frac{2}{3})$ stacking of the $16$-atom $C2/c$ approximant. \textbf{Middle:} ABAC $(0,\frac{1}{3},0,\frac{2}{3})$ stacking of the $64$-atom $Pcca$ approximant. \textbf{Bottom:} AB $(0,\frac{2}{3})$ stacking of the $32$-atom $P4/ncc$ approximant. The relative stability of the stackings is shown beneath the structures; the $Fdd2$ and bcc phases become stable immediately off to the left and right of the plot, respectively.}
    \label{stacking_comparisons}
\end{figure}

After establishing the $Fdd2$ structure and $C2/c$ approximant as competitive phases, we conducted auxiliary searches with peturbed supercells of the $C2/c$ approximant containing $32$, $64$, $128$ and $288$ atoms. These searches produced two new HG approximants of $P4/ncc$ symmetry ($32$ atoms) and $Pcca$ symmetry ($64$ atoms) with slightly lower enthalpies. These approximants have the same commensurate ratio as the $C2/c$ approximant ($\gamma = 4/3$) and feature the same host structure, but possess different guest structures. The bottom of Fig. \ref{stacking_comparisons} shows the relative enthalpies of these different host-guest approximants. It can be seen that the $16$-atom $C2/c$ approximant is actually favoured only in a narrow pressure range ($384$-$400$ GPa), before the $64$-atom $Pcca$ approximant becomes favourable between $400$-$590$ GPa. The $32$-atom $P4/ncc$ approximant is then the ground state between $590$-$679$ GPa, with the bcc $Im\bar{3}m$ phase becoming stable thereafter.

Each of the $C2/c$, $Pcca$ and $P4/ncc$ approximants represent a different ordering of the guest atom chains, in which adjacent chains are displaced by a different amount parallel to the direction of the chain axes (i.e. have different heights in the $z$ direction). We found that this displacement was always either $1/3$ or $2/3$ of the guest lattice $c$ axis length ($c_{G}$). Fig. \ref{stacking_comparisons} shows the different orderings as viewed along, and perpendicular to, the axes of the chains. In each case, the chains can clearly be described as having been stacked in a certain order, with the $C2/c$, $Pcca$ and $P4/ncc$ approximants possessing a stacking order of ABC, ABAC and AB respectively. Fig. \ref{stacking_comparisons} also shows that the aforementioned distortions/modulations of the host and guest structures are slightly different in each case, being largest for the $C2/c$ structure and smallest for the $P4/ncc$ structure. This suggests a coupling between the chain orderings and the distortions; an observation that we shall return to later.

The lowest-enthalpy stacking ordering at a given pressure is a trade off between minimisation of the total electronic energy and the $pV$ term. The AB ordering of the $P4/ncc$ approximant stacks most efficiently to give the lowest $pV$ term at all pressures, but has the largest total electronic energy of all approximants. This situation is reversed in the ABC stacking of the $C2/c$ approximant, which has the lowest electronic energy but the largest $pV$ term. The $Pcca$ approximant of ABAC stacking is a compromise between these two motifs and explains why it is favoured over a comparatively large pressure range. At lower pressures ($<400$ GPa), the $pV$ term is less important than the electronic term, and the $C2/c$ approximant becomes the ground state. At much larger pressures ($>590$ GPa), the $pV$ term becomes dominant, and the $P4/ncc$ approximant is the lowest enthalpy phase.
\par
Chain orderings that differ from the `ideal' all-aligned case have been discovered in other host-guest phases. The AB-ordered HG phase (our $P4/ncc$ approximant) that S adopts above $590$ GPa is also the ground state of the group-V elements As, Sb and Bi at much lower pressures \cite{HG_group_V,Haussermann_2002}. The Ca-VII phase of Calcium \cite{chain_ordered_calcium} is well-described by a $128$-atom approximant in which the guest chains have different $z$-heights in both the $a$ and $b$ directions (as opposed to S, where the $z$-heights change only along the $a$ direction). In the case of Ba-IVc, the structure adopted by Barium at pressures a few GPa higher than the Ba-IVa phase, the chain displacements are arranged in such a way as to produce an extremely complicated, mesoscopic-scale, interlocking S-shape patterning, which is well-represented by a $768$-atom cell \cite{complicated_barium}.
\par
Whilst the enthalpy differences between the ABC, ABAC and AB stackings are mostly ${\sim}1$ meV per atom, we note that the `ideal' stacking (of $I4/mcm$ symmetry), in which all guest chains are perfectly aligned (i.e., AAA...), is higher in enthalpy at all pressures by at least $20$ meV per atom and up to $45$ meV, indicating that perfect chain alignment is strongly disfavoured in S. This raises the question of why any stacking that differs from the all-aligned case has such a drastically reduced enthalpy. Surprisingly, we find that explicitly offsetting the chains causes a only a very small enthalpy reduction in itself. Instead, we show that the answer lies in the distortions/modulations of the host and guest atoms, which cannot occur unless the chains are first offset.
\par
\begin{figure}[!htbp]
    
    \hspace{-0.69cm}
    \includegraphics[width=0.49\textwidth]{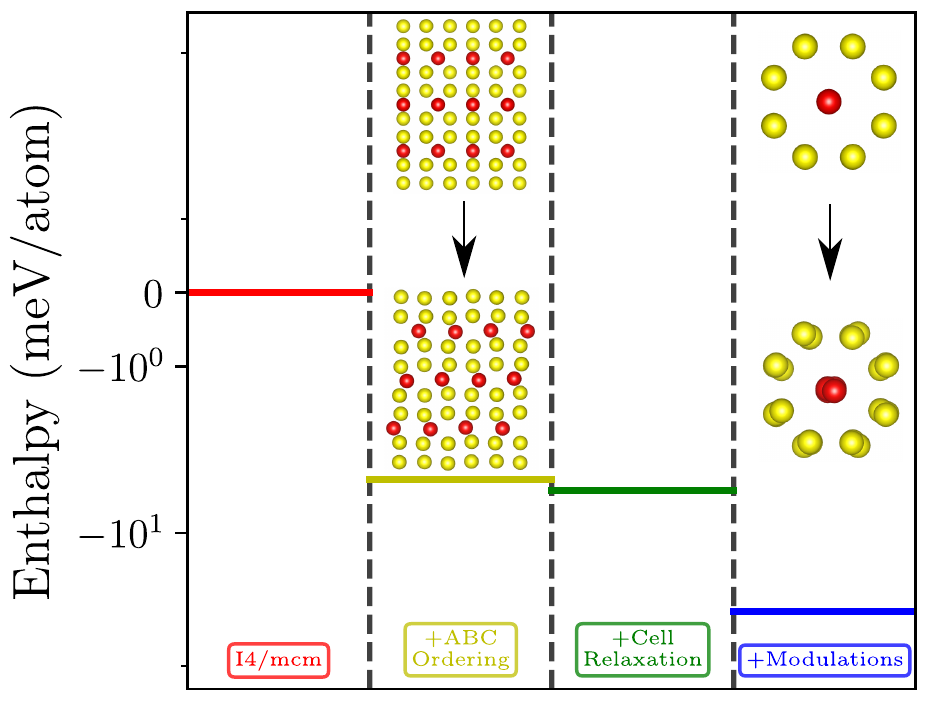}

    \hspace{-0.35cm}
    \includegraphics[width=0.49\textwidth]{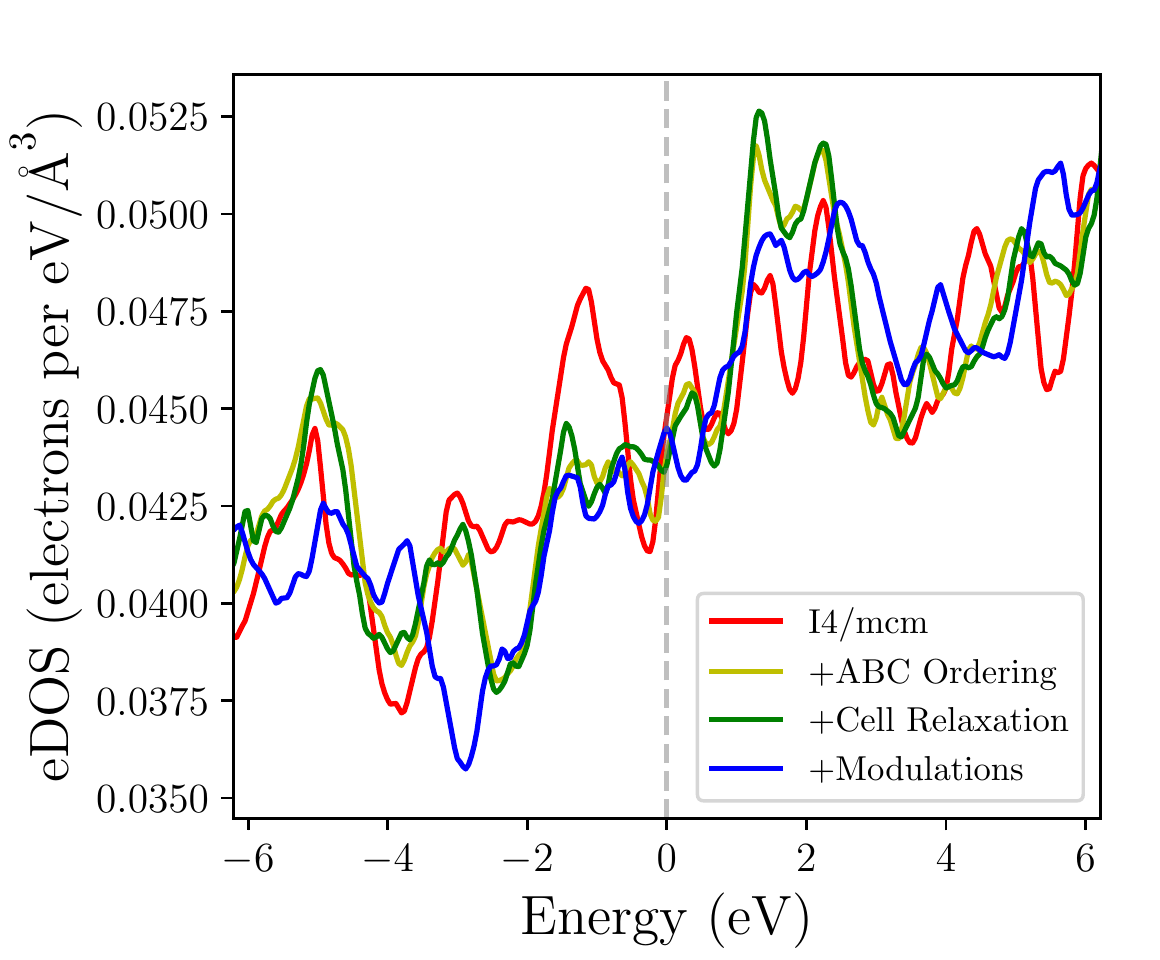}

    \caption{\textbf{Top:} Schematic decomposing the $39$ meV enthalpy reduction that occurs going from the all-aligned approximant to the ABC ordered approximant at $387$ GPa. The enthalpy axis is logarithmic. \textbf{Bottom:} The corresponding changes in the eDOS. The Fermi level is set at $0$ eV.}
    \label{enthalpy_evolution}
\end{figure}

Fig. \ref{enthalpy_evolution} details the situation for the ABC case at $387$ GPa, although the ABAC and AB cases are analogous. Starting from the all-aligned $I4/mcm$ approximant, we manually offset the chains such as to have an ABC ordering, keeping all else fixed. This gives rise to an enthalpy reduction of $4$ meV per atom, which can be attributed to the opening of several directional band gaps at the Fermi level (see Supplemental Fig. 5 \cite{supp}). We subsequently relax the cell vectors (with the atomic positions fixed), which gives rise to a further enhalpy reduction of $1$ meV. Finally, we add in the modulation of the host and guest atoms, which further reduces the enthalpy by $34$ meV; an order of magnitude larger than the reductions caused by the chain offsets and the cell vector relaxation. The bottom of Fig. \ref{enthalpy_evolution} shows that these energetic reductions arise as a result of shifting eDOS weight to lower energies, whilst little changes at the Fermi level itself.

Since the modulations provide the greatest contribution by far to the total enthalpy reduction, it is natural to ask why offsetting the guest chains - a change that causes only a very small enthalpy reduction in itself - is necessary to see these modulations appear. When the guest chains are offset, the host atoms are exposed to an asymmetric environment which creates forces that cause the host atoms to distort, which in turn distorts the guest atoms. If the chains are all-aligned, then the environment on either side of the host atoms is the same; symmetry implies that these forces do not arise, and there are therefore no distortions. In the ABC and ABAC stackings, both the host and guest atoms are distorted in all of the $x$, $y$ and $z$ directions (see Fig. \ref{stacking_comparisons}). However, in the case of AB stacking, symmetry dictates that the guests are displaced only in the $z$ direction. These $z$ distortions longitudinally pinch together pairs of guest atoms along the chain axes and dimerize them such that the covalent bonds alternate in a short-long-short-long pattern, with the short bond ${\sim}3.2 \%$ smaller than the long bond. This 1D dimerization is referred to as \textit{quasi}-pairing in the literature \cite{modulated_Bi_and_Sb}, and the same behaviour has recently been predicted to occur in the CDW phase of sulfur at lower pressures \cite{Whaley_Baldwin}. Detailed experimental \cite{modulated_Bi_and_Sb} and theoretical \cite{Kartoon_Bismuth} analyses of both the host and guest distortions in the Bi-III structure confirm that the modulations play an indispensable role in the stability of HG phases.
\par
In addition to the structures found in our structure search, we calculated the enthalpy of various complex structures that appear in the vicinity of HG phases in other elements, such as oC52 in rubidium \cite{oC52_rubidium}, oC84 in caesium \cite{oC84_caesium}, oP8 in potassium \cite{oP8_potassium}, and A7 in the group-V elements \cite{Haussermann_2002}. These structures were found to be at least $0.1$ eV per atom above the ground state at all pressures considered here. We also constructed some $\gamma=4/3$ approximants with longer-period chain stackings such as ABAAC, ABCAB, ABCACB and permutations thereof, as well as configurations where the chain heights changed in the $b$- direction also (see Supplemental Material for a table of all the stackings that were tested \cite{supp}). Most of these structures collapsed to the 64-atom $Pcca$ ABAC stacking, which seems to have a particularly large basin of attraction in the configuration space of chain orderings. Those that did not collapse to the ABAC stacking simply had a higher enthalpy at all pressures, although we note that an ABCACB stacking was particularly close to becoming the ground state ($+0.2$ meV above the ABAC stacking at $550$ GPa). We further considered the effect of spin polarisation in our calculations, and seeded our structures with varying ferromagnetic and anti-ferromagnetic orderings, but in all cases, the density relaxed to a spin-unpolarised configuration.

\subsection{\label{charge_density}Charge Density}

\begin{figure}[!htbp]
    \includegraphics[width=0.368\textwidth]{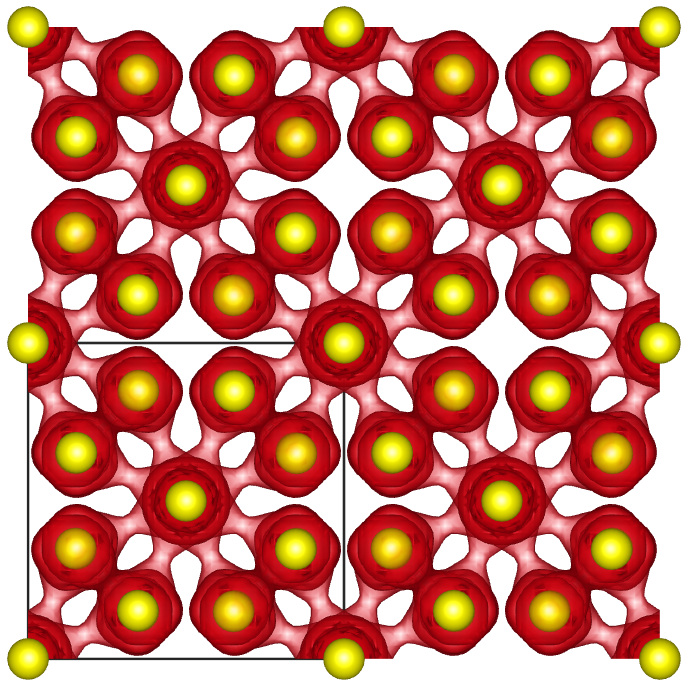}
    
    \vspace{0.125cm}
    
    \includegraphics[width=0.368\textwidth]{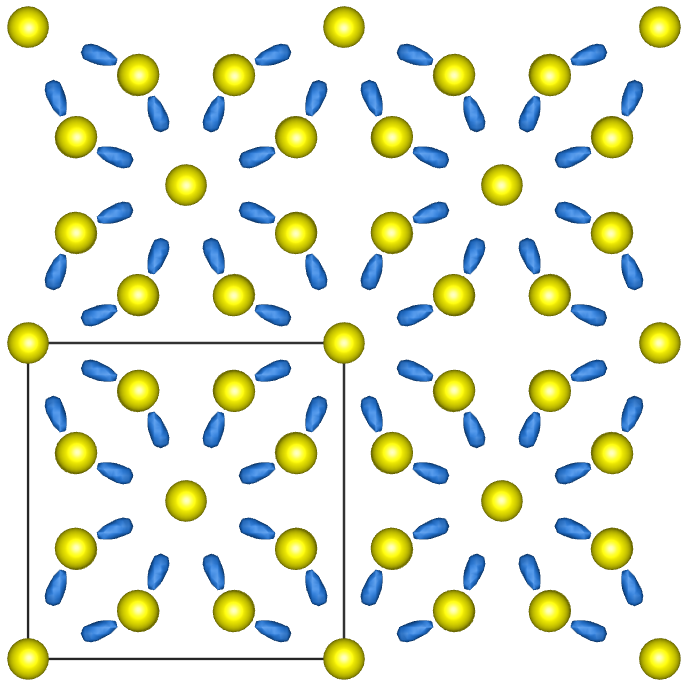}
    
    \vspace{0.15cm}
    
    \includegraphics[width=0.46\textwidth]{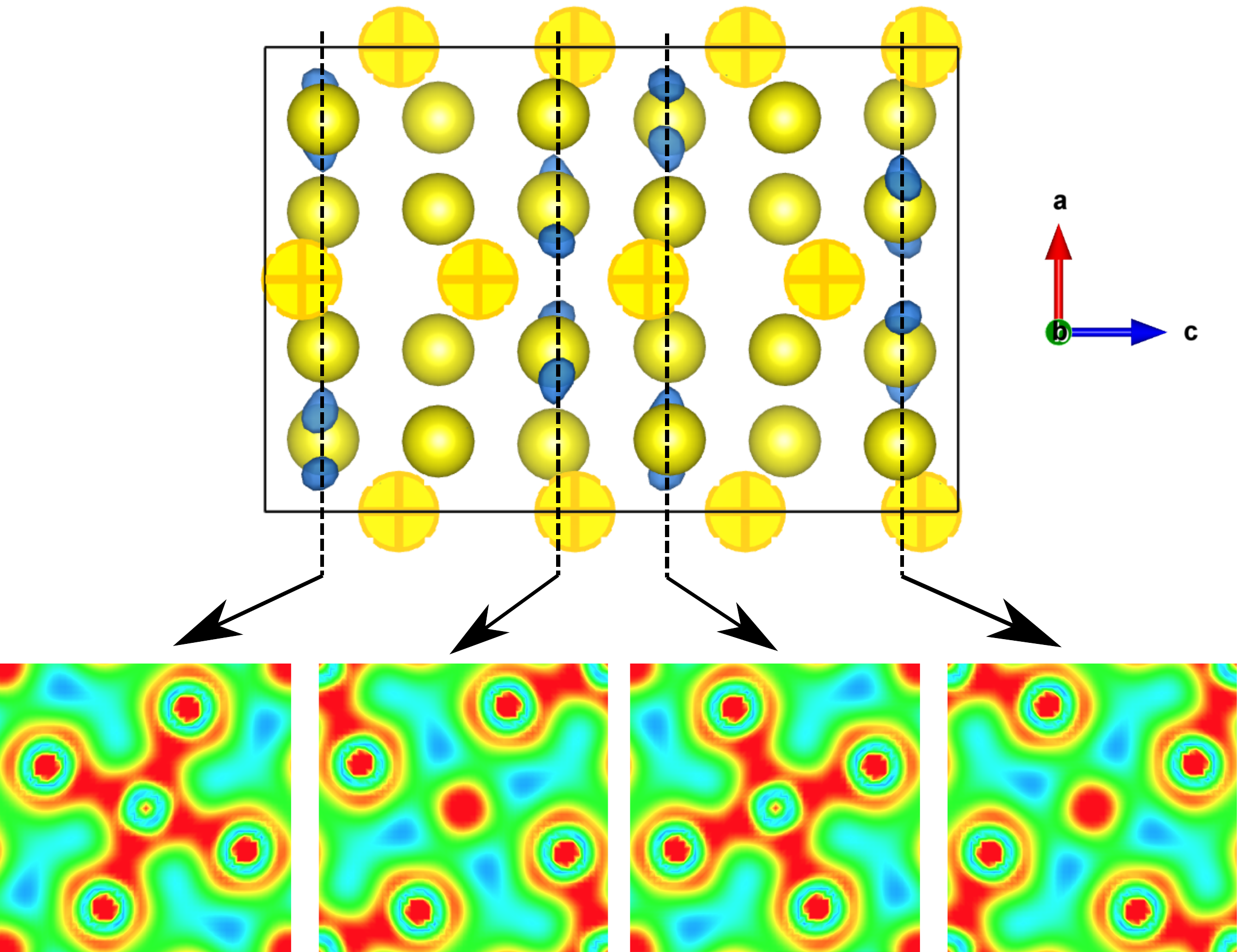}
    
    \caption{Charge density in HG sulfur. \textbf{Top:} Isosurface at $35\%$ of maximum, displaying clear covalent bonds. \textbf{Middle:} Isosurface at $0.3\%$ of maximum, showing the charge-depleted `voids'. \textbf{Bottom:} 2D slices of the charge density taken perpendicular to the guest chains for the AB ordering (guest atoms highlighted). The ABC and ABAC stackings give similar charge distributions, but the voids have different $z$ coordinates in each case (as they are tied to the guest structure).}
    \label{voids_and_covalents}
\end{figure}

We compare charge density isosurfaces of HG sulfur in Fig. \ref{voids_and_covalents} at moderate ($35\%$ of maximum) and very small ($0.3\%$ of maximum) values, alongside several 2D slices of the charge density taken perpendicular to the guest chains. The charge density of S is peaked near atomic sites and along lines connecting neighbouring atoms, which is representative of a covalently bonded solid. S exhibits no significant interstitial localisation of charge and therefore does not constitute a conventional electride, unlike the HG phases of the alkali metals \cite{HG_alkali_metals} or aluminium \cite{HG_Aluminium}. However, S displays curious behaviour when studied at very small isosurface values, which correspond to charge-depleted regions. The middle of Fig. \ref{voids_and_covalents} shows that these charge-depleted regions form well localised lobes, which we refer to as charge `voids', and co-incide with minima in the Electron Localisation Function (ELF) \cite{ELF_paper}. These voids form a structure equivalent to the host lattice of the alkali metal HG phases \cite{HG_alkali_metals}, with one host unit situated at the midpoint of every `short' intra-chain covalent bond. The bottom of Fig. \ref{voids_and_covalents} shows the 2D charge density taken at the midpoints of these `short' bonds for the AB case; when these slices are superimposed, the full set of voids (as seen in the middle of Fig. \ref{voids_and_covalents}) is recovered.

Remarkably, these voids are the exact locations of charge density \textit{maxima} in the HG phase of aluminium \cite{HG_Aluminium}. The duality between the charge densities of S and Al can be explained in terms of the intra-chain bonding between the guest atoms, which is present in S but absent in Al. In S, the short intra-chain covalent bonds contain significantly more charge than the weaker long bonds; this excess charge is drawn from the charge voids, and explains why there is precisely one set of voids per short bond.
\par
We note that in S there is no opening of a pseudogap in the electronic Density of States (eDOS) at the Fermi level (see Fig. \ref{enthalpy_evolution} and Supplemental Fig. 6 \cite{supp}). This is in contrast to the behaviour of high-pressure electrides \cite{HG_alkali_metals,HG_Aluminium}, in which there is often a dramatic reduction in the eDOS at the Fermi level (in Al, for example, the eDOS reduces by a factor of ${\sim}4$). Interestingly, we find that there is still significant $s \rightarrow d$ charge transfer in S; for the HG phase at $500$ GPa, the Fermi-level eDOS comprises $11\%$ $s$ character, $42\%$ $p$ character and $47\%$ $d$ character.

\subsection{\label{TCs}Superconductivity and Phason Modes}

We have calculated the superconducting critical temperatures of our high-pressure sulfur phases using Migdal-Eliashberg theory within Density Functional Perturbation Theory (DFPT) using the \verb|QUANTUM ESPRESSO| code \cite{QE}. We used the McMillan-Allen-Dynes formula \cite{McMillan_Allen_Dynes} to calculate Tc:

\begin{equation}
    T_c = \frac{\omega_{_\mathrm{log}}}{1.2} \mathrm{exp} \bigg( \frac{-1.04(1+\lambda)}{\lambda - \mu^{*} (1+0.62 \lambda) } \bigg)    
\end{equation}

Where $\omega_{_\mathrm{log}}$ is the logarithmically averaged phonon frequency, $\lambda$ is the electron-phonon coupling constant, and $\mu^{*}$ is the Coulomb pseudopotential parameter, rescaled from the bare value $\mu$ to account for retardation \cite{Morel_Anderson}:

\begin{equation}
    \mu^{*} = \frac{\mu}{1 + \mu \ \mathrm{ln}\big( \frac{\epsilon_F}{\Theta_D} \big)}
\label{renormalised_mu}
\end{equation}

Where $\epsilon_F$ is the Fermi energy, and $\Theta_D$ is the phonon Debye frequency. We note that a relatively recent publication \cite{SCDFT_sulfur} has employed Superconducting DFT (SCDFT) \cite{SCDFT_1,SCDFT_2} to calculate the $T_c$ of the $R\bar{3}m$ phase at pressures up to $200$ GPa, and shows that the value of $\mu$ for S in this regime remains nearly constant at $0.20$. Rescaling this using Eqn. \ref{renormalised_mu}, we obtain a $\mu^{*}$ value of $0.09$ that is essentially constant across all the phases, since the ratio $\epsilon_F / \Theta_D$ changes very little between the phases and with increasing pressure. Further, a previous publication \cite{zakharov_and_cohen_sulfur} derived a $\mu^{*}$ value of $0.08$ for the bcc phase at $500$ GPa using the static Thomas-Fermi dielectric function. We therefore choose $\mu^{*} \in [0.08,0.10]$ as a reasonable range of representative $\mu^{*}$ values.

\begin{figure}[h!]
    
    \hspace{-0.5cm}
    \includegraphics[width=0.45\textwidth]{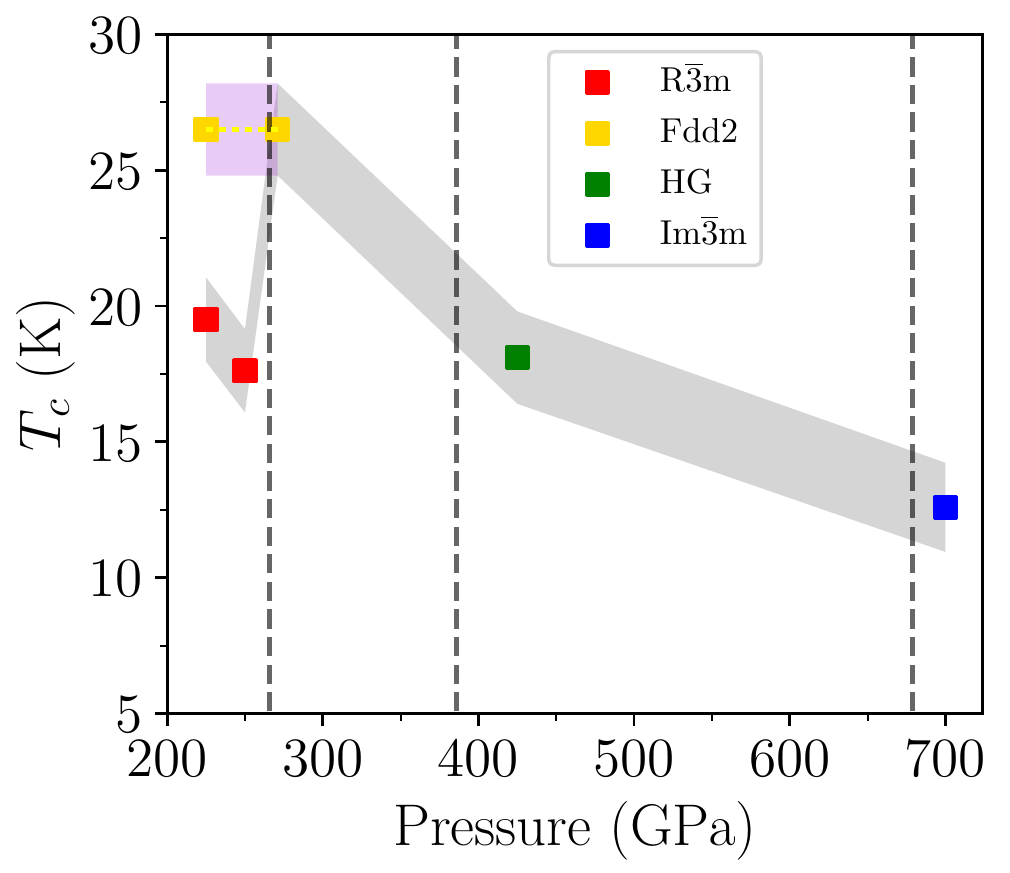}
    
    \caption{Superconducting critical temperatures of the sulfur phases as a function of pressure. Dotted vertical lines denote the structural phase transitions discussed in this work. The shading corresponds to $\mu^{*} \in [0.08,0.10]$; the grey region connects the lowest-enthalpy structures at each pressure, and the purple region in the top left is a possible metastable path if S can be trapped in the $Fdd2$ symmetry below $270$ GPa. The data has been linearly interpolated.}
    \label{Tc_vs_pressure}
\end{figure}

Fig. \ref{Tc_vs_pressure} shows our calculated $T_c$ values as a function of pressure. It can be seen that $T_c$ is expected to peak shortly after the transition to the $Fdd2$ phase, where we predict $T_c$ to be $24.8$ - $28.2$ K at $271$ GPa. For a purely elemental solid, this is a high critical temperature and is close to the record among the elements (excluding hydrogen) of $29$ K in high-pressure Ca-VII \cite{SC_elements_review,Calcium_experimental,chain_ordered_calcium}. Phonon calculations show that although it is not the lowest enthalpy phase at such pressures, the $Fdd2$ structure is dynamically stable down to pressures as low as $215$ GPa (below here, it develops a soft mode that pushes it into the $Fddd$ structure, as previously mentioned). Since we find the $T_c$ of the $Fdd2$ phase to be nearly constant in this region, it is possible that upon reducing pressure from $>270$ GPa, hysteresis effects may allow $T_c$ to take a path similar to that indicated in purple in the top left of Fig. \ref{Tc_vs_pressure}. We note that whilst $T_c$ falls monotonically with pressure after the transition to the $Fdd2$ phase, we predict that S will remain superconducting up to at least $700$ GPa.
\par
Whilst we used \textit{isotropic} Migdal-Eliashberg theory to calculate $T_c$, we note that the aforementioned SCDFT calculations in Ref. \cite{SCDFT_sulfur} account fully for the Fermi surface anisotropy. Our $T_c$ results for the $R\bar{3}m$ phase at $250$ GPa are slightly higher (by ${\sim}3$ K) than the $T_c$ obtained from the use of SCDFT in Ref. \cite{SCDFT_sulfur}, although our Eliashberg function $\alpha^2F(\omega)$ is very similar (see Fig. \ref{eliashberg_function}). Conversely, our $T_c$ value for the $R\bar{3}m$ structure is actually slightly lower (by ${\sim}4$ K) than the experimental value obtained by Drozdov et. al. \cite{drozdov_hydrogen_sulfide_experiment}.

\begin{figure}[h!]
    
    \vspace{-0.25cm}
    \hspace{-0.4cm}
    \includegraphics[width=0.48\textwidth]{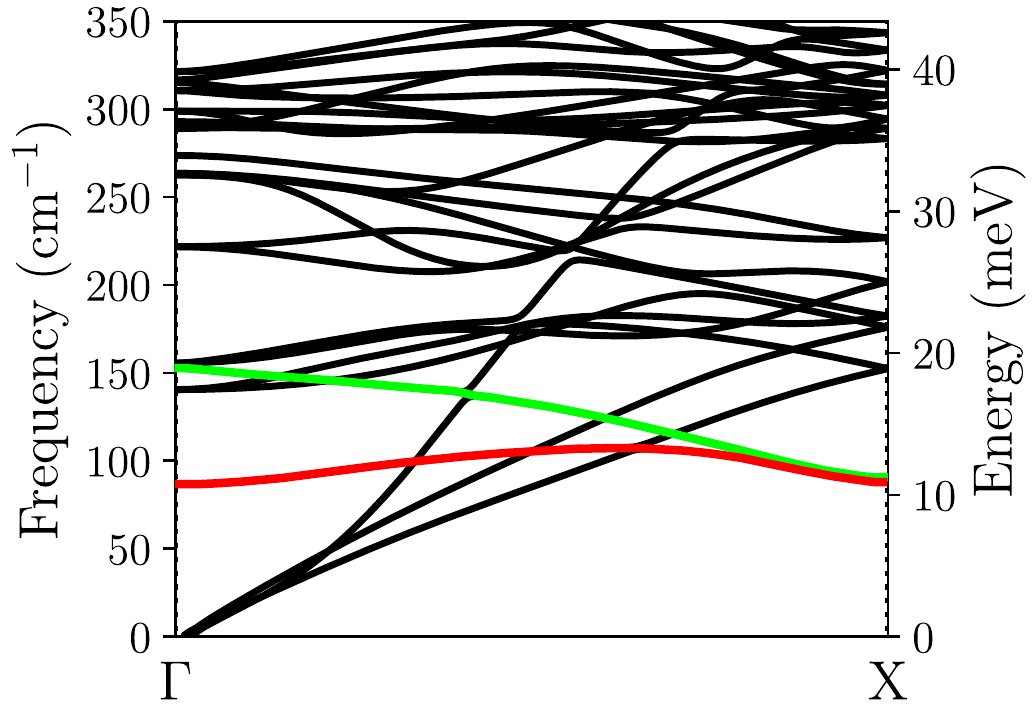}
        
    \caption{Phonon dispersion relation for the AB-ordered $P4/ncc$ approximant at $375$ GPa (commensurate ratio $\gamma = 4/3$). The $\Gamma \rightarrow$ X direction runs across the tetragonal face of the HG structure (perpendicular to the guest chains). Two low-energy phason modes, which correspond to relative movement of the host and guest lattices, have been highlighted in green and red. In the limit of a truly incommensurate structure, the frequencies of these modes vanish.}
    \label{phason_mode}
\end{figure}

In the absence of complications such as the formation of finite-sized domains, the energy of a \textit{truly incommensurate} HG structure must necessarily be invariant with respect to relative translation of the host and guest lattices. Such a relative translation can be achieved with an optical phonon mode running perpendicular to the guest chains, which displaces the host and guest structures in opposite directions. The energy of such a phonon mode should be exactly zero for a true \textit{quasi}-crystal, although the use of commensurate approximants in \textit{ab}-\textit{initio} studies such as this work means that they appear to have a small positive frequency in a phonon calculation. Fig. \ref{phason_mode} shows the phonon dispersion relation for the AB-ordered $P4/ncc$ approximant, in which two low-energy optical modes can be seen. These `phason' modes naturally lead to large values of the mode-specific electron-phonon coupling constant $\lambda_{\mathrm{\textbf{q}},\nu}$, which varies with phonon frequency as $\sim \omega^{-1}_{\mathrm{\textbf{q}},\nu}$, and give rise to strong-coupling superconductivity in HG phases \cite{HG_Bismuth_SC}; however, as mentioned, our use of a commensurate approximant forces the phason modes in Fig. \ref{phason_mode} to have a nonzero frequency. Consequently in Table \ref{superconducting_properties}, which summarises relevant superconducting properties of interest for several of the structures, the value of the electron-phonon coupling constant $\lambda$ for the HG phase is \textit{not} in the strong-coupling regime where $\lambda >> 1$ \cite{McMillan_Allen_Dynes}. It is probable that the use of a larger approximant (with accompanying lower phason frequencies) would give rise to a larger value of  $\lambda$, although such a calculation is beyond the scope of this work and would be computationally expensive. We also note that solving the Eliashberg equations (necessary for the true strong coupling case) gave very similar $T_c$ values to those obtained using the McMillan-Allen-Dynes formula for all of the structures discussed in this work. Interestingly, we find that $\lambda$ peaks in the $Fdd2$ phase, which by virtue of being a commensurate structure, cannot possess a phason mode. Nonetheless, we still predict a relatively high $T_c$ of $16.4$ - $19.8$ K for the HG phase at $425$ GPa, which we partially attribute to the lack of a significant Fermi-level pseudogap in the eDOS (see Table \ref{superconducting_properties}).

\addtolength{\tabcolsep}{-0.85pt} % decrease column spacing
\begin{table}[!htbp]
\begin{tabular}{c c c c c c}
\hline\hline
\centering
Structure & Pressure (GPa) & $T_c$ (K) & $\lambda$ & $\langle \omega_{log} \rangle$ (K) & $\mathcal{N}(\epsilon_F)$ \\ \hline
$R\bar{3}m$ & $250$ & $17.6$ & $0.689$  & $469$ & $0.0390$ \\
$Fdd2$ & $271$ & $26.5$ & $0.847$  & $439$ & $0.0414$ \\
HG $(\gamma=\frac{4}{3})$ & $425$ & $18.1$ & $0.738$  & $417$ & $0.0445$ \\
$Im\bar{3}m$ & $700$ & 12.6 & $0.543$  & $660$ & $0.0470$ \\
\hline\hline
\end{tabular}
\caption{Superconducting critical temperature $T_c$, electron-phonon coupling constant $\lambda$, logarithmic average frequency $\langle \omega_{log} \rangle$ and Fermi-level electronic density of states $\mathcal{N}(\epsilon_F)$ (in units of electrons per eV per \AA $^{3}$) for several of the structures discussed in this work at various pressures.}
\label{superconducting_properties}
\end{table}
\addtolength{\tabcolsep}{+0.85pt} % increase column spacing

\begin{figure}[h!]

    \begin{subfigure}[t]{0.236\textwidth}
        \hspace{-0.2cm}
        \includegraphics[width=\linewidth]{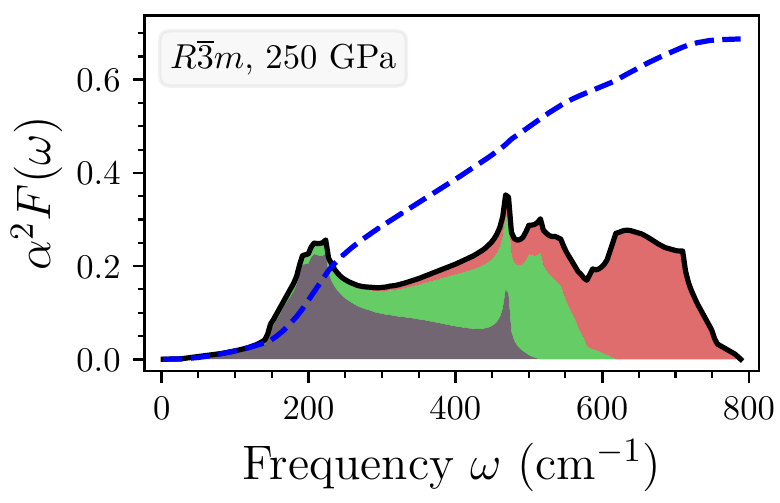}
    \end{subfigure}
    \hfill
    \begin{subfigure}[t]{0.236\textwidth}
        \hspace{-0.3cm}
        \includegraphics[width=\linewidth]{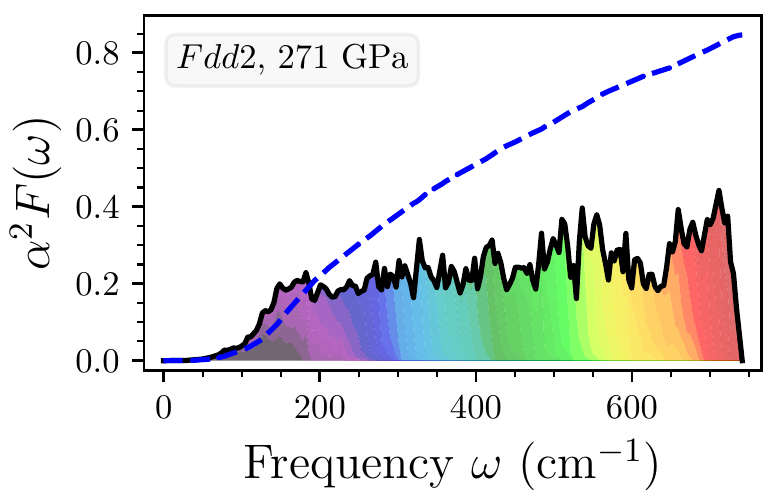}
    \end{subfigure}
    
    \vspace{0.4cm}
    
    \begin{subfigure}[t]{0.236\textwidth}
        \hspace{-0.2cm}
        \includegraphics[width=\linewidth]{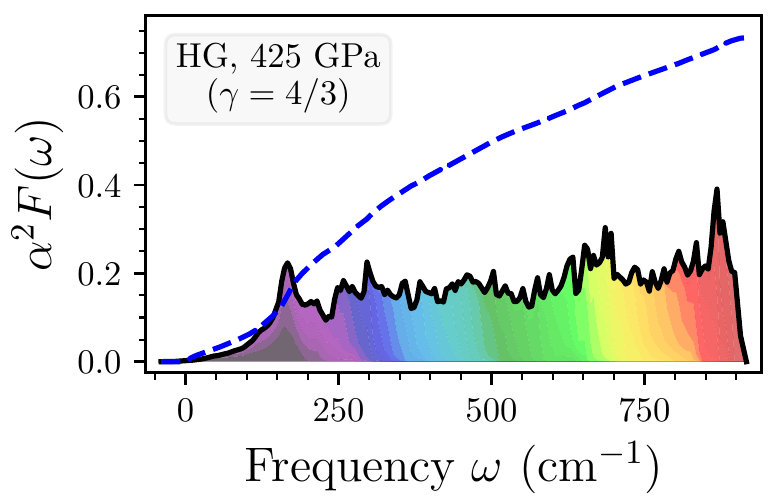}
    \end{subfigure}
    \hfill
    \begin{subfigure}[t]{0.236\textwidth}
        \hspace{-0.3cm}
        \includegraphics[width=\linewidth]{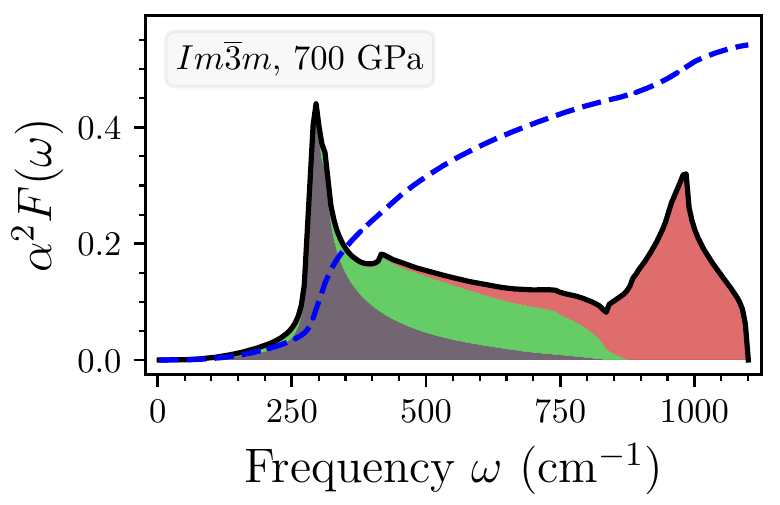}
    \end{subfigure}
    
    \caption{Eliashberg function $\alpha^2F(\omega)$ for selected structures at given pressures. The contribution of each phonon branch to $\alpha^2F(\omega)$ has been shaded with a different colour. The (cumulative) electron-phonon coupling constant $\lambda(\omega)$ has been plotted with a dashed blue line.}
    \label{eliashberg_function}
\end{figure}

\section{\label{conclusion}Conclusions}

We have shown that upon increasing pressure beyond the experimentally-known $R\bar{3}m$ ($\beta$-Po) phase, sulfur adopts a commensurate structure of $Fdd2$ symmetry in the range $266$-$386$ GPa that resembles a strongly distorted Ba-IVa phase. Between $386$ and $679$ GPa, S possesses a truly incommensurate HG structure of the Ba-IVa type. Above $679$ GPa, we find that S adopts a more closely packed bcc structure. We have accounted for the effects of finite temperatures on the phase boundaries, and have shown that all of the transition pressures can be lowered with an increase in temperature, as well as predicting the existence of an $Fddd$ phase below $215$ GPa and at temperatures $\gtrsim 1260$ K.
\par
We have studied the properties of the incommensurate HG phase in detail, and have shown that the ideal incommensurate ratio $\gamma_{0}$ decreases monotonically with increasing pressure. Over the stability range of the HG phase, S transforms between several competing chain-ordered variants of the incommensurate HG structure, with an ABAC chain stacking stable over most of this pressure range. These different chain orderings are intimately coupled to modulations of the host and guest atoms, which contribute substantially to the stability of the HG phases. We find that there is no significant interstitial localisation of charge in S, instead showing that there are notable localised \textit{absences} of charge (`voids'). We have calculated the superconducting critical temperature of S for several of the structures, and expect $T_c$ to peak between $24.8$ and $28.2$ K in the $Fdd2$ phase at $271$ GPa. We predict that S remains a superconductor up to at least $700$ GPa.

\section{\label{acknowledgements}Acknowledgements}

We wish to thank Malte Grosche for useful discussions. The computational resources for this project were provided by the Cambridge Service for Data Driven Discovery (CSD3), and we are further grateful for computational support from the UK national high performance computing service, ARCHER, for which access was obtained via the UKCP consortium and funded by EPSRC grant ref EP/P022561/1.

\nocite{*} % include references not cited in main text body

\bibliography{references.bib}

%%%%%%%%%% Merge with supplemental materials %%%%%%%%%%
\pagebreak
\begin{center}
\textbf{\color{red}\large SUPPLEMENTAL MATERIAL}
\end{center}
%%%%%%%%%% Merge with supplemental materials %%%%%%%%%%
%%%%%%%%%% Prefix a "S" to all equations, figures, tables and reset the counter %%%%%%%%%%
\setcounter{equation}{0}
\setcounter{figure}{0}
\setcounter{table}{0}
\setcounter{page}{1}
\makeatletter
\renewcommand{\theequation}{S\arabic{equation}}
\renewcommand{\thefigure}{S\arabic{figure}}
%\renewcommand{\bibnumfmt}[1]{[S#1]}
%\renewcommand{\citenumfont}[1]{S#1}
%%%%%%%%%% Prefix a "S" to all equations, figures, tables and reset the counter %%%%%%%%%%

\maketitle

\section{\label{pseudo}Pseudopotentials}
We used a sulfur pseudopotential designed using the OTFG code bundled with \verb|CASTEP|. Our pseudopotential had a cutoff radius of $0.74$ \AA $ \ =1.4$ Bohr (far smaller than half the smallest interatomic separation encountered in this study) and included the $n=3$ valence shell only. Our tests show that inclusion of the core $n=2$ states has no effect on resolving enthalpy differences at these pressures. The local channel was chosen as $l_{loc}=3$. This pseudopotential was used for all \verb|CASTEP| calculations in the manuscript.
\\\\
The OTFG code string used the generate the pseudopotential was:

\begin{verbatim}
    S 3|1.4|4|10|12|30:31:32(qc=7)
\end{verbatim}

For the \verb|QUANTUM ESPRESSO| (QE) superconducting critical temperature calculations, the default scalar relativistic, ultrasoft PBE pseudopotential for S from  \href{https://dalcorso.github.io/pslibrary/}{pslibrary} were found to be sufficient in all cases, except for the $C2/c$ approximant at $425$ GPa, which required a smaller $R_c$. For this calculation, a custom pseudopotential with cutoff radius $R_c=1.6$ Bohr was generated via the \verb|ld1.x| code bundled with QE, using the following input file:

\begin{verbatim}
&input
   title='S',
   zed=16.,
   rel=1,
   config='[Ne] 3s2 3p4 3d-1',
   iswitch=3,
   dft='pbe'
 /
 &inputp
   lpaw=.false.,
   pseudotype=3,
   file_pseudopw='S.UPF',
   author='Jack Whaley-Baldwin',
   lloc=-1,
   rcloc=1.6,
   which_augfun='PSQ',
   rmatch_augfun_nc=.true.,
   nlcc=.true.,
   new_core_ps=.true.,
   rcore=1.4,
   tm=.true.
 /
6
3S  1  0  2.00  0.00  1.40  1.60  0.0
3S  1  0  0.00  6.00  1.40  1.60  0.0
3P  2  1  4.00  0.00  1.40  1.60  0.0
3P  2  1  0.00  3.00  1.40  1.60  0.0
3D  3  2  0.00  0.10  1.40  1.60  0.0
3D  3  2  0.00  0.30  1.40  1.60  0.0
\end{verbatim}

This custom pseudopotential was tested against its respective default pslibrary pseudopotential for a few structures, and was found to agree in volume to within $0.1 \%$, and relative enthalpies agreed to $\pm 0.1$ meV.

\section{\label{conv_calcs}Convergence Calculations}

\begin{figure}[h!]
    \hspace{-0.5cm}
    \includegraphics[width=0.45\textwidth]{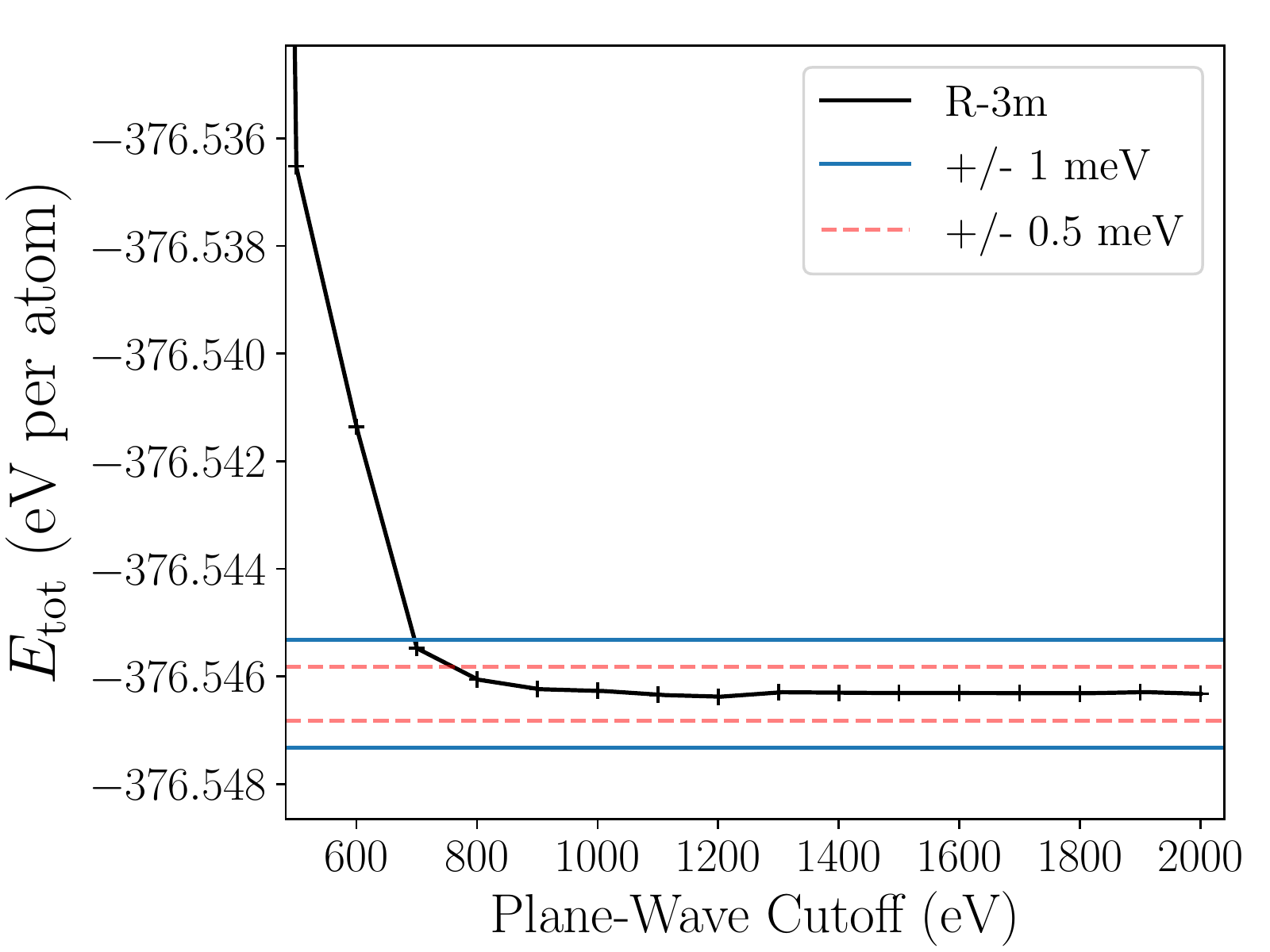}
    
    \vspace{0.25cm}
    
    \hspace{-0.75cm}
    \includegraphics[width=0.43\textwidth]{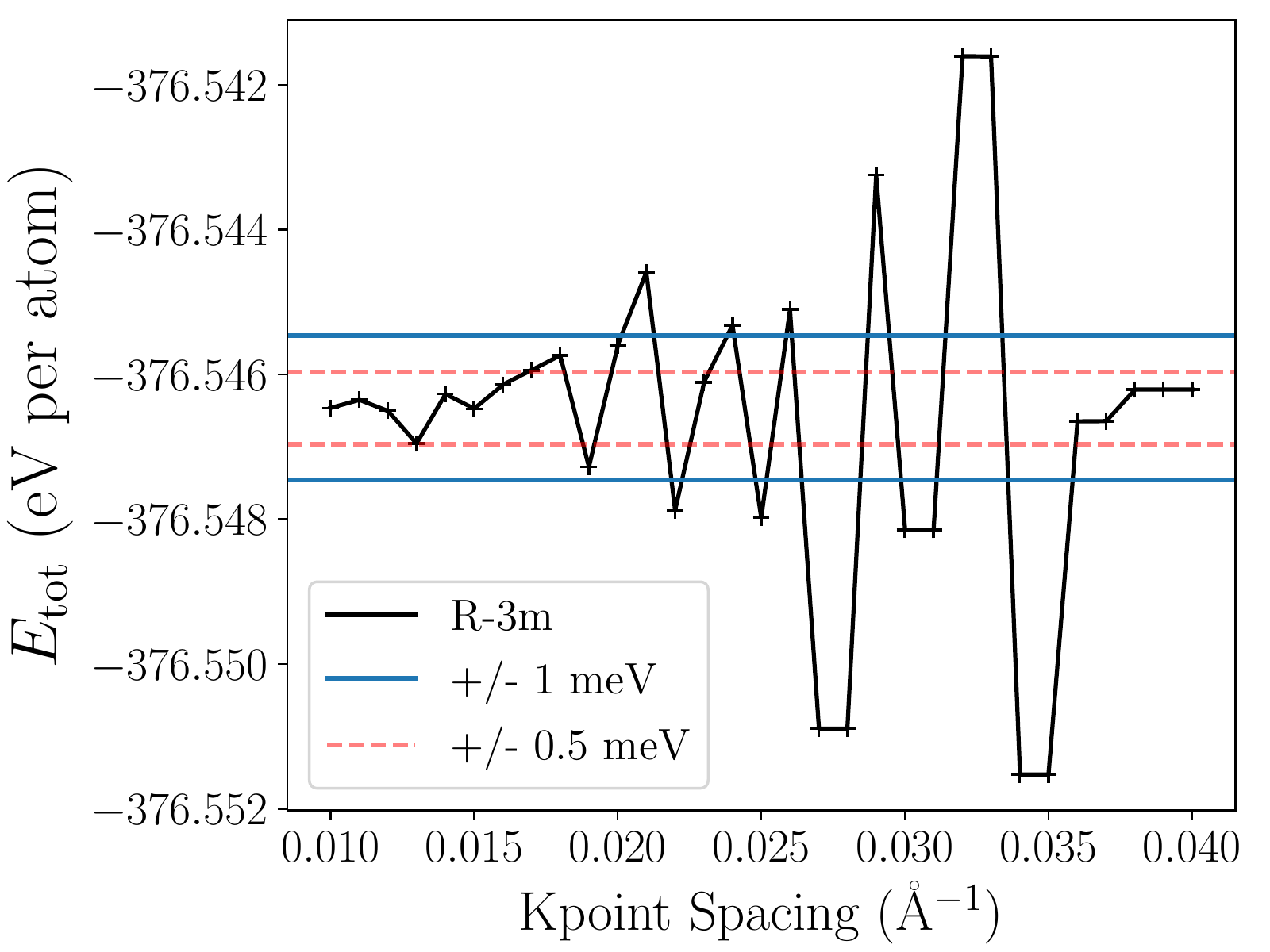}
    
    \vspace{0.25cm}
    
    \hspace{-0.95cm}
    \includegraphics[width=0.45\textwidth]{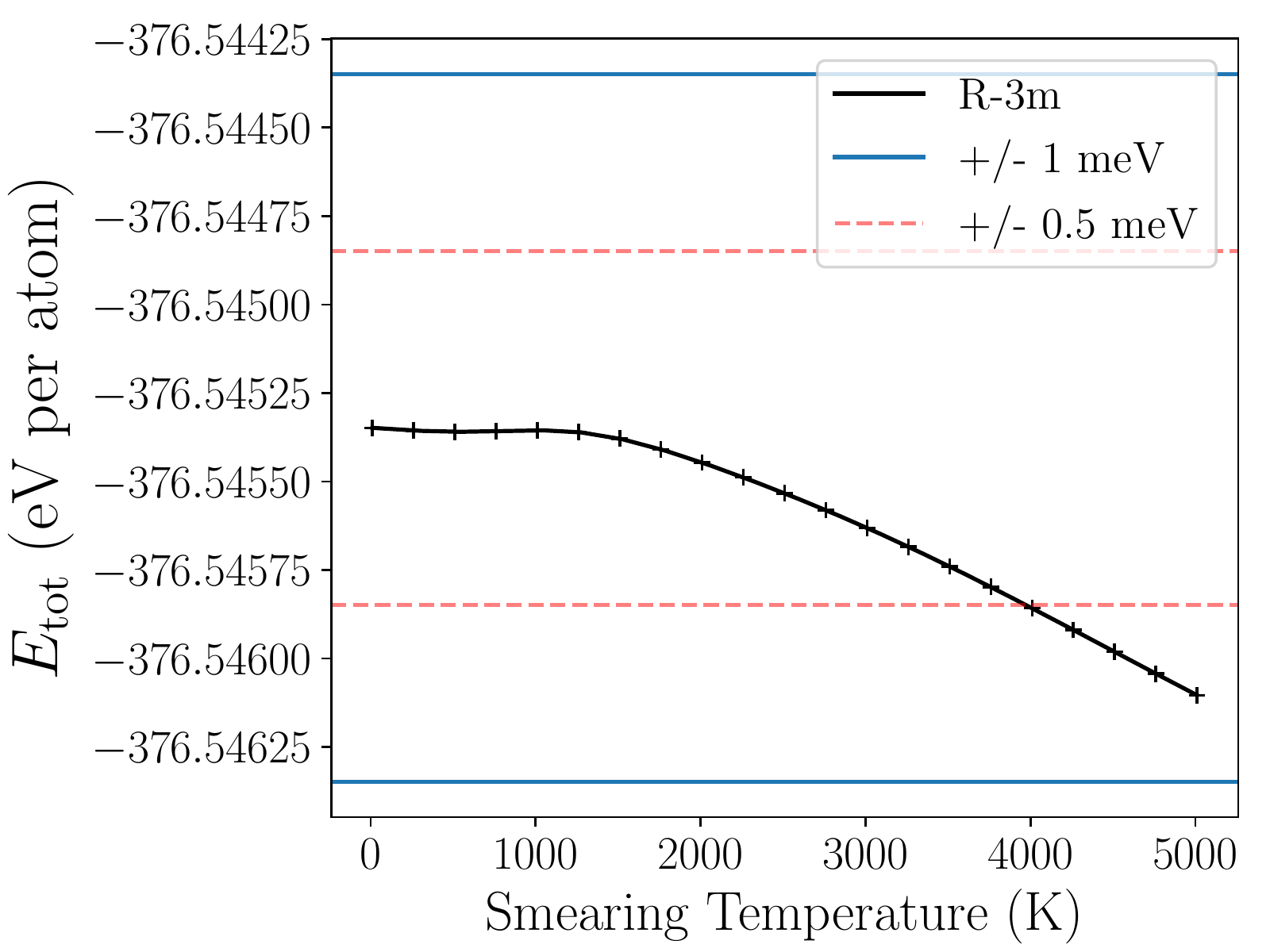}
    \caption{Absolute convergence of the total electronic energy of the $R\bar{3}m$ structure at $400$ GPa, for the plane-wave cutoff (top), \textbf{k}-point spacing (middle) and electronic smearing temperature (bottom). $\pm 1$ meV and $\pm 0.5$ meV lines are shown relative to the most expensive result.}
    \label{400gpa_convergence}
\end{figure}

\begin{figure}[h!]
    \hspace{-0.5cm}
    \includegraphics[width=0.45\textwidth]{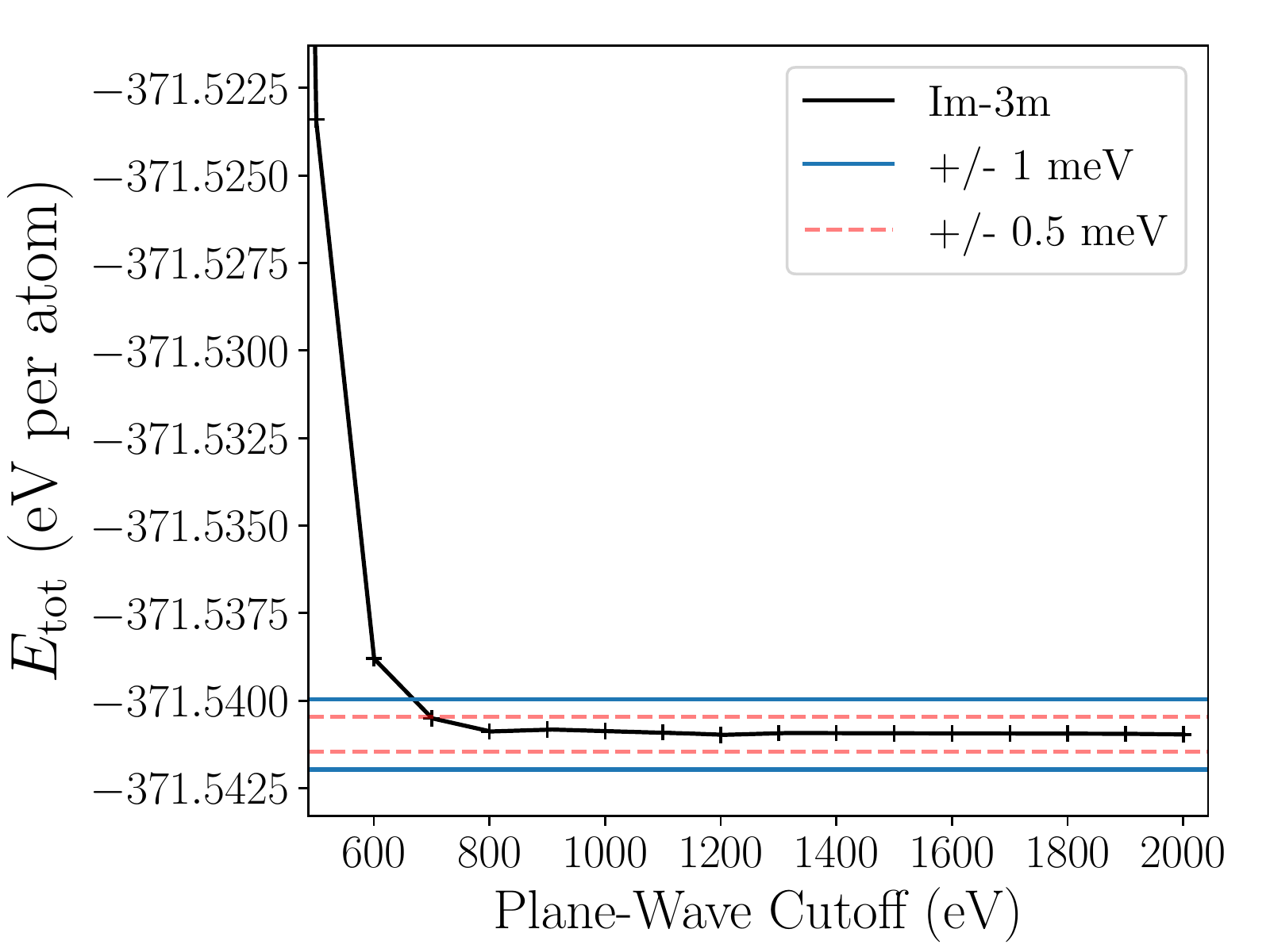}
    
    \vspace{0.25cm}
    
    \hspace{-0.7cm}
    \includegraphics[width=0.43\textwidth]{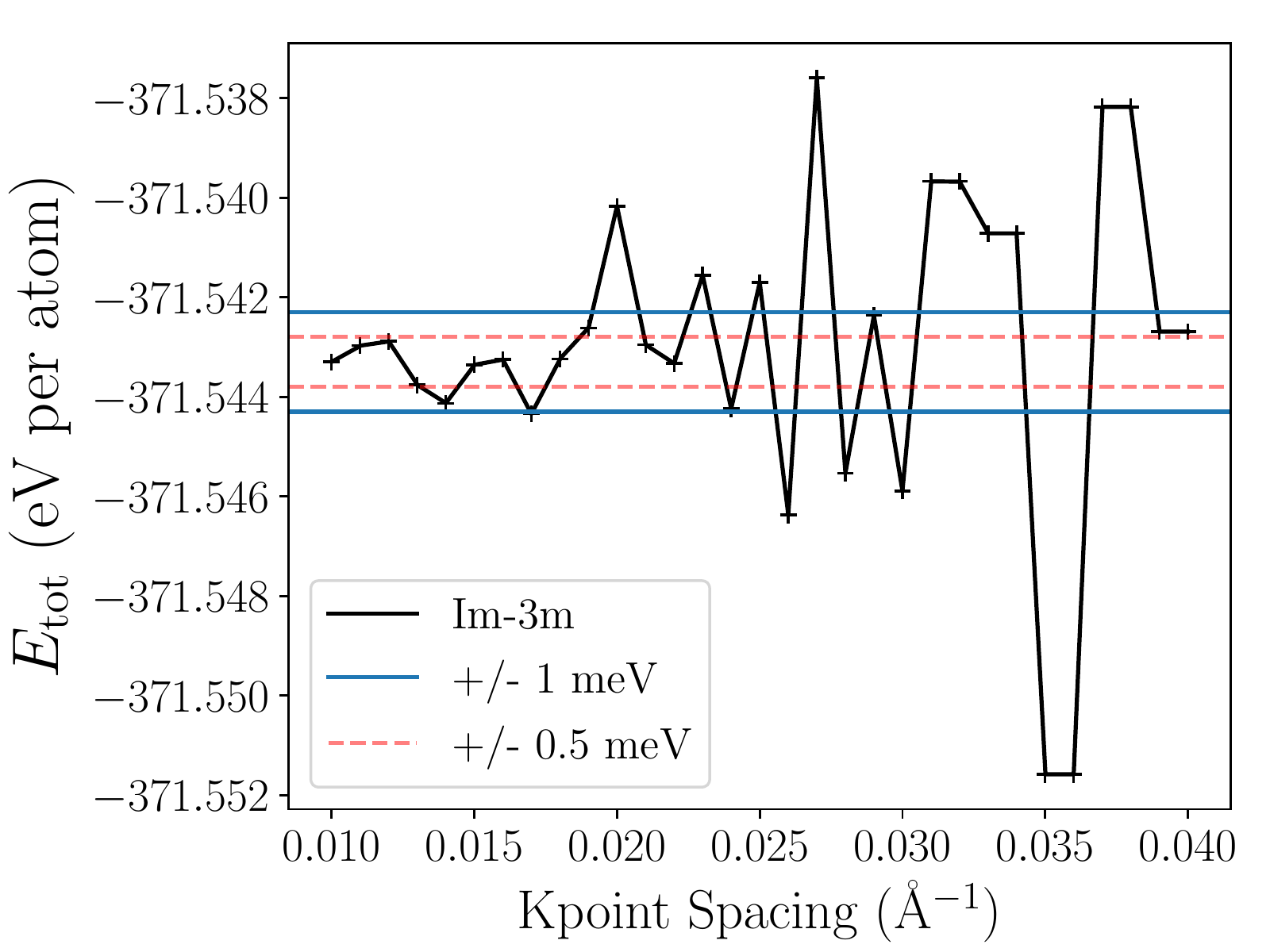}
    
    \vspace{0.25cm}
    
    \hspace{-0.7cm}
    \includegraphics[width=0.45\textwidth]{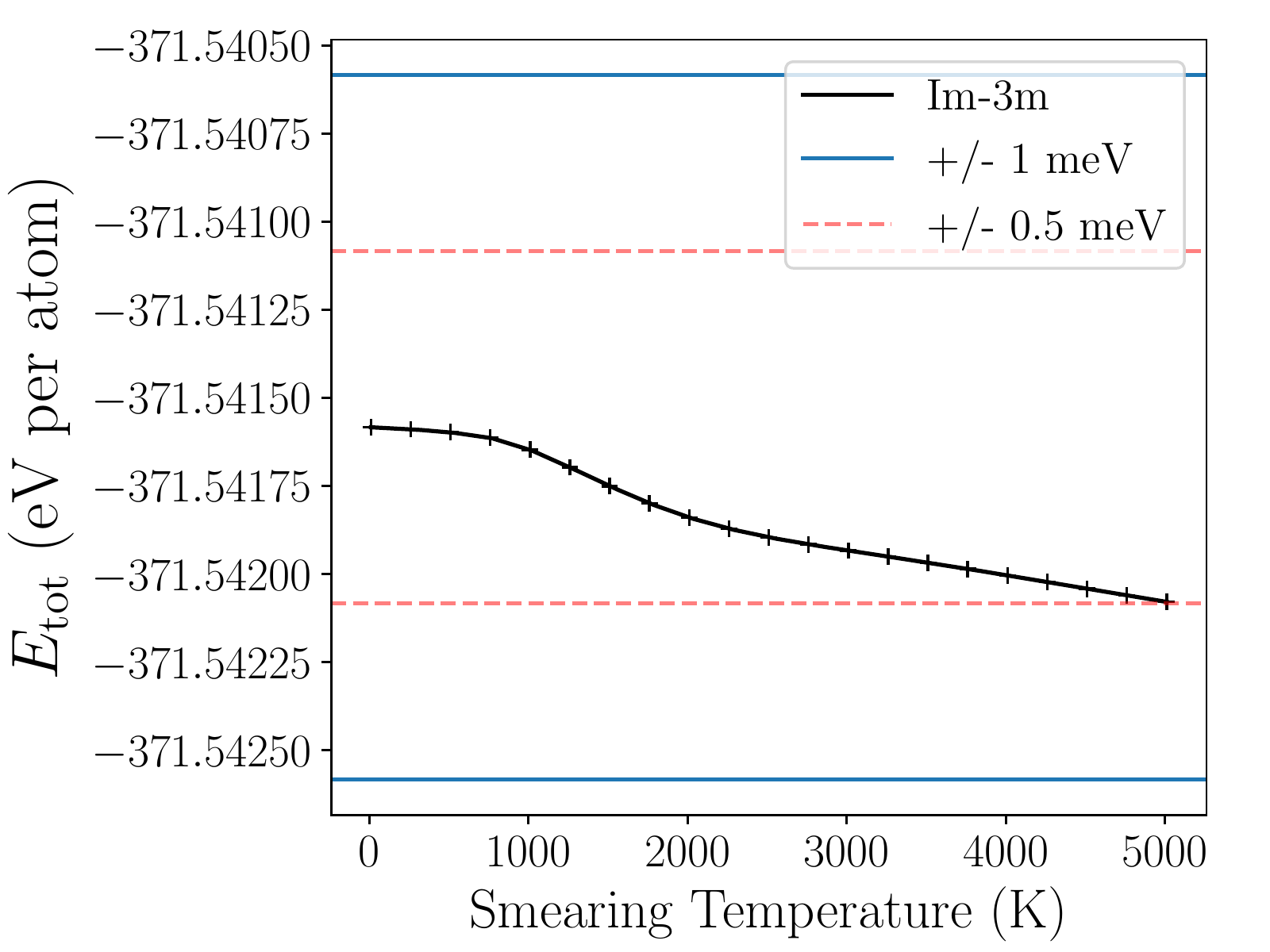}
    \caption{Absolute convergence of the total electronic energy of the $Im\bar{3}m$ structure at $800$ GPa, for the plane-wave cutoff (top), \textbf{k}-point spacing (middle) and electronic smearing temperature (bottom). $\pm 1$ meV and $\pm 0.5$ meV lines are shown relative to the most expensive result.}
    \label{800gpa_convergence}
\end{figure}

Here, we present convergence calculations at $400$ and $800$ GPa, which span the majority of the pressure range covered. We consider as representative structures the $R\bar{3}m$ stucture at $400$ GPa and the $Im\bar{3}m$ structure at $800$ GPa. We show only the convergence results for the PBE pseudopotential here, because the converged cutoffs and \textbf{k}-point spacings for the the LDA and PBEsol pseudopotentials are nearly identical.
\\\\
Supplemental Figs. \ref{400gpa_convergence} and \ref{800gpa_convergence} show that our chosen parameters of a plane-wave cutoff of $850$ eV, a \textbf{k}-point spacing of $2\pi \times 0.01 \ $\AA$^{-1}$, and the \verb|CASTEP| default electronic smearing temperature ($\approx 1200$ K) were sufficient to achieve absolute convergence of at least $\pm 0.2$ meV throughout the pressure range of interest; the relative energies of the phases are converged to better than this.
\\\\
Such stringent convergence was necessary as the different chain-ordered host-guest phases are separated by enthalpy differences of order $\approx 0.5$ meV.

\clearpage

\section{\label{ratio_fitting}Ideal Host-Guest Ratio Fitting Procedure}

\begin{figure}[h!]
    \hspace{-0.7cm}
    \includegraphics[width=0.475\textwidth]{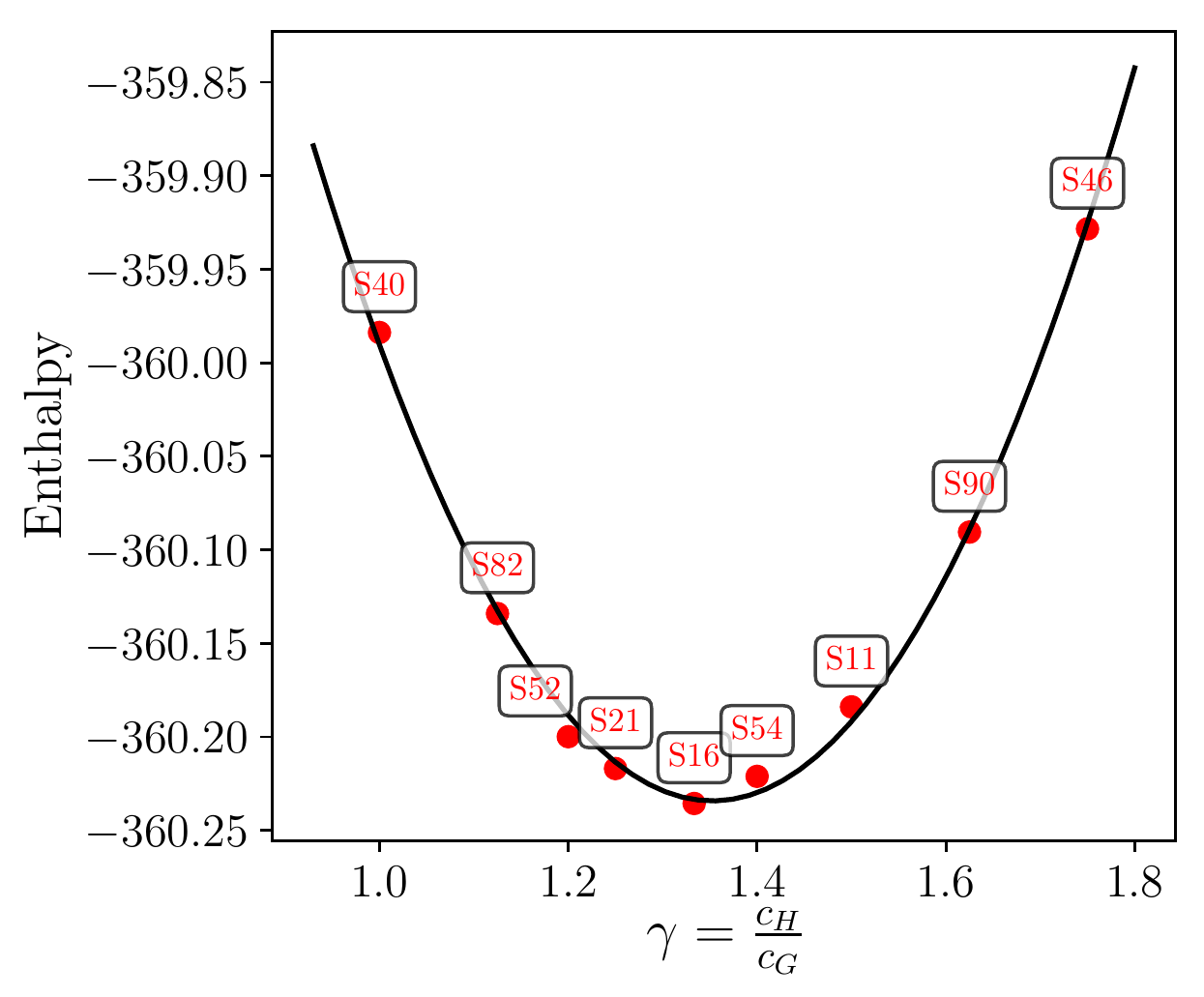}
    \caption{Dependence of the enthalpy of the HG phase on the $c_H / c_G$ ratio at $400$ GPa using the PBE functional. The black line is a least-squares fit to a quadratic curve, and the `ideal' value $x_0$ is taken from the minimum of the curve. Each point represents a particular host-guest approximant, and the label shows the number of atoms in that approximant. By constructing similar curves at different pressures, we can obtain $\gamma_0$ at each pressure and thus plot Fig. 4 in the manuscript.}
    \label{gamma_parabola}
\end{figure}

Supplemental Fig. \ref{gamma_parabola} details how the `ideal' host-guest $\gamma$-value was obtained for a given pressure.
\\\\
In addition to those approximants found in the search, several approximants with different $\gamma$-values were constructed `by hand', and the resulting enthalpies were fitted to a quadratic curve. The ideal value $\gamma_0$ is simply read off from the minimum of this curve.

\newpage
\section{\label{C2c_and_Fdd2}Enthalpies of C2/c and Fdd2 Structures}

\begin{figure}[h!]
    \hspace{-0.275cm}
    \includegraphics[width=0.46\textwidth]{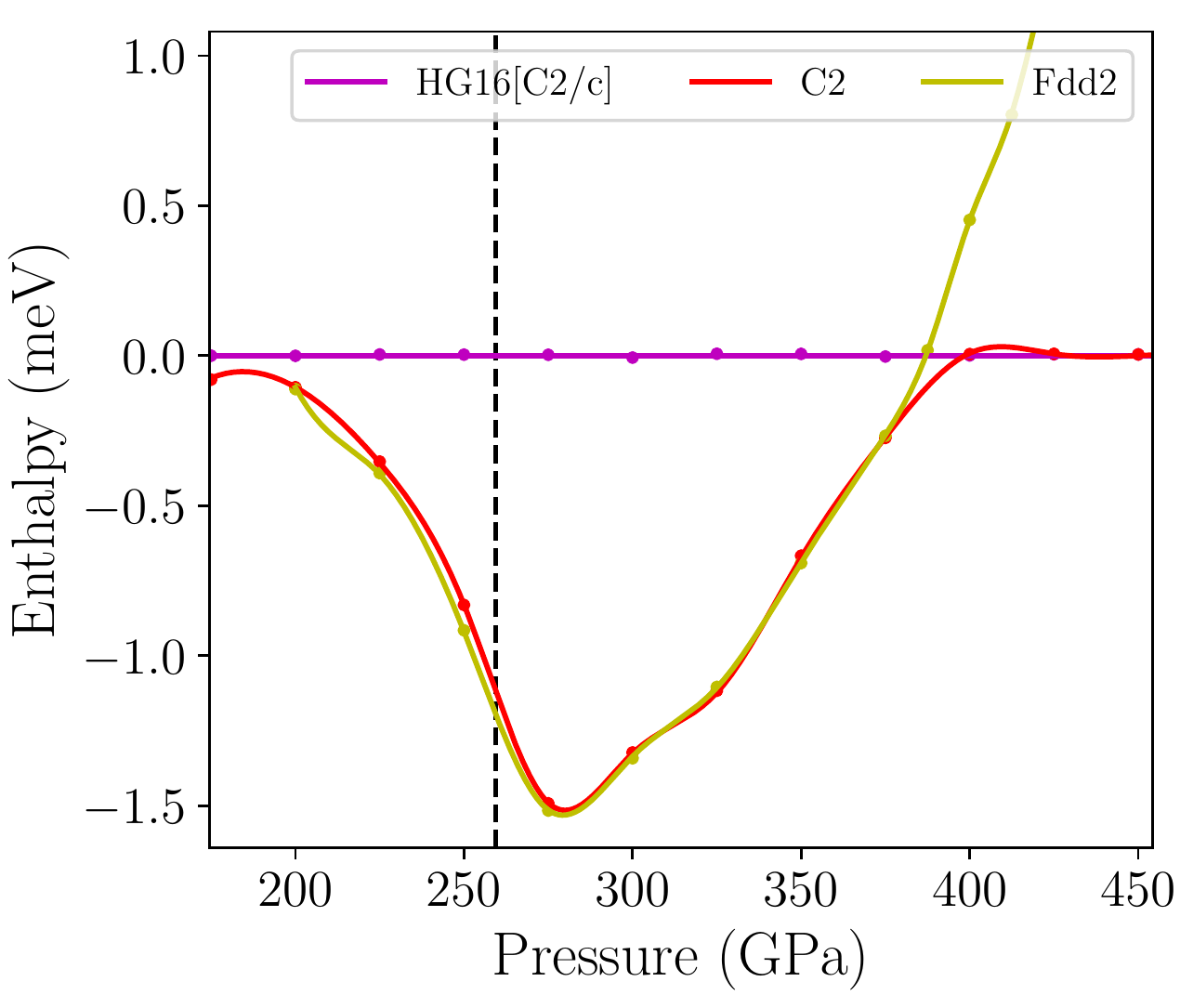}
    \caption{Relative enthalpies of the $C2$, $C2/c$ and $Fdd2$ structures. $C2$ is a symmetry subgroup of both $C2/c$ and $Fdd2$.}
    \label{C2_vs_C2c_vs_Fdd2}
\end{figure}

The $C2/c$ HG approximant exhibits an imaginary phonon mode at the $\Gamma$-point below $380$ GPa, which distorts it into the $Fdd2$ structure.
\\\\
The $C2/c$ and $Fdd2$ structures are very nearly identical, both structurally and in enthalpy. The enthalpy difference between them is at most $\approx 1.5$ meV at $275$ GPa. Supplemental Fig. \ref{C2_vs_C2c_vs_Fdd2} compares the enthalpies of the two structures, in addition to another enthalpy curve that detects only the $C2$ symmetry group, which is a subgroup of both $C2/c$ and $Fdd2$. This shows that sulfur `selects' the $Fdd2$ symmetry below $380$ GPa and the $C2/c$ HG approximant above this pressure.

\newpage
\section{\label{other_chain_orderings}Enthalpies of Other Stacking Orderings}

In the manuscript, we state that we tested various other stacking orderings that turned out to either be not energetically competitive, or simply collapsed into another chain ordering. Supplemental Table \ref{extra_chains} details these other stacking orderings.

\addtolength{\tabcolsep}{+5pt} % increase column spacing
\begin{table}[!htbp]
\begin{tabular}{l l l l}
\hline\hline
Stacking Sequence & Chain Heights & Smallest Enthalpy (meV/atom) & Pressure (GPa) \\ \hline
ABCACB & $(0,1,2,0,2,1)$  & $+0.3$ & 550 \\
ABACBCBC$^{\dagger}$ & $(0,1,0,2,1,2,1,2)$  & $+0.8$ & $500$ \\
AB$^{\dagger}$ & $(0,1)$  & $+1.3$ & $450$ \\
ABB & $(0,1,1)$  & $+16.9$ & $480$ \\
ABACB & $(0,1,0,2,1)$  & collapsed to ABC : $(0,1,2)$ & N/A \\
ABB & $(0,2,2)$  & collapsed to ABB : $(0,1,1)$ & N/A \\
ABBB & $(0,1,1,1)$  & collapsed to ABAC : $(0,1,0,2)$ & N/A \\
ABBC & $(0,1,1,2)$  & collapsed to ABAC : $(0,1,0,2)$ & N/A \\
ABBC & $(0,2,2,1)$  & collapsed to ABAC : $(0,1,0,2)$ & N/A \\
\hline\hline
\end{tabular}
\caption{Explicitly tested stacking sequences that were constructed by hand. The `Chain Heights' ($c$-axis displacement as one moves along the $a$-direction) are given in units of a third of the intra-chain atomic spacing, e.g. `2' corresponds to $\frac{2}{3}$. The `Smallest Enthalpy' is measured relative to the lowest-enthalpy stacking sequence as given in the manuscript, which is ABC in the range $384$-$400$ GPa, ABAC in the range $400$-$590$ GPa, and AB in the range $590$-$679$ GPa. The `Pressure' column in this table is the pressure at which the enthalpy of that sequence was lowest. Where the stacking sequence collapsed into another ordering, `N/A' (Not Applicable) has been placed in the Pressure column. Sequences marked with a dagger ($^{\dagger}$) developed (after relaxation) $c$-axis displacements that changed when moving along the $b$-axis also.}
\label{extra_chains}
\end{table}
\addtolength{\tabcolsep}{-5pt} % decrease column spacing

\clearpage

\section{\label{band_gap_chain_ordering}Band Gaps Opened by Chain Ordering}

Supplemental Fig. \ref{bandstructures} below shows the bandstructures of the $\gamma=4/3$ approximant with all chains aligned (i.e. AAAAAA...) and then with  an ABC ordering. It can be seen that adding the ABC ordering opens up several band gaps at the Fermi level, and this contributes to the lowering of electronic energy as described in Fig. 6 of the main manuscript.

\begin{figure}[!htbp]

    \hspace{-0.4cm}
    \includegraphics[width=0.435\textwidth]{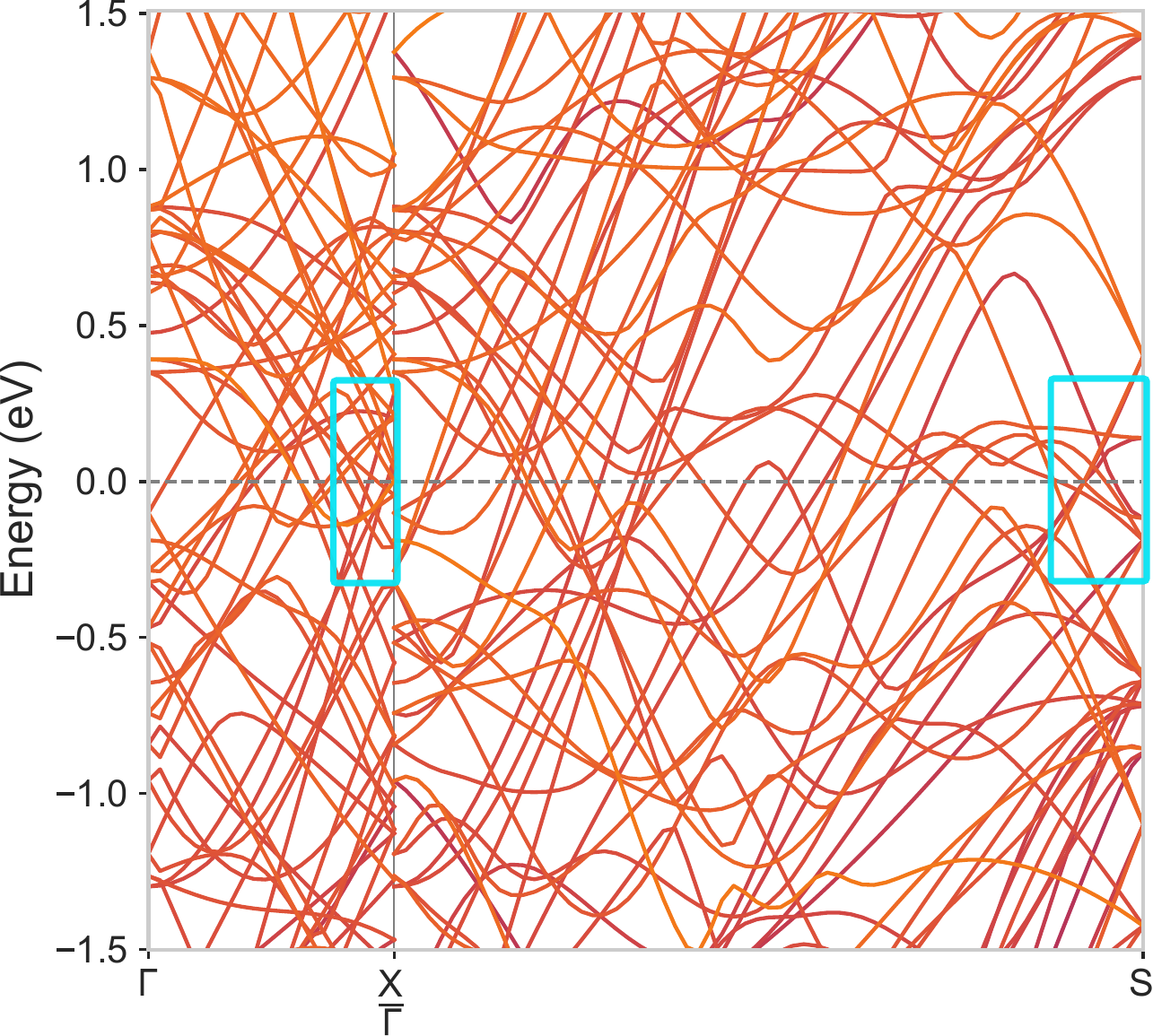}
    
    \vspace{0.4cm}
    
    \hspace{-0.4cm}
    \includegraphics[width=0.435\textwidth]{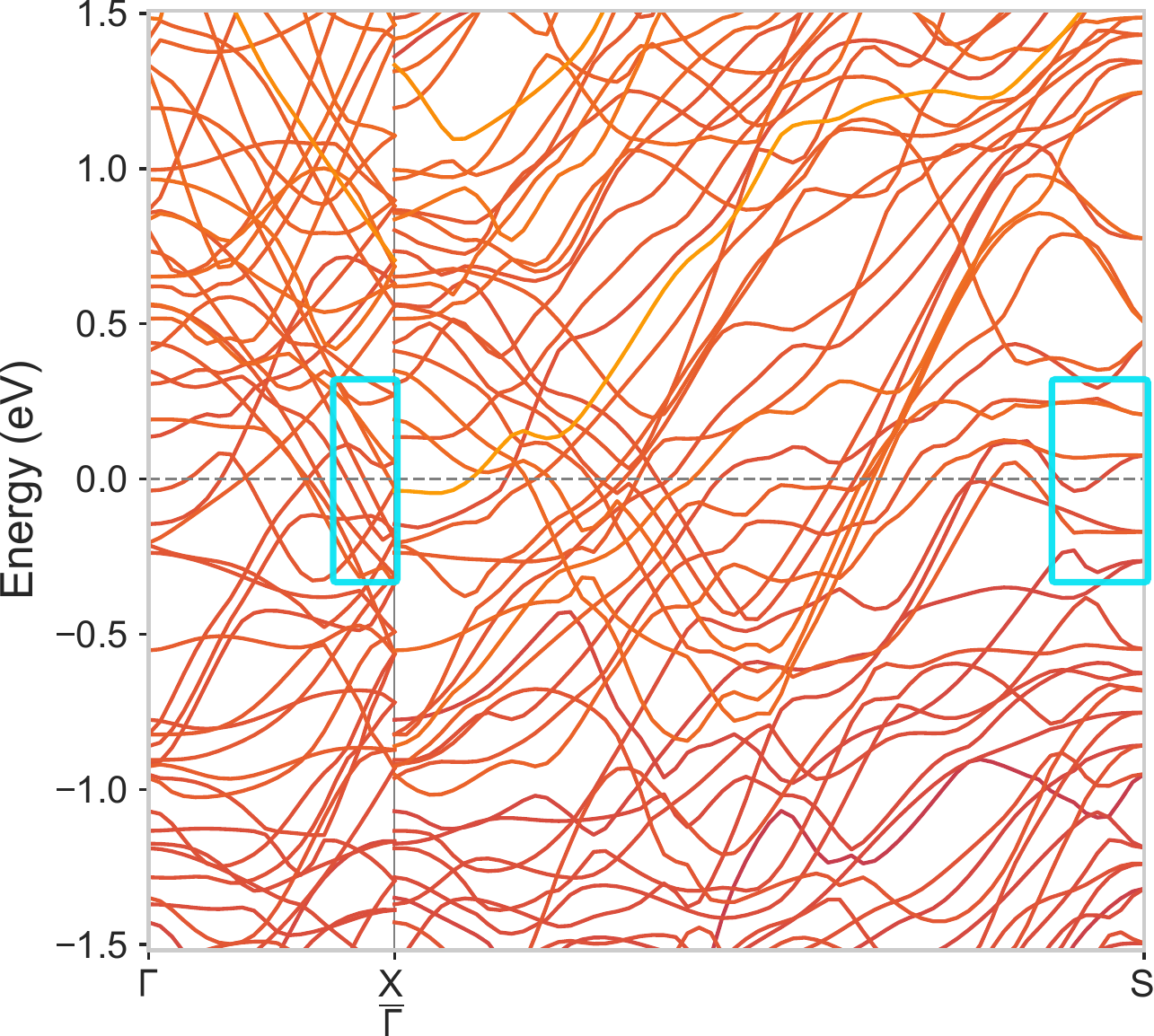}

    \caption{Electronic bandstructures at $387$ GPa along $\Gamma \rightarrow$ X:$(0.5,0,0)$ and $\Gamma \rightarrow$ S:$(0.5,0.5,0)$ high symmetry directions for a ($96$-atom) $3\times1\times1$ tiling of the ideal $I4/mcm$ structure. Band gap openings are highlighted with cyan boxes. \textbf{Top:} With all chains aligned. \textbf{Bottom:} With the chains offset `by hand' as to have an ABC ordering along the $a$-axis.}
    \label{bandstructures}
\end{figure}

\clearpage

\section{\label{eDOS}Electronic Density of States}

The top of Supplemental Fig. \ref{eDOSs} compares the electronic Densities of States (eDOS) of the $Im\bar{3}m$ phase and 64-atom $Pcca$ HG approximant. The HG phase features a relative reduction in total electronic energy as a result of transferring its eDOS weight to lower energies. A small reduction in the Fermi level eDOS is also visible.
\\\\
For comparison, the bottom of Supplemental Fig. \ref{eDOSs} compares the eDOSs of HG aluminium at $5.20$ TPa and HG sulfur at $500$ GPa (where the 64-atom $Pcca$ approximant was used to model the HG phase). It can clearly be seen that a drastic drop in the Fermi level eDOS that is present in Al does not occur for S, where the eDOS at $E_F$ is in fact relatively flat.

\begin{figure}[h!]
    \hspace{-0.3cm}
    \includegraphics[width=0.45\textwidth]{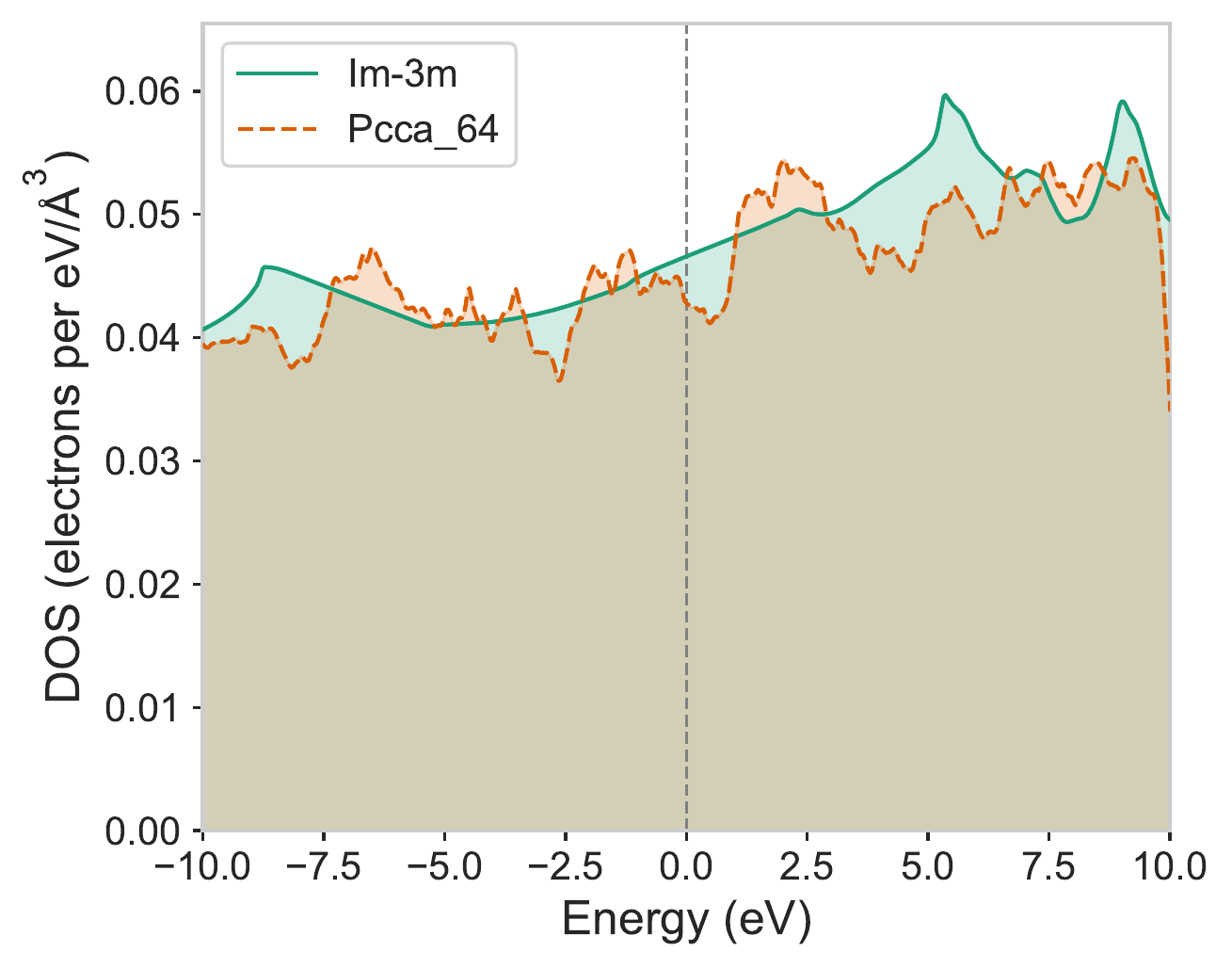}
    
    \vspace{0.35cm}
    
    \includegraphics[width=0.45\textwidth]{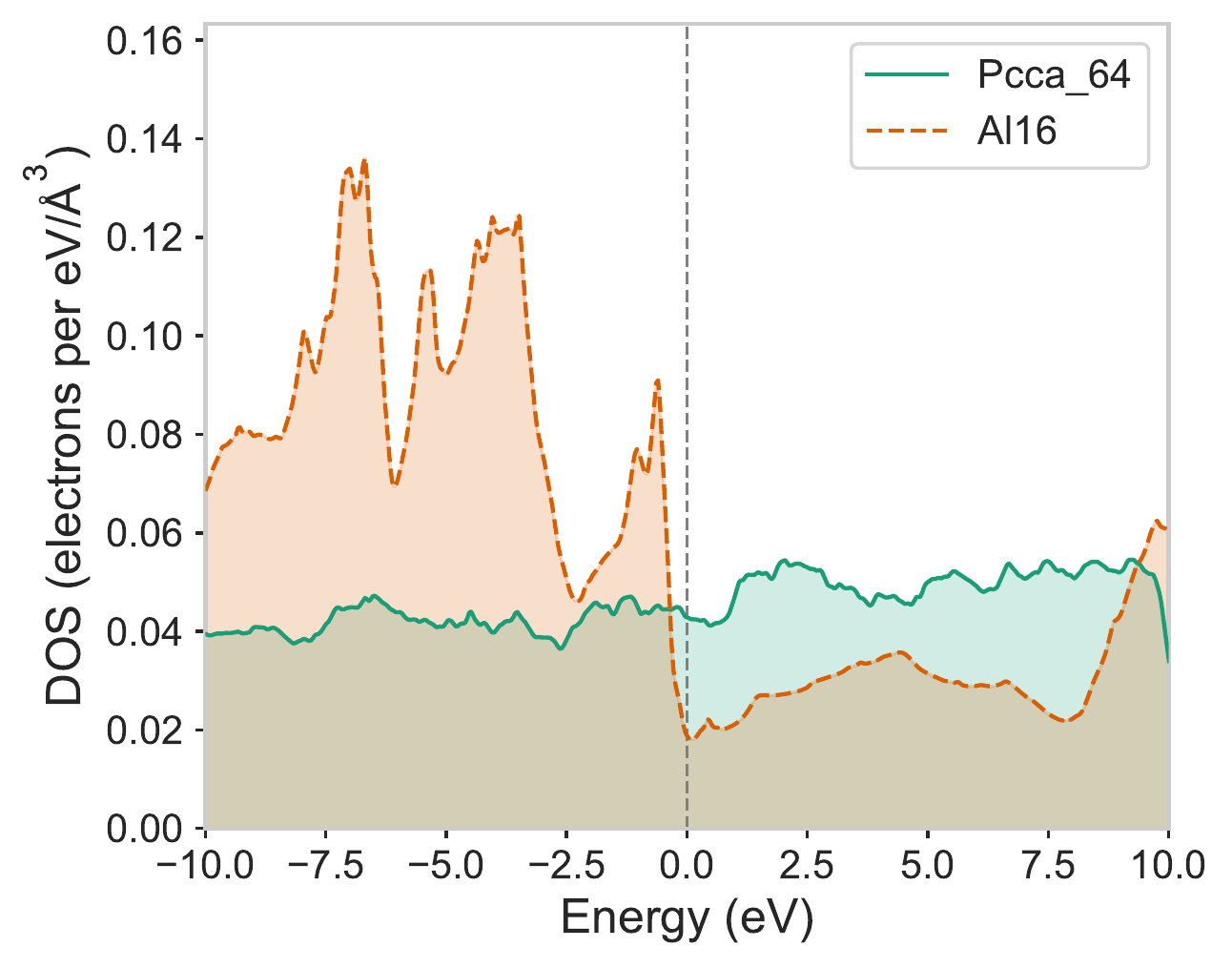}
    \caption{Electronic Densities of States (eDOS). \textbf{Top:} For the 64-atom $Pcca$ approximant and $Im\bar{3}m$ phase at $500$ GPa. \textbf{Bottom:} For HG Aluminium at $5.20$ TPa and HG sulfur at $500$ GPa (64-atom $Pcca$ approximant) - These pressures roughly correspond to the midpoint of the respective HG stability ranges.}
    \label{eDOSs}
\end{figure}

\section{\label{eos_discos}Volume Discontinuity}

\begin{figure}[h!]
    \includegraphics[width=0.45\textwidth]{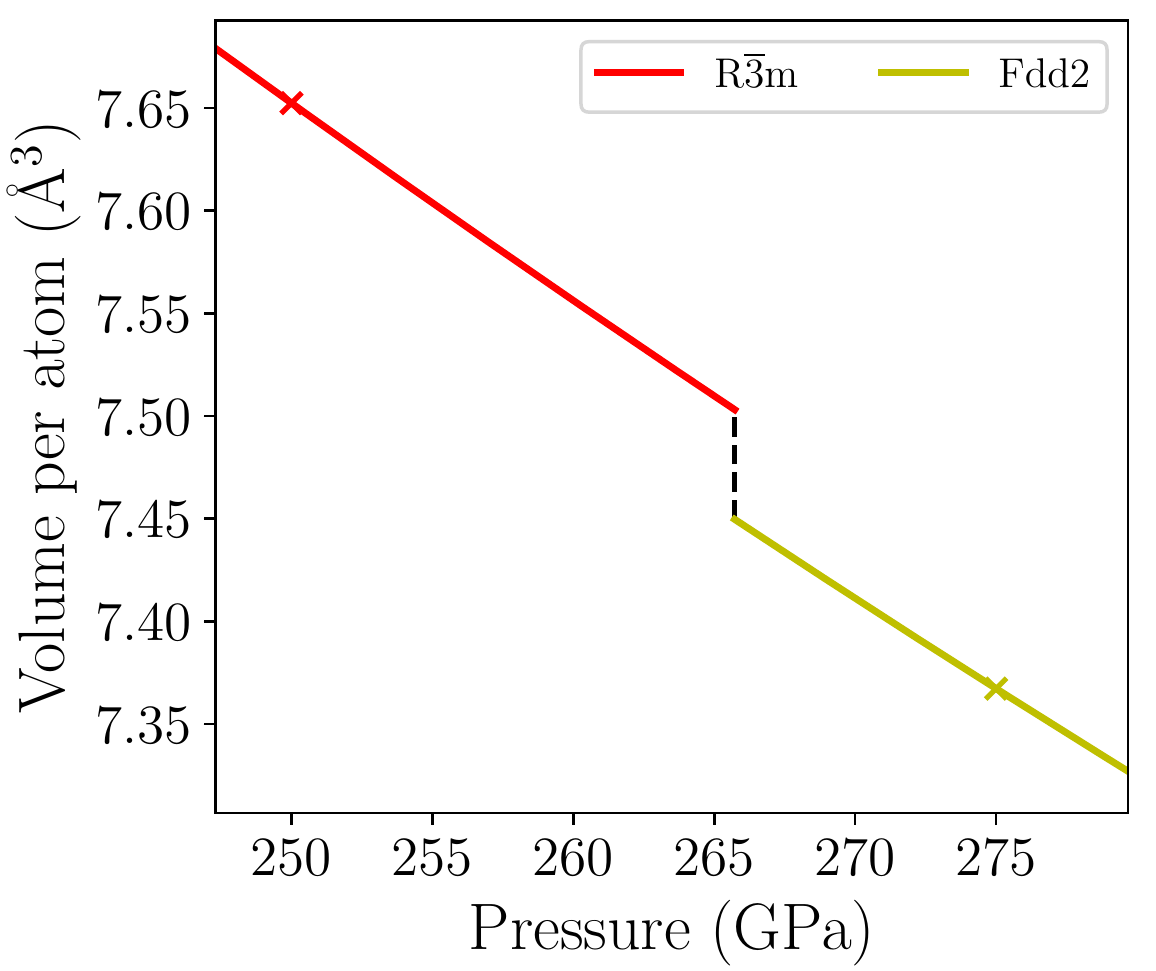}
    
    \vspace{0.5cm}
    
    \hspace{0.05cm}
    \includegraphics[width=0.45\textwidth]{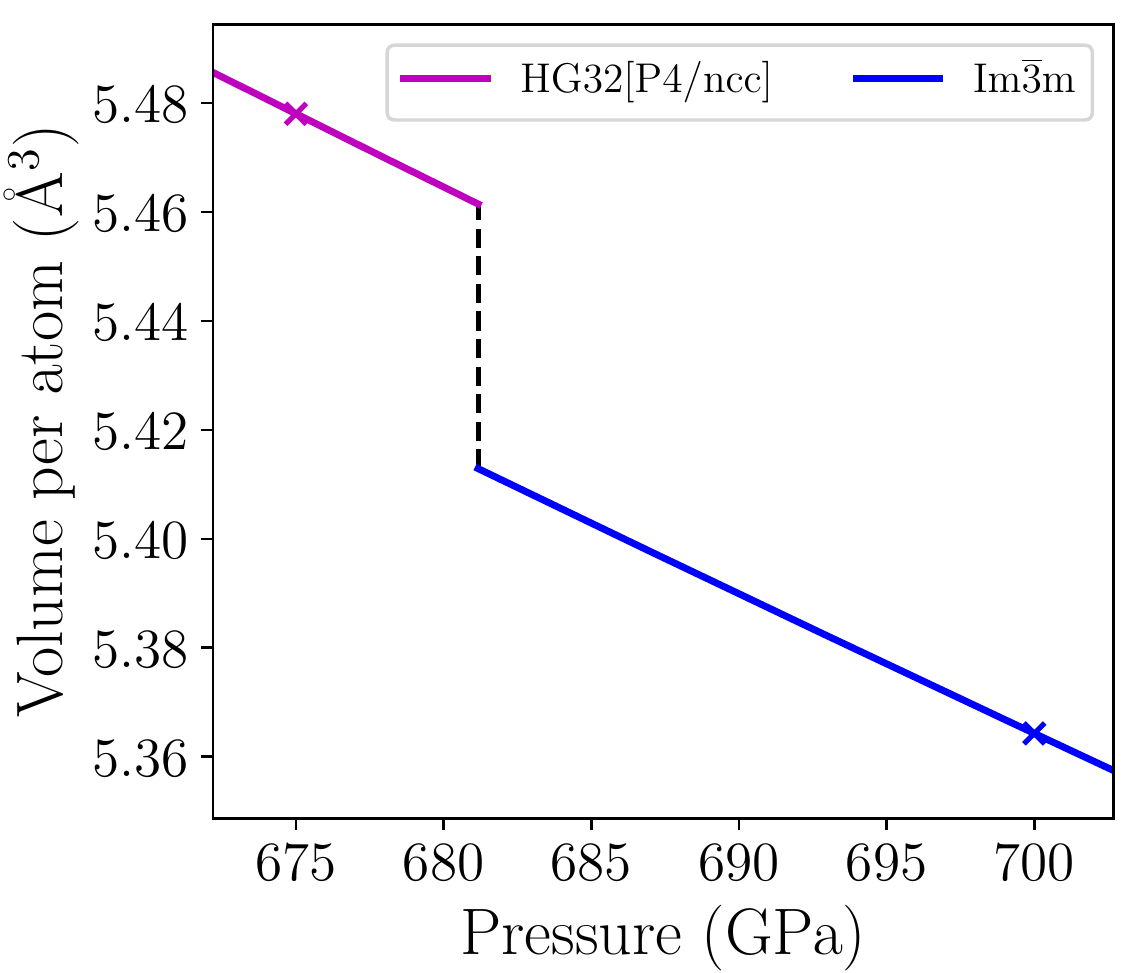}
    
    \caption{Volume discontinuities with pressure for the $R\bar{3}m \rightarrow Fdd2$ transition (top) and the HG $\rightarrow Im\bar{3}m$ transition (bottom).}
    \label{volume_discontinuities}
\end{figure}

Supplemental Fig. \ref{volume_discontinuities} shows that the $R\bar{3}m \rightarrow Fdd2$ and HG $\rightarrow Im\bar{3}m$ transitions introduce volume discontinuities into the equation of state of high-pressure sulfur. The fractional volume change $\frac{|\Delta V|}{V_0}$ is $ \approx 0.71 \%$ upon transition to the $Fdd2$ phase, and $ \approx 0.99 \%$ upon transition to the $Im\bar{3}m$ phase. The magnitude of the volume change at the $Fdd2 \rightarrow$ HG transition (not shown here) was less than $0.01 \%$.

\clearpage

\section{\label{diffraction} Powder Diffraction Patterns}

\begin{figure}[h!]
    \includegraphics[width=0.43\textwidth]{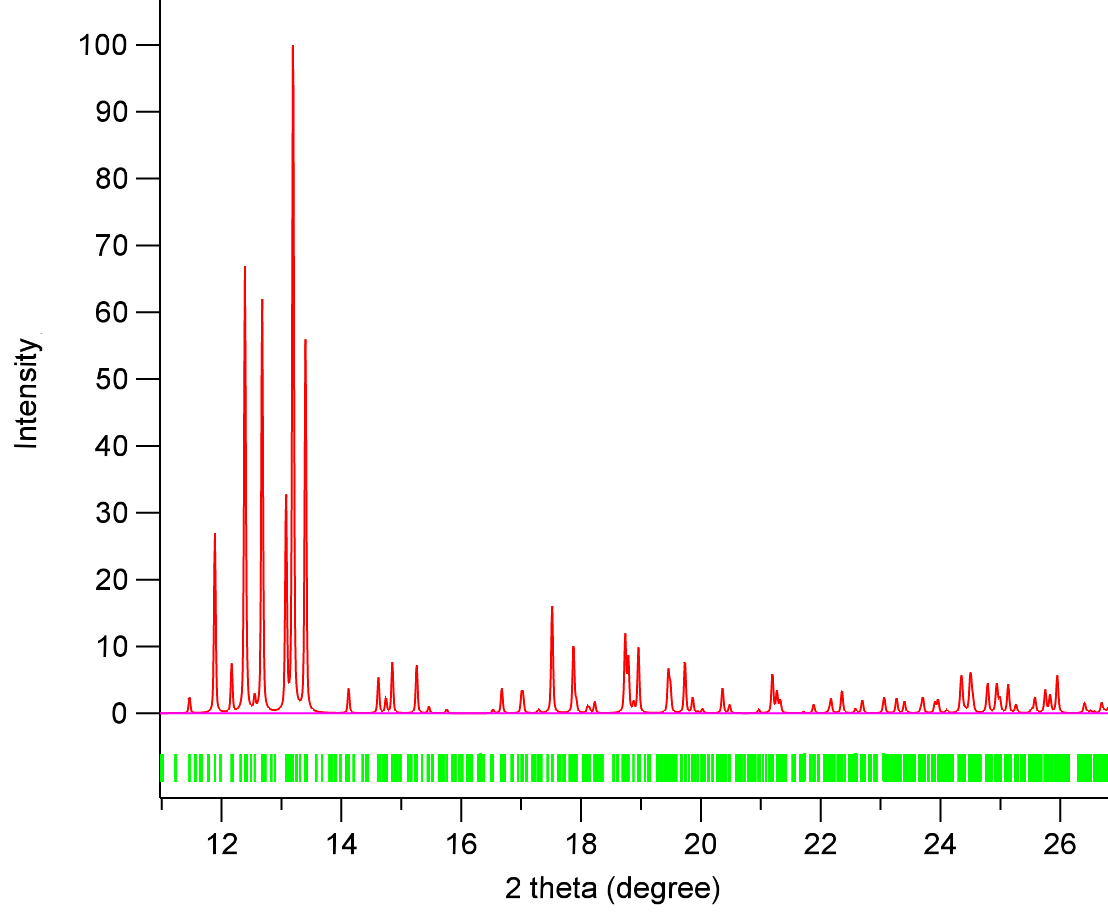}
    
    \vspace{0.4cm}
    
    \includegraphics[width=0.43\textwidth]{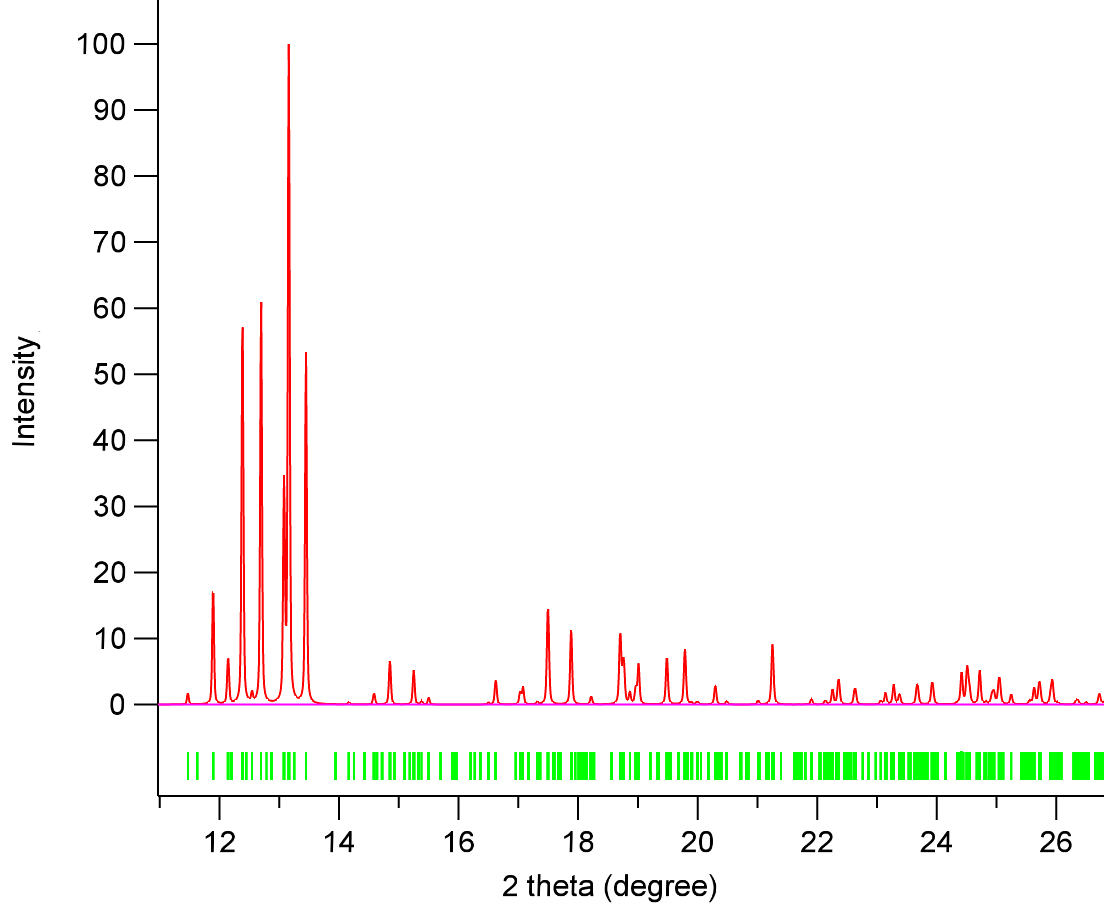}
    
    \vspace{0.4cm}
    
    \includegraphics[width=0.43\textwidth]{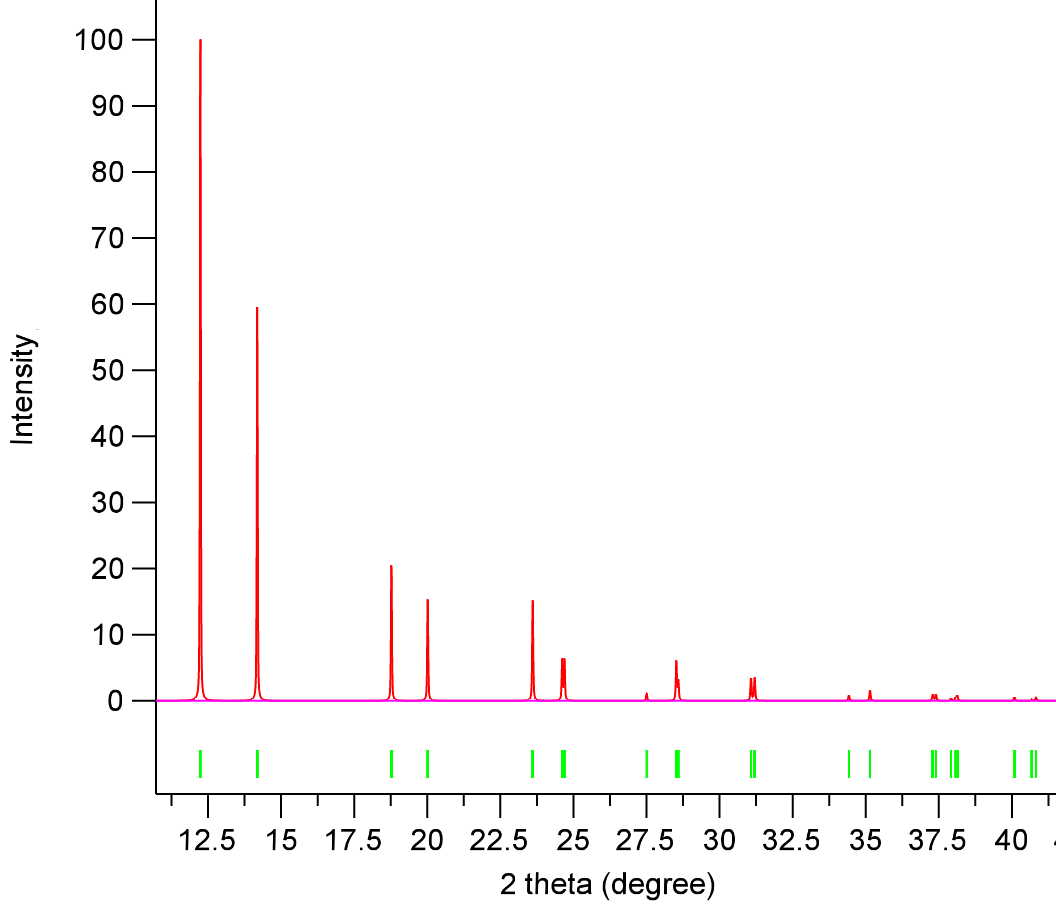}
    
    \caption{Simulated powder diffraction patterns for the $Fdd2$ phase at $275$ GPa (top), the $C2/c$ ABC-ordered HG approximant at $275$ GPa (middle) and $R\bar{3}m$ phase at 250 GPa (bottom). The incident radiation wavelength was set to $0.4$ \AA.}
    \label{powder_diffraction_patterns}
\end{figure}

Supplemental Fig. \ref{powder_diffraction_patterns} shows simulated powder diffraction patterns for the $Fdd2$, $C2/c$ approximant ($\gamma = 4/3$) and $R\bar{3}m$ ($\beta$-Po) phases. At $275$ GPa, the $Fdd2$ phase should be the ground state structure. However, we present diffraction patterns for both the $C2/c$ approximant \textbf{\textit{and}} the $Fdd2$ structure at $275$ GPa, in order to assist experimentalists in identifying any difference between the diffraction patterns of these two (very) similar structures.
\\\\
We note that the most significant difference is the absence of a peak at $2 \theta \sim 14.1 ^{\circ}$ in the pattern of the $C2/c$ approximant, which is present in the $Fdd2$ pattern. In fact, there are several peaks present in the $Fdd2$ pattern that are missing in the $C2/c$. We also note that the relative heights of a few peaks are changed between the two patterns. We expect the real host-guest diffraction pattern (to which the $C2/c$ is an approximation) to possess satellite peaks, which will not be present on the pattern presented here.

\clearpage

\section{\label{Al_and_As}Aluminium and Arsenic Phase Diagrams}

In the manuscript, we mention that the $Fdd2$ structure does not become the ground state in other elements that possess high-pressure HG phases. In Supplemental Fig. \ref{Al_and_As_phase_diagram} below, we present, as examples, the phase diagrams of Al and As with the $Fdd2$ structure added in.

\begin{figure}[h!]
    \hspace{-0.45cm}
    \includegraphics[width=0.46\textwidth]{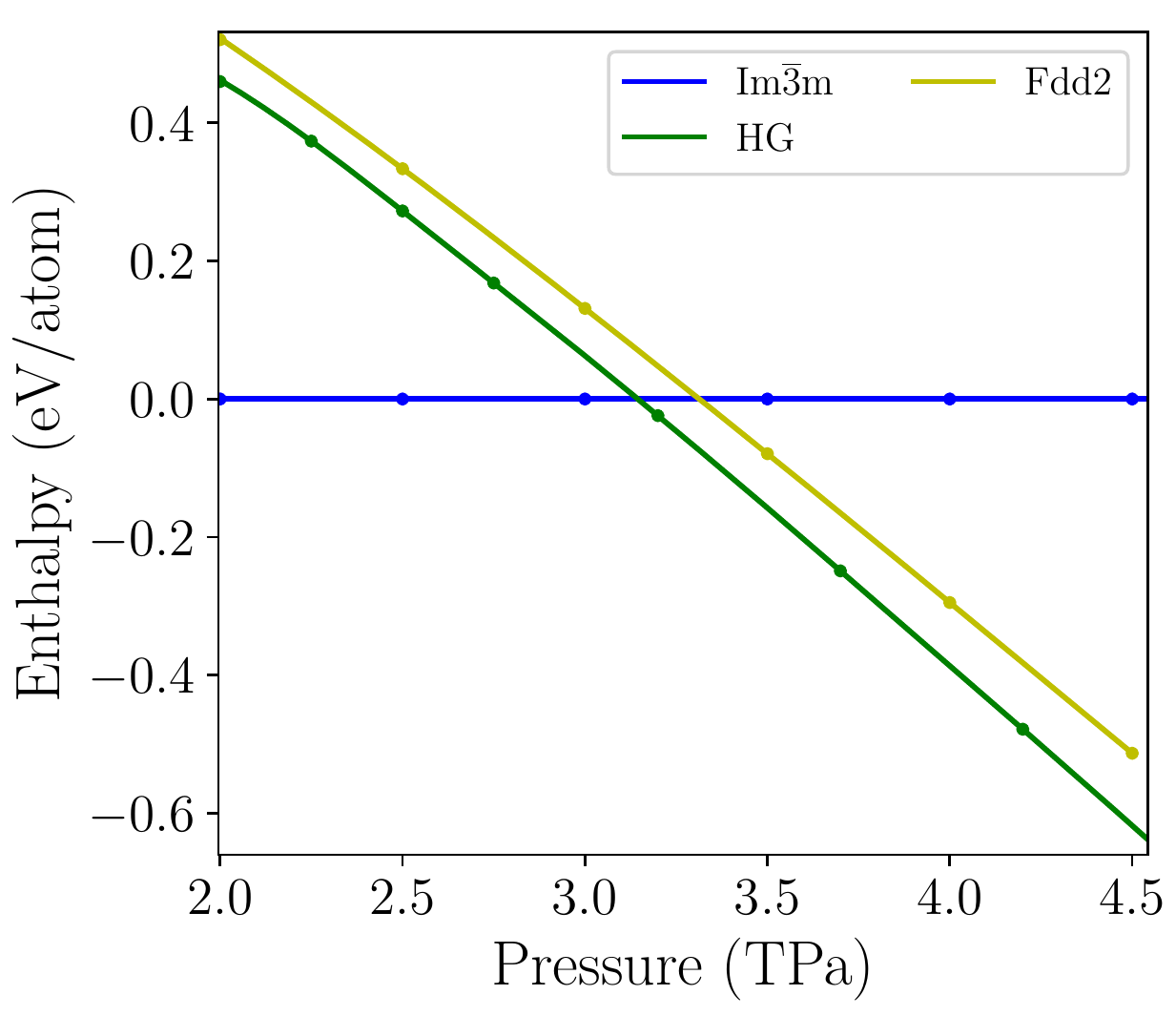}
    
    \vspace{0.5cm}
    
    \hspace{-0.45cm}
    \includegraphics[width=0.46\textwidth]{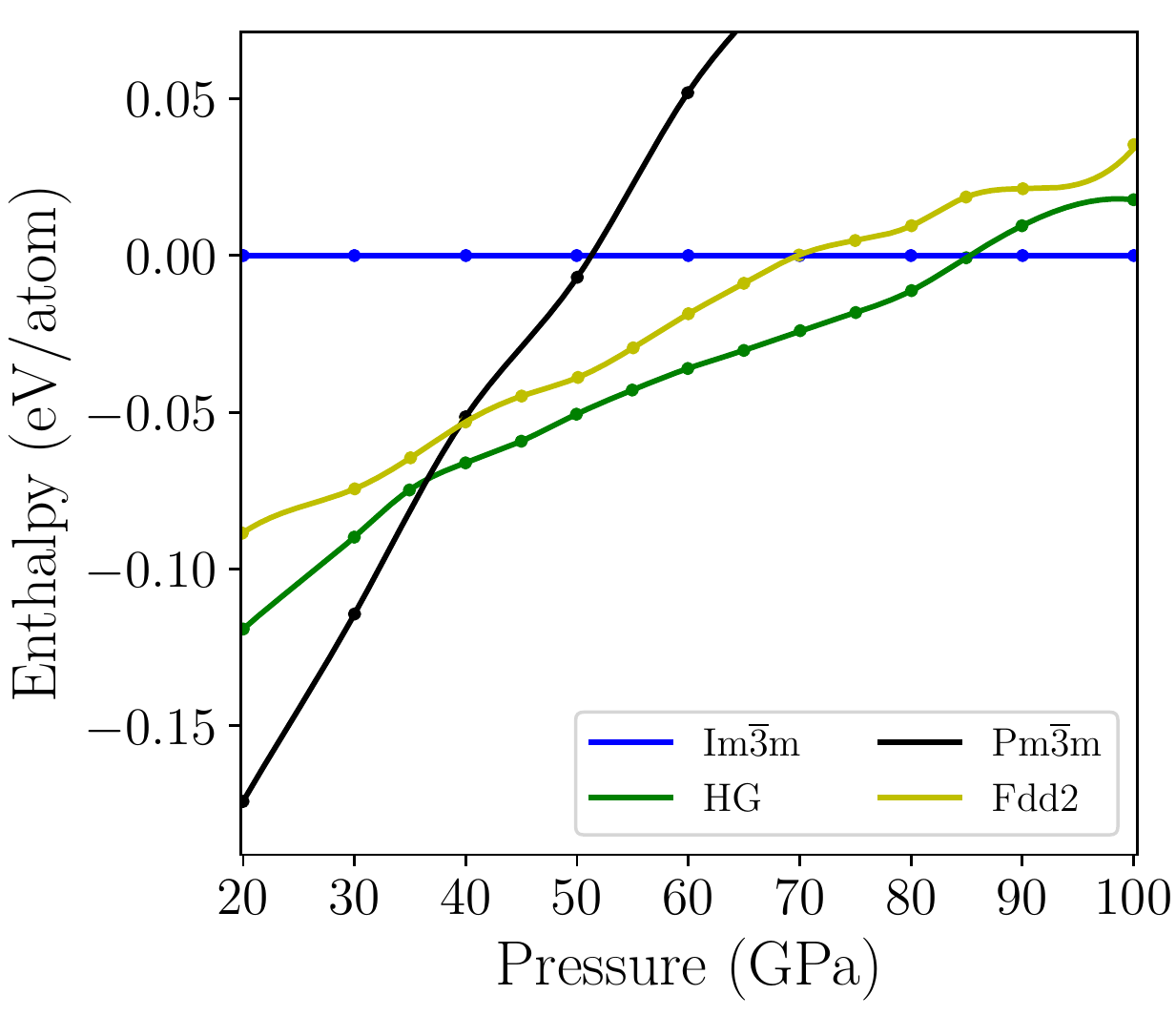}
    
    \caption{Phase diagrams of aluminium in the terapascal regime (top) and arsenic between $20$-$100$ GPa (bottom). For Al, the $C2/c$ approximant was used to represent the HG phase (as per Ref. \cite{HG_Aluminium}), and for As, the 32-atom $P4/ncc$ approximant was used (as per Ref. \cite{Haussermann_2002}).}
    \label{Al_and_As_phase_diagram}
\end{figure}

Whilst certainly competitive, the $Fdd2$ structure does not become the ground state in either of these cases.

\newpage

\section{\label{TC} Superconducting Critical Temperature Calculations}

We used the plane-wave DFT code \verb|QUANTUM ESPRESSO| \cite{QE} for electron-phonon coupling calculations and subsequent calculation of the superconducting critical temperature $T_c$ within Migdal-Eliashberg theory and using the McMillan-Allen-Dynes formula \cite{McMillan_Allen_Dynes}. Density Functional Perturbation Theory (DFPT) was used for all phonon calculations.
\\\\
A crucial consideration in accurate $T_c$ calculations is the choice of the double-delta smearing parameter $\sigma$ in the calculation of the Eliashberg function:

\begin{align}
    \alpha^2 F(\omega) = \frac{1}{g(\epsilon_F)} \sum_{m,n} \sum_{\vec{q},\nu} \delta(\omega - \omega_{\vec{q},\nu}) \sum_{\vec{k}} |g^{\vec{q},\nu,m,n}_{\vec{k}+\vec{q},\vec{k}}|^2
    \notag\\
    \times \ \delta(\epsilon_{\vec{k}+\vec{q},m}-\epsilon_F) \  \delta(\epsilon_{\vec{k},n}-\epsilon_F)
\label{eliashberg}
\end{align}

Where the delta functions are replaced by Gaussians of width $\sigma$:

\begin{align}
    \delta(\epsilon_{\vec{q},\nu}-\epsilon_F) \rightarrow \mathrm{exp} \bigg( \frac{(\epsilon_{\vec{q},\nu} - \epsilon_F)^2}{\sigma^2} \bigg)
\label{gaussian}
\end{align}

This replacement is necessary, as Eqn. \ref{eliashberg} can only be used as-is in the infinite \textbf{q}- and \textbf{k}- kpoint sampling limit. The use of finite \textbf{q}- and \textbf{k}- point grids in electronic structure calculations necessitates the replacement of the delta functions by Gaussians.
\\\\
Careful selection of the $\sigma$ value is very important in elemental systems, where $T_c$ values are generally of order $10$ K, and thus an error of a few Kelvin represents a large relative error. The optimal value of $\sigma$ can be determined by calculating $T_c(\sigma)$ for two similarly-sized \textbf{k}-point grids and finding the point at which the critical temperatures of the two grids start to diverge. It is a necessary requirement that the \textbf{k}-point grid be an integer multiple of the \textbf{q}-point grid. As an example, if we use a 6x6x6 \textbf{q}-point grid, we might calculate $T_c(\sigma)$ for a 24x24x24 and a 30x30x30 $k$-point grid, corresponding to $kpq$ (`\textbf{k}-points per \textbf{q}-point') values of 4 and 5, respectively. This procedure is described in detail in  ref. \cite{niobium_Tc}, and we show an example of such a $T_c$ calculation for the $Fdd2$ structure in Supplemental Fig. \ref{tc_method} below.

\begin{figure}[h!]
    \includegraphics[width=0.45\textwidth]{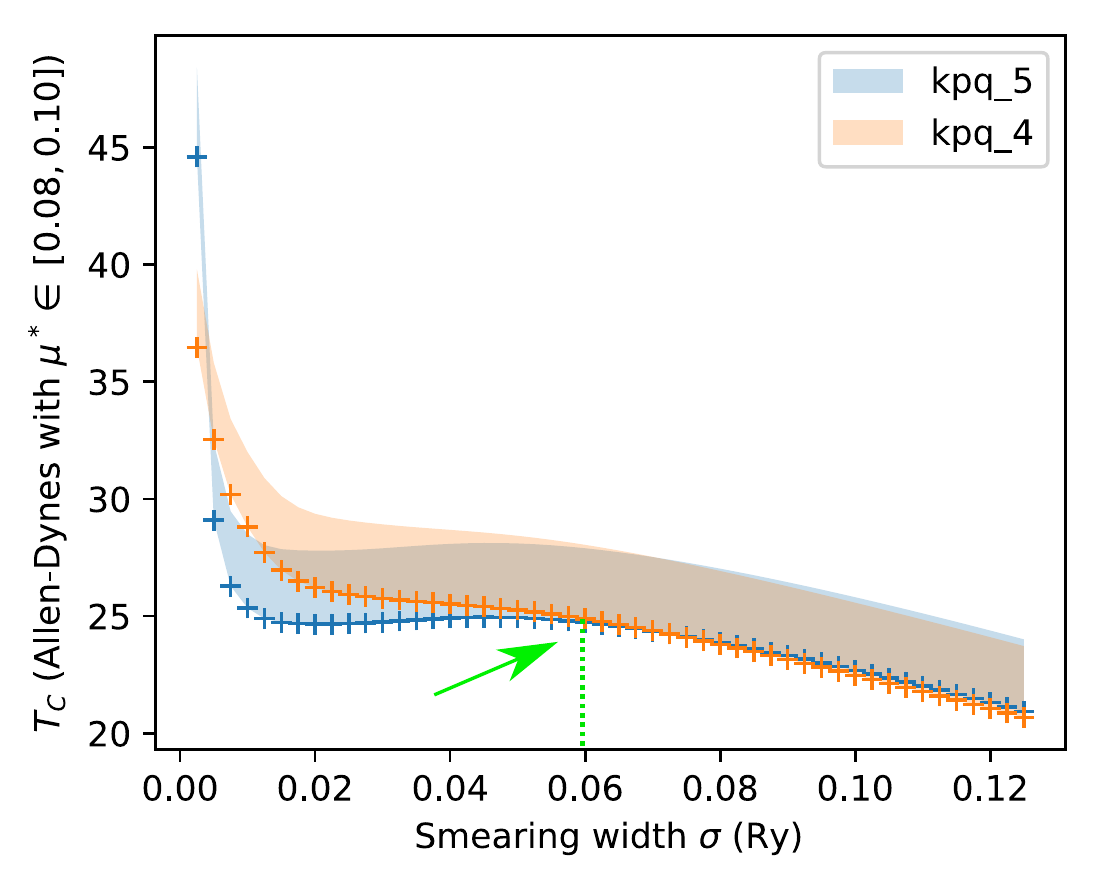}
    
    \caption{$T_c$ of the $Fdd2$ structure at $271$ GPa as a function of smearing width $\sigma$ for two \textbf{k}-point grids - one with $4$ \textbf{k}-points per \textbf{q}-point, and one with $5$ \textbf{k}-points per \textbf{q}-point. A green arrow indicates the point where the two curves start to diverge, and the green dotted line shows that the ideal smearing width is $\sim{0.06}$ Rydberg in this case.}
    \label{tc_method}
\end{figure}

%\clearpage

\newpage

\section{\label{structures} Structural Information}

We present (primitive) \verb|.cell| files for various structures below.
\\\\

\color{red}
$Fddd \ @ \ 200$ GPa
\color{black}
\begin{verbatim}
    %block LATTICE_CART
    ang
      0.0000000   3.0716500   7.5560978
      2.8136036   0.0000000   7.5560978
      2.8136036   3.0716500   0.0000000
    %endblock LATTICE_CART
    
    %block POSITIONS_FRAC
      S  0.581656103  0.791591252  0.050624316
      S  0.071197879  0.428802121  0.071197879
      S  0.397001150  0.397001150  0.102998850
      S  0.852998850  0.852998850  0.147001150
      S  0.178802121  0.821197879  0.178802121
      S  0.668343897  0.458408748  0.199375684
      S  0.102998850  0.102998850  0.397001150
      S  0.428802121  0.071197879  0.428802121
      S  0.673871671  0.199375684  0.458408748
      S  0.791591252  0.581656103  0.576128329
      S  0.050624316  0.576128329  0.581656103
      S  0.199375684  0.673871671  0.668343897
      S  0.458408748  0.668343897  0.673871671
      S  0.576128329  0.050624316  0.791591252
      S  0.821197879  0.178802121  0.821197879
      S  0.147001150  0.147001150  0.852998850
    %endblock POSITIONS_FRAC
\end{verbatim}

\newpage

\color{red}
$R\bar{3}m \ @ \ 250$ GPa
\color{black}
\begin{verbatim}
    %block LATTICE_CART
    ang
    1.6211882   0.9359935   0.8405054
    -1.6211882   0.9359935   0.8405054
    0.0000000  -1.8719869   0.8405054
    %endblock LATTICE_CART
    
    %block POSITIONS_FRAC
    S  0.500000000  0.500000000  0.500000000
    %endblock POSITIONS_FRAC
\end{verbatim}

\color{red}
$Fdd2 \ @ \ 300$ GPa
\color{black}
\begin{verbatim}
    %block LATTICE_CART
    ang
      0.0000000   7.2469132   2.6760045
      2.9544756   0.0000000   2.6760045
      2.9544756   7.2469132   0.0000000
    %endblock LATTICE_CART
    
    %block POSITIONS_FRAC
      S  0.460350598  0.792807744  0.055443387
      S  0.731418288  0.482390943  0.064087521
      S  0.058060264  0.051661061  0.071035871
      S  0.178964129  0.430757196  0.191939736
      S  0.430757196  0.178964129  0.198338939
      S  0.850284770  0.573046329  0.274747620
      S  0.573046329  0.850284770  0.301921281
      S  0.975252380  0.948078719  0.399715230
      S  0.558601729  0.194556613  0.457192256
      S  0.185912479  0.527896752  0.518581712
      S  0.948078719  0.975252380  0.676953671
      S  0.792807744  0.460350598  0.691398271
      S  0.482390943  0.731418288  0.722103248
      S  0.527896752  0.185912479  0.767609057
      S  0.194556613  0.558601729  0.789649402
      S  0.051661061  0.058060264  0.819242804
    %endblock POSITIONS_FRAC
\end{verbatim}

\color{red}
$C2/c \ @ \ 400$ GPa
\color{black}
\begin{verbatim}
    %block LATTICE_CART
    ang
      2.8740970  -2.5844681   0.0000000
      2.8740970   2.5844681   0.0000000
     -2.5688715   0.0000000   7.0105305
    %endblock LATTICE_CART
    
    %block POSITIONS_FRAC
      S  0.257215059  0.026326505  0.048178790
      S  0.578217142  0.431995613  0.119203117
      S  0.095583293  0.639743506  0.124195854
      S  0.709544949  0.927654278  0.230208302
      S  0.072345722  0.290455051  0.269791698
      S  0.360256494  0.904416707  0.375804146
      S  0.568004387  0.421782858  0.380796883
      S  0.973673495  0.742784941  0.451821210
      S  0.026326505  0.257215059  0.548178790
      S  0.431995613  0.578217142  0.619203117
      S  0.639743506  0.095583293  0.624195854
      S  0.927654278  0.709544949  0.730208302
      S  0.290455051  0.072345722  0.769791698
      S  0.904416707  0.360256494  0.875804146
      S  0.421782858  0.568004387  0.880796883
      S  0.742784941  0.973673495  0.951821210
    %endblock POSITIONS_FRAC
\end{verbatim}

\color{red}
$Pcca \ @ \ 500$ GPa
\color{black}
\begin{verbatim}
    %block LATTICE_CART
    ang
     10.1310268   0.0000000   0.0000000
      0.0000000   5.0612578   0.0000000
      0.0000000   0.0000000   7.5490969
    %endblock LATTICE_CART
    
    %block POSITIONS_FRAC
      S  0.921154366  0.141917384  0.001201932
      S  0.578845634  0.858082616  0.001201932
      S  0.178845563  0.342070917  0.004277600
      S  0.321154437  0.657929083  0.004277600
      S  0.750000000  0.000000000  0.027404472
      S  0.003489518  0.491437926  0.122581360
      S  0.496510482  0.508562074  0.122581360
      S  0.570435274  0.163949693  0.162606534
      S  0.929564726  0.836050307  0.162606534
      S  0.828512529  0.359431377  0.164791163
      S  0.671487471  0.640568623  0.164791163
      S  0.326611856  0.346111455  0.168767685
      S  0.173388144  0.653888545  0.168767685
      S  0.076817651  0.146729300  0.171665485
      S  0.423182349  0.853270700  0.171665485
      S  0.250000000  0.000000000  0.227082257
      S  0.750000000  0.000000000  0.272917743
      S  0.923182349  0.146729300  0.328334515
      S  0.576817651  0.853270700  0.328334515
      S  0.673388144  0.346111455  0.331232315
      S  0.826611856  0.653888545  0.331232315
      S  0.171487471  0.359431377  0.335208837
      S  0.328512529  0.640568623  0.335208837
      S  0.429564726  0.163949693  0.337393466
      S  0.070435274  0.836050307  0.337393466
      S  0.996510482  0.491437926  0.377418640
      S  0.503489518  0.508562074  0.377418640
      S  0.250000000  0.000000000  0.472595528
      S  0.821154437  0.342070917  0.495722400
      S  0.678845563  0.657929083  0.495722400
      S  0.078845634  0.141917384  0.498798068
      S  0.421154366  0.858082616  0.498798068
      S  0.578845634  0.141917384  0.501201932
      S  0.921154366  0.858082616  0.501201932
      S  0.321154437  0.342070917  0.504277600
      S  0.178845563  0.657929083  0.504277600
      S  0.750000000  0.000000000  0.527404472
      S  0.496510482  0.491437926  0.622581360
      S  0.003489518  0.508562074  0.622581360
      S  0.929564726  0.163949693  0.662606534
      S  0.570435274  0.836050307  0.662606534
      S  0.671487471  0.359431377  0.664791163
      S  0.828512529  0.640568623  0.664791163
      S  0.173388144  0.346111455  0.668767685
      S  0.326611856  0.653888545  0.668767685
      S  0.423182349  0.146729300  0.671665485
      S  0.076817651  0.853270700  0.671665485
      S  0.250000000  0.000000000  0.727082257
      S  0.750000000  0.000000000  0.772917743
      S  0.576817651  0.146729300  0.828334515
      S  0.923182349  0.853270700  0.828334515
      S  0.826611856  0.346111455  0.831232315
      S  0.673388144  0.653888545  0.831232315
      S  0.328512529  0.359431377  0.835208837
      S  0.171487471  0.640568623  0.835208837
      S  0.070435274  0.163949693  0.837393466
      S  0.429564726  0.836050307  0.837393466
      S  0.503489518  0.491437926  0.877418640
      S  0.996510482  0.508562074  0.877418640
      S  0.250000000  0.000000000  0.972595528
      S  0.678845563  0.342070917  0.995722400
      S  0.821154437  0.657929083  0.995722400
      S  0.421154366  0.141917384  0.998798068
      S  0.078845634  0.858082616  0.998798068
    %endblock POSITIONS_FRAC
    
\end{verbatim}

\color{red}
$P4/ncc \ @ \ 650$ GPa
\color{black}
\begin{verbatim}
    %block LATTICE_CART
    ang
      4.9149353   0.0000000   0.0000000
      0.0000000   4.9149353   0.0000000
      0.0000000   0.0000000   7.3482019
    %endblock LATTICE_CART
    
    %block POSITIONS_FRAC
      S  0.500000000  0.000000000  0.052805013
      S  0.154488689  0.145587719  0.083832564
      S  0.645587719  0.345511311  0.083832564
      S  0.354412281  0.654488689  0.083832564
      S  0.845511311  0.854412281  0.083832564
      S  0.000000000  0.500000000  0.193048082
      S  0.848729494  0.151270506  0.250000000
      S  0.348729494  0.348729494  0.250000000
      S  0.651270506  0.651270506  0.250000000
      S  0.151270506  0.848729494  0.250000000
      S  0.500000000  0.000000000  0.306951918
      S  0.145587719  0.154488689  0.416167436
      S  0.654488689  0.354412281  0.416167436
      S  0.345511311  0.645587719  0.416167436
      S  0.854412281  0.845511311  0.416167436
      S  0.000000000  0.500000000  0.447194987
      S  0.500000000  0.000000000  0.552805013
      S  0.845511311  0.145587719  0.583832564
      S  0.354412281  0.345511311  0.583832564
      S  0.645587719  0.654488689  0.583832564
      S  0.154488689  0.854412281  0.583832564
      S  0.000000000  0.500000000  0.693048082
      S  0.151270506  0.151270506  0.750000000
      S  0.651270506  0.348729494  0.750000000
      S  0.348729494  0.651270506  0.750000000
      S  0.848729494  0.848729494  0.750000000
      S  0.500000000  0.000000000  0.806951918
      S  0.854412281  0.154488689  0.916167436
      S  0.345511311  0.354412281  0.916167436
      S  0.654488689  0.645587719  0.916167436
      S  0.145587719  0.845511311  0.916167436
      S  0.000000000  0.500000000  0.947194987
    %endblock POSITIONS_FRAC
\end{verbatim}

\newpage

\color{red}
$Im\bar{3}m \ @ \ 700$ GPa
\color{black}
\begin{verbatim}
    %block LATTICE_CART
    ang
     -1.1027680   1.1027680   1.1027680
      1.1027680  -1.1027680   1.1027680
      1.1027680   1.1027680  -1.1027680
    %endblock LATTICE_CART
    
    %block POSITIONS_FRAC
      S  0.000000000  0.000000000  0.000000000
    %endblock POSITIONS_FRAC
\end{verbatim}

@MISC{supp,note="Supplemental material has been appended to the end of this document, and contains pseudopotential information, convergence calculations, ideal host-guest ratio fitting procedure, enthalpy comparisons of other chain orderings, electronic bandstructure plots, simulated diffraction patterns, comparisons of S to aluminium and arsenic, details of the superconducting critical temperature calculations and structural information."}

@article{Degtyareva_and_Hemley_PRB,
  title = {Crystal structure of the superconducting phases of S and Se},
  author = {Degtyareva, Olga and Gregoryanz, Eugene and Somayazulu, Maddury and Mao, Ho-kwang and Hemley, Russell J.},
  journal = {Phys. Rev. B},
  volume = {71},
  issue = {21},
  pages = {214104},
  numpages = {6},
  year = {2005},
  month = {Jun},
  publisher = {American Physical Society},
  doi = {10.1103/PhysRevB.71.214104},
  url = {https://link.aps.org/doi/10.1103/PhysRevB.71.214104}
}

@Article{chain_structures_sulfur,
author={Degtyareva, Olga
and Gregoryanz, Eugene
and Somayazulu, Maddury
and Dera, Przemyslaw
and Mao, Ho-kwang
and Hemley, Russell J.},
title={Novel chain structures in group VI elements},
journal={Nature Materials},
year={2005},
month={Feb},
day={01},
volume={4},
number={2},
pages={152-155},
issn={1476-4660},
doi={10.1038/nmat1294},
url={https://doi.org/10.1038/nmat1294}
}

@article{CDW_sulfur,
  title = {Competition of Charge-Density Waves and Superconductivity in Sulfur},
  author = {Degtyareva, O. and Magnitskaya, M. V. and Kohanoff, J. and Profeta, G. and Scandolo, S. and Hanfland, M. and McMahon, M. I. and Gregoryanz, E.},
  journal = {Phys. Rev. Lett.},
  volume = {99},
  issue = {15},
  pages = {155505},
  numpages = {4},
  year = {2007},
  month = {Oct},
  publisher = {American Physical Society},
  doi = {10.1103/PhysRevLett.99.155505},
  url = {https://link.aps.org/doi/10.1103/PhysRevLett.99.155505}
}

@article{Luo_et_al,
  title = {\ensuremath{\beta}-Po phase of sulfur at 162 GPa: X-ray diffraction study to 212 GPa},
  author = {Luo, Huan and Greene, Raymond G. and Ruoff, Arthur L.},
  journal = {Phys. Rev. Lett.},
  volume = {71},
  issue = {18},
  pages = {2943--2946},
  numpages = {0},
  year = {1993},
  month = {Nov},
  publisher = {American Physical Society},
  doi = {10.1103/PhysRevLett.71.2943},
  url = {https://link.aps.org/doi/10.1103/PhysRevLett.71.2943}
}

@article{Hejny_et_al,
  title = {Incommensurate sulfur above $100\phantom{\rule{0.3em}{0ex}}\mathrm{GPa}$},
  author = {Hejny, C. and Lundegaard, L. F. and Falconi, S. and McMahon, M. I. and Hanfland, M.},
  journal = {Phys. Rev. B},
  volume = {71},
  issue = {2},
  pages = {020101},
  numpages = {4},
  year = {2005},
  month = {Jan},
  publisher = {American Physical Society},
  doi = {10.1103/PhysRevB.71.020101},
  url = {https://link.aps.org/doi/10.1103/PhysRevB.71.020101}
}

@article{Whaley_Baldwin,
        doi = {10.1088/1367-2630/ab6068},
        url = {https://doi.org/10.1088%2F1367-2630%2Fab6068},
        year = 2020,
        month = {Mar},
        publisher = {{IOP} Publishing},
        volume = {22},
        number = {2},
        pages = {023020},
        author = {Jack Whaley-Baldwin and Richard Needs},
        title = {First-principles high pressure structure searching, longitudinal-transverse mode coupling and absence of simple cubic phase in sulfur.},
        journal = {New Journal of Physics},
}

@article{USPEX,
author = {Gavryushkin, Pavel N. and Litasov, Konstantin D. and Dobrosmislov, Sergey S. and Popov, Zakhar I.},
title = {High-pressure phases of sulfur: Topological analysis and crystal structure prediction},
journal = {physica status solidi (b)},
volume = {254},
number = {7},
pages = {1600857},
keywords = {crystal structure, density functional theory, high pressure, phases, sulfur},
doi = {https://doi.org/10.1002/pssb.201600857},
url = {https://onlinelibrary.wiley.com/doi/abs/10.1002/pssb.201600857},
year = {2017},
}

@article{HG_Aluminium,
  title = {Aluminium at terapascal pressures},
  author = {Chris J. Pickard, R. J. Needs}, 
  year = 2010, month = {July},journal = {Nature Materials}, volume = {9}, pages = {624-627},
  url = {https://doi.org/10.1038/nmat2796},
  doi = {10.1038/nmat2796},
}

@article {Jeanloz,
        author = {Jeanloz, Raymond and Celliers, Peter M. and Collins, Gilbert W. and Eggert, Jon H. and Lee, Kanani K. M. and McWilliams, R. Stewart and Brygoo, St{\'e}phanie and Loubeyre, Paul},
        title = {Achieving high-density states through shock-wave loading of precompressed samples},
        volume = {104},
        number = {22},
        pages = {9172--9177},
        year = {2007},
        doi = {10.1073/pnas.0608170104},
        publisher = {National Academy of Sciences},
        issn = {0027-8424},
        URL = {https://www.pnas.org/content/104/22/9172},
        journal = {Proceedings of the National Academy of Sciences}
}

@article{oxygen_terapascal,
  title = {Persistence and Eventual Demise of Oxygen Molecules at Terapascal Pressures},
  author = {Sun, Jian and Martinez-Canales, Miguel and Klug, Dennis D. and Pickard, Chris J. and Needs, Richard J.},
  journal = {Phys. Rev. Lett.},
  volume = {108},
  issue = {4},
  pages = {045503},
  numpages = {5},
  year = {2012},
  month = {Jan},
  publisher = {American Physical Society},
  doi = {10.1103/PhysRevLett.108.045503},
  url = {https://link.aps.org/doi/10.1103/PhysRevLett.108.045503}
}

@article{airss_PRL,
  title = {High-Pressure Phases of Silane},
  author = {Pickard, Chris J. and Needs, R. J.},
  journal = {Phys. Rev. Lett.},
  volume = {97},
  issue = {4},
  pages = {045504},
  numpages = {4},
  year = {2006},
  month = {Jul},
  publisher = {American Physical Society},
  doi = {10.1103/PhysRevLett.97.045504},
  url = {https://link.aps.org/doi/10.1103/PhysRevLett.97.045504}
}

@article{airss_JPCM,
        doi = {10.1088/0953-8984/23/5/053201},
        url = {https://doi.org/10.1088/0953-8984/23/5/053201},
        year = 2011,
        month = {jan},
        publisher = {{IOP} Publishing},
        volume = {23},
        number = {5},
        pages = {053201},
        author = {Chris J Pickard and R J Needs},
        title = {Ab initio random structure searching},
        journal = {Journal of Physics: Condensed Matter},
}

@article{ice_terapascal,
  title = {Decomposition and Terapascal Phases of Water Ice},
  author = {Pickard, Chris J. and Martinez-Canales, Miguel and Needs, Richard J.},
  journal = {Phys. Rev. Lett.},
  volume = {110},
  issue = {24},
  pages = {245701},
  numpages = {5},
  year = {2013},
  month = {Jun},
  publisher = {American Physical Society},
  doi = {10.1103/PhysRevLett.110.245701},
  url = {https://link.aps.org/doi/10.1103/PhysRevLett.110.245701}
}

@article{castep,
author = {Stewart J. Clark and Matthew D. Segall and Chris J. Pickard and Phil J. Hasnip and Matt I. J. Probert and Keith Refson and Mike C. Payne},
doi = {doi:10.1524/zkri.220.5.567.65075},
url = {https://doi.org/10.1524/zkri.220.5.567.65075},
title = {First principles methods using CASTEP},
journal = {Zeitschrift f{\"u}r Kristallographie - Crystalline Materials},
number = {5-6},
volume = {220},
year = {2005},
pages = {567--570}
}

@article{HG_Barium,
  title = {Self-Hosting Incommensurate Structure of Barium IV},
  author = {Nelmes, R. J. and Allan, D. R. and McMahon, M. I. and Belmonte, S. A.},
  journal = {Phys. Rev. Lett.},
  volume = {83},
  issue = {20},
  pages = {4081--4084},
  numpages = {0},
  year = {1999},
  month = {Nov},
  publisher = {American Physical Society},
  doi = {10.1103/PhysRevLett.83.4081},
  url = {https://link.aps.org/doi/10.1103/PhysRevLett.83.4081}
}

@article{HG_Lithium,
  title = {Structural prediction of host-guest structure in lithium at high pressure},
  author = {Prutthipong Tsuppayakorn-aek and Wei Luo and Teeraphat Watcharatharapong and Rajeev Ahuja and Thiti Bovornratanaraks},
  journal = {Nature Scientific Reports},
  volume = {8},
  year = {2018},
  month = {March},
  publisher = {Nature},
  doi = {https://doi.org/10.1038/s41598-018-23473-5},
}

@article {HG_Calcium,
        author = {Arapan, Sergiu and Mao, Ho-kwang and Ahuja, Rajeev},
        title = {Prediction of incommensurate crystal structure in Ca at high pressure},
        volume = {105},
        number = {52},
        pages = {20627--20630},
        year = {2008},
        doi = {10.1073/pnas.0810813105},
        publisher = {National Academy of Sciences},
        issn = {0027-8424},
        URL = {https://www.pnas.org/content/105/52/20627},
        journal = {Proceedings of the National Academy of Sciences}
}

@article{RT_supercond,
  title = {Room-temperature superconductivity in a carbonaceous sulfur hydride},
  author = {Elliot Snider and Nathan Dasenbrock-Gammon and Raymond McBride and Mathew Debessai and Hiranya Vindana and Kevin Vencatasamy and Keith V. Lawler and Ashkan Salamat and Ranga P. Dias},
  journal = {Nature},
  volume = {586},
  pages = {373–377},
  year = {2020},
  month = {October},
  publisher = {Nature},
  url = {https://doi.org/10.1038/s41586-020-2801-z},
  doi = {10.1038/s41586-020-2801-z},
}

@article{Meyer_sulfur,
author = {Meyer, Beat},
title = {Elemental sulfur},
journal = {Chemical Reviews},
volume = {76},
number = {3},
pages = {367-388},
year = {1976},
doi = {10.1021/cr60301a003},
URL = {https://doi.org/10.1021/cr60301a003},
}

@article{sulfur_allotropes,
author = {Meyer, Beat},
title = {Solid Allotropes of Sulfur},
journal = {Chemical Reviews},
volume = {64},
number = {4},
pages = {429-451},
year = {1964},
doi = {10.1021/cr60230a004},
URL = {https://doi.org/10.1021/cr60230a004},
}

@article{HG_Sulfur_Hydrides,
  title = {Novel Cooperative Interactions and Structural Ordering in ${\mathrm{H}}_{2}\mathrm{S}\mathrm{\text{\ensuremath{-}}}{\mathrm{H}}_{2}$},
  author = {Strobel, Timothy A. and Ganesh, P. and Somayazulu, Maddury and Kent, P. R. C. and Hemley, Russell J.},
  journal = {Phys. Rev. Lett.},
  volume = {107},
  issue = {25},
  pages = {255503},
  numpages = {4},
  year = {2011},
  month = {Dec},
  publisher = {American Physical Society},
  doi = {10.1103/PhysRevLett.107.255503},
  url = {https://link.aps.org/doi/10.1103/PhysRevLett.107.255503}
}

@article {HG_Bismuth_SC,
        author = {Brown, Philip and Semeniuk, Konstantin and Wang, Diandian and Monserrat, Bartomeu and Pickard, Chris J. and Grosche, F. Malte},
        title = {Strong coupling superconductivity in a quasiperiodic host-guest structure},
        volume = {4},
        number = {4},
        elocation-id = {eaao4793},
        year = {2018},
        doi = {10.1126/sciadv.aao4793},
        publisher = {American Association for the Advancement of Science},
        URL = {https://advances.sciencemag.org/content/4/4/eaao4793},
        journal = {Science Advances}
}

@article{Kartoon_Bismuth,
  title = {Structural and electronic properties of the incommensurate host-guest Bi-III phase},
  author = {Kartoon, D. and Makov, G.},
  journal = {Phys. Rev. B},
  volume = {100},
  issue = {1},
  pages = {014104},
  numpages = {8},
  year = {2019},
  month = {Jul},
  publisher = {American Physical Society},
  doi = {10.1103/PhysRevB.100.014104},
  url = {https://link.aps.org/doi/10.1103/PhysRevB.100.014104}
}

@article{SC_elements_review,
title = "Superconducting elements under high pressure",
journal = "Physica C: Superconductivity and its Applications",
volume = "552",
pages = "30 - 33",
year = "2018",
issn = "0921-4534",
doi = "https://doi.org/10.1016/j.physc.2018.05.012",
url = "http://www.sciencedirect.com/science/article/pii/S0921453418302119",
author = "Katsuya Shimizu",
keywords = "Pressure, Element, Superconductivity, Diamond anvil cell",
}

@article{Calcium_experimental,
  title = {Superconducting state of Ca-VII below a critical temperature of 29 K at a pressure of 216 GPa},
  author = {Sakata, Masafumi and Nakamoto, Yuki and Shimizu, Katsuya and Matsuoka, Takahiro and Ohishi, Yasuo},
  journal = {Phys. Rev. B},
  volume = {83},
  issue = {22},
  pages = {220512},
  numpages = {4},
  year = {2011},
  month = {Jun},
  publisher = {American Physical Society},
  doi = {10.1103/PhysRevB.83.220512},
  url = {https://link.aps.org/doi/10.1103/PhysRevB.83.220512}
}

@article{ELF_paper,
author = {Becke,A. D.  and Edgecombe,K. E. },
title = {A simple measure of electron localization in atomic and molecular systems},
journal = {The Journal of Chemical Physics},
volume = {92},
number = {9},
pages = {5397-5403},
year = {1990},
doi = {10.1063/1.458517},
URL = {https://doi.org/10.1063/1.458517},
}

@Article{electride_review,
author ="Liu, Chang and Nikolaev, Sergey A. and Ren, Wei and Burton, Lee A.",
title  ="Electrides: a review",
journal  ="J. Mater. Chem. C",
year  ="2020",
volume  ="8",
issue  ="31",
pages  ="10551-10567",
publisher  ="The Royal Society of Chemistry",
doi  ="10.1039/D0TC01165G",
url  ="http://dx.doi.org/10.1039/D0TC01165G",
}

@article{HG_scandium,
author = {Tsuppayakorn-aek,Prutthipong  and Luo,Wei  and Pungtrakoon,Wirunti  and Chuenkingkeaw,Kittana  and Kaewmaraya,Thanayut  and Ahuja,Rajeev  and Bovornratanaraks,Thiti },
title = {The ideal commensurate value of Sc and the superconducting phase under high pressure},
journal = {Journal of Applied Physics},
volume = {124},
number = {22},
pages = {225901},
year = {2018},
doi = {10.1063/1.}
}

@article{novel_sulfur_hydrides,
  title = {Novel sulfur hydrides synthesized at extreme conditions},
  author = {Laniel, Dominique and Winkler, Bjoern and Bykova, Elena and Fedotenko, Timofey and Chariton, Stella and Milman, Victor and Bykov, Maxim and Prakapenka, Vitali and Dubrovinsky, Leonid and Dubrovinskaia, Natalia},
  journal = {Phys. Rev. B},
  volume = {102},
  issue = {13},
  pages = {134109},
  numpages = {8},
  year = {2020},
  month = {Oct},
  publisher = {American Physical Society},
  doi = {10.1103/PhysRevB.102.134109},
  url = {https://link.aps.org/doi/10.1103/PhysRevB.102.134109}
}

@article{Jones_mechanism,
author = {Jones, Harry  and Tyndall, Arthur Mannering },
title = {Applications of the Bloch theory to the study of alloys and of the properties of bismuth},
journal = {Proceedings of the Royal Society of London. Series A - Mathematical and Physical Sciences},
volume = {147},
number = {861},
pages = {396-417},
year = {1934},
doi = {10.1098/rspa.1934.0224},
URL = {https://royalsocietypublishing.org/doi/abs/10.1098/rspa.1934.0224},
}

@article{HG_alkali_metals,
  title = {Structural and electronic properties of the alkali metal incommensurate phases},
  author = {Woolman, Gavin and Naden Robinson, Victor and Marqu\'es, Miriam and Loa, Ingo and Ackland, Graeme J. and Hermann, Andreas},
  journal = {Phys. Rev. Materials},
  volume = {2},
  issue = {5},
  pages = {053604},
  numpages = {12},
  year = {2018},
  month = {May},
  publisher = {American Physical Society},
  doi = {10.1103/PhysRevMaterials.2.053604},
  url = {https://link.aps.org/doi/10.1103/PhysRevMaterials.2.053604}
}

@article{hydrogen_sulfide_Goncharov,
  title = {Hydrogen sulfide at high pressure: Change in stoichiometry},
  author = {Goncharov, Alexander F. and Lobanov, Sergey S. and Kruglov, Ivan and Zhao, Xiao-Miao and Chen, Xiao-Jia and Oganov, Artem R. and Kon\^opkov\'a, Zuzana and Prakapenka, Vitali B.},
  journal = {Phys. Rev. B},
  volume = {93},
  issue = {17},
  pages = {174105},
  numpages = {7},
  year = {2016},
  month = {May},
  publisher = {American Physical Society},
  doi = {10.1103/PhysRevB.93.174105},
  url = {https://link.aps.org/doi/10.1103/PhysRevB.93.174105}
}

@article{HG_Strontium,
  title = {Observation of the incommensurate barium-IV structure in strontium phase V},
  author = {McMahon, M. I. and Bovornratanaraks, T. and Allan, D. R. and Belmonte, S. A. and Nelmes, R. J.},
  journal = {Phys. Rev. B},
  volume = {61},
  issue = {5},
  pages = {3135--3138},
  numpages = {0},
  year = {2000},
  month = {Feb},
  publisher = {American Physical Society},
  doi = {10.1103/PhysRevB.61.3135},
  url = {https://link.aps.org/doi/10.1103/PhysRevB.61.3135}
}

@article{HG_Antimony,
  title = {Pressure-induced incommensurate-to-incommensurate phase transition in antimony},
  author = {Degtyareva, O. and McMahon, M. I. and Nelmes, R. J.},
  journal = {Phys. Rev. B},
  volume = {70},
  issue = {18},
  pages = {184119},
  numpages = {5},
  year = {2004},
  month = {Nov},
  publisher = {American Physical Society},
  doi = {10.1103/PhysRevB.70.184119},
  url = {https://link.aps.org/doi/10.1103/PhysRevB.70.184119}
}

@article{HG_group_V,
  title = {Ba-IV-Type Incommensurate Crystal Structure in Group-V Metals},
  author = {McMahon, M. I. and Degtyareva, O. and Nelmes, R. J.},
  journal = {Phys. Rev. Lett.},
  volume = {85},
  issue = {23},
  pages = {4896--4899},
  numpages = {0},
  year = {2000},
  month = {Dec},
  publisher = {American Physical Society},
  doi = {10.1103/PhysRevLett.85.4896},
  url = {https://link.aps.org/doi/10.1103/PhysRevLett.85.4896}
}

@article{oC52_rubidium,
  title = {Structure of Rb-III: Novel Modulated Stacking Structures in Alkali Metals},
  author = {Nelmes, R. J. and McMahon, M. I. and Loveday, J. S. and Rekhi, S.},
  journal = {Phys. Rev. Lett.},
  volume = {88},
  issue = {15},
  pages = {155503},
  numpages = {4},
  year = {2002},
  month = {Apr},
  publisher = {American Physical Society},
  doi = {10.1103/PhysRevLett.88.155503},
  url = {https://link.aps.org/doi/10.1103/PhysRevLett.88.155503}
}

@article{oC84_caesium,
  title = {Complex Crystal Structure of Cesium-III},
  author = {McMahon, M. I. and Nelmes, R. J. and Rekhi, S.},
  journal = {Phys. Rev. Lett.},
  volume = {87},
  issue = {25},
  pages = {255502},
  numpages = {4},
  year = {2001},
  month = {Nov},
  publisher = {American Physical Society},
  doi = {10.1103/PhysRevLett.87.255502},
  url = {https://link.aps.org/doi/10.1103/PhysRevLett.87.255502}
}

@article{modulated_Bi_and_Sb,
  title = {Incommensurate modulations of Bi-III and Sb-II},
  author = {McMahon, M. I. and Degtyareva, O. and Nelmes, R. J. and van Smaalen, S. and Palatinus, L.},
  journal = {Phys. Rev. B},
  volume = {75},
  issue = {18},
  pages = {184114},
  numpages = {5},
  year = {2007},
  month = {May},
  publisher = {American Physical Society},
  doi = {10.1103/PhysRevB.75.184114},
  url = {https://link.aps.org/doi/10.1103/PhysRevB.75.184114}
}

@article{highest_experimental_pressure_sulfur,
  title = {Superconductivity in the chalcogens up to multimegabar pressures},
  author = {Gregoryanz, Eugene and Struzhkin, Viktor V. and Hemley, Russell J. and Eremets, Mikhail I. and Mao, Ho-kwang and Timofeev, Yuri A.},
  journal = {Phys. Rev. B},
  volume = {65},
  issue = {6},
  pages = {064504},
  numpages = {6},
  year = {2002},
  month = {Jan},
  publisher = {American Physical Society},
  doi = {10.1103/PhysRevB.65.064504},
  url = {https://link.aps.org/doi/10.1103/PhysRevB.65.064504}
}

@article{dissociation_H2S,
  title = {Dissociation products and structures of solid ${\mathrm{H}}_{2}\mathrm{S}$ at strong compression},
  author = {Li, Yinwei and Wang, Lin and Liu, Hanyu and Zhang, Yunwei and Hao, Jian and Pickard, Chris J. and Nelson, Joseph R. and Needs, Richard J. and Li, Wentao and Huang, Yanwei and Errea, Ion and Calandra, Matteo and Mauri, Francesco and Ma, Yanming},
  journal = {Phys. Rev. B},
  volume = {93},
  issue = {2},
  pages = {020103},
  numpages = {5},
  year = {2016},
  month = {Jan},
  publisher = {American Physical Society},
  doi = {10.1103/PhysRevB.93.020103},
  url = {https://link.aps.org/doi/10.1103/PhysRevB.93.020103}
}

@article{chain_ordered_calcium,
  title = {Ca-VII: A Chain Ordered Host-Guest Structure of Calcium above 210 GPa},
  author = {Fujihisa, Hiroshi and Nakamoto, Yuki and Sakata, Masafumi and Shimizu, Katsuya and Matsuoka, Takahiro and Ohishi, Yasuo and Yamawaki, Hiroshi and Takeya, Satoshi and Gotoh, Yoshito},
  journal = {Phys. Rev. Lett.},
  volume = {110},
  issue = {23},
  pages = {235501},
  numpages = {5},
  year = {2013},
  month = {Jun},
  publisher = {American Physical Society},
  doi = {10.1103/PhysRevLett.110.235501},
  url = {https://link.aps.org/doi/10.1103/PhysRevLett.110.235501}
}

@article{complicated_barium,
  title = {Extraordinarily complex crystal structure with mesoscopic patterning in barium at high pressure},
  author = {Loa, I and Nelmes, R. J. and Lundegaard, L. F. and McMahon, M. I.},
  journal = {Nature Materials},
  volume = {11},
  pages = {627-632},
  year = {2012},
  month = {Jun},
  doi = {10.1038/nmat3342},
}

@article{oP8_potassium,
  title = {Observation of the $oP8$ crystal structure in potassium at high pressure},
  author = {Lundegaard, L. F. and Marqu\'es, M. and Stinton, G. and Ackland, G. J. and Nelmes, R. J. and McMahon, M. I.},
  journal = {Phys. Rev. B},
  volume = {80},
  issue = {2},
  pages = {020101},
  numpages = {4},
  year = {2009},
  month = {Jul},
  publisher = {American Physical Society},
  doi = {10.1103/PhysRevB.80.020101},
  url = {https://link.aps.org/doi/10.1103/PhysRevB.80.020101}
}

@article{McMillan_Allen_Dynes,
  title = {Transition temperature of strong-coupled superconductors reanalyzed},
  author = {Allen, P. B. and Dynes, R. C.},
  journal = {Phys. Rev. B},
  volume = {12},
  issue = {3},
  pages = {905--922},
  numpages = {0},
  year = {1975},
  month = {Aug},
  publisher = {American Physical Society},
  doi = {10.1103/PhysRevB.12.905},
  url = {https://link.aps.org/doi/10.1103/PhysRevB.12.905}
}

@article{QE,
	doi = {10.1088/0953-8984/21/39/395502},
	year = 2009,
	month = {sep},
	publisher = {{IOP} Publishing},
	volume = {21},
	number = {39},
	pages = {395502},
	author = {Paolo Giannozzi et al.},
	title = {{QUANTUM} {ESPRESSO}: a modular and open-source software project for quantum simulations of materials},
	journal = {Journal of Physics: Condensed Matter},
}

@article{sulfur_no_SC,
        author={Jack Whaley-Baldwin},
        title={Zero-Point Energies prevent a Trigonal to Simple Cubic Transition in High-Pressure Sulfur},
        journal={Electronic Structure},
        url={http://iopscience.iop.org/article/10.1088/2516-1075/abd487},
        year={2020},
}

@article{hydrogen_sulfide_Cui,
        doi = {10.1088/1367-2630/ab0a87},
        url = {https://doi.org/10.1088/1367-2630/ab0a87},
        year = 2019,
        month = {mar},
        publisher = {{IOP} Publishing},
        volume = {21},
        number = {3},
        pages = {033023},
        author = {Ting Ting Cui and Da Chen and Jian Chen Li and Wang Gao and Qing Jiang},
        title = {Favored decomposition paths of hydrogen sulfide at high pressure},
        journal = {New Journal of Physics},
}

@article{SCDFT_sulfur,
  title = {Origin of the critical temperature discontinuity in superconducting sulfur under high pressure},
  author = {Monni, M. and Bernardini, F. and Sanna, A. and Profeta, G. and Massidda, S.},
  journal = {Phys. Rev. B},
  volume = {95},
  issue = {6},
  pages = {064516},
  numpages = {5},
  year = {2017},
  month = {Feb},
  publisher = {American Physical Society},
  doi = {10.1103/PhysRevB.95.064516},
  url = {https://link.aps.org/doi/10.1103/PhysRevB.95.064516}
}

@article{SCDFT_1,
  title = {Ab initio theory of superconductivity. I. Density functional formalism and approximate functionals},
  author = {L\"uders, M. and Marques, M. A. L. and Lathiotakis, N. N. and Floris, A. and Profeta, G. and Fast, L. and Continenza, A. and Massidda, S. and Gross, E. K. U.},
  journal = {Phys. Rev. B},
  volume = {72},
  issue = {2},
  pages = {024545},
  numpages = {17},
  year = {2005},
  month = {Jul},
  publisher = {American Physical Society},
  doi = {10.1103/PhysRevB.72.024545},
  url = {https://link.aps.org/doi/10.1103/PhysRevB.72.024545}
}

@article{SCDFT_2,
  title = {Ab initio theory of superconductivity. II. Application to elemental metals},
  author = {Marques, M. A. L. and L\"uders, M. and Lathiotakis, N. N. and Profeta, G. and Floris, A. and Fast, L. and Continenza, A. and Gross, E. K. U. and Massidda, S.},
  journal = {Phys. Rev. B},
  volume = {72},
  issue = {2},
  pages = {024546},
  numpages = {13},
  year = {2005},
  month = {Jul},
  publisher = {American Physical Society},
  doi = {10.1103/PhysRevB.72.024546},
  url = {https://link.aps.org/doi/10.1103/PhysRevB.72.024546}
}

@Article{record_Tc_sulfur,
author={Struzhkin, Viktor V.
and Hemley, Russell J.
and Mao, Ho-kwang
and Timofeev, Yuri A.},
title={Superconductivity at 10--17{\thinspace}K in compressed sulphur},
journal={Nature},
year={1997},
month={Nov},
day={01},
volume={390},
number={6658},
pages={382-384},
issn={1476-4687},
doi={10.1038/37074},
url={https://doi.org/10.1038/37074}
}

@article{incommensurate_mercury_compound_axe,
  title = {Long-wavelength excitations in incommensurate intergrowth compounds with application to ${\mathrm{Hg}}_{3\ensuremath{-}\ensuremath{\delta}}\mathrm{As}{\mathrm{F}}_{6}$},
  author = {Axe, J. D. and Bak, Per},
  journal = {Phys. Rev. B},
  volume = {26},
  issue = {9},
  pages = {4963--4973},
  numpages = {0},
  year = {1982},
  month = {Nov},
  publisher = {American Physical Society},
  doi = {10.1103/PhysRevB.26.4963},
  url = {https://link.aps.org/doi/10.1103/PhysRevB.26.4963}
}

@article{incommensurate_mercury_compound_buiting,
        doi = {10.1088/0305-4608/14/10/013},
        url = {https://doi.org/10.1088/0305-4608/14/10/013},
        year = 1984,
        month = {oct},
        publisher = {{IOP} Publishing},
        volume = {14},
        number = {10},
        pages = {2343--2358},
        author = {J J M Buiting and M Weger and F M Mueller},
        title = {Electronic structure of the incommensurate metal Hg$_{3-\delta}${AsF}6},
        journal = {Journal of Physics F: Metal Physics} 
}

@article{incommensurate_mercury_compound_degroot,
  title = {Self-consistent electronic-band-structure calculation for ${\mathrm{Hg}}_{3} {\mathrm{AsF}}_{6}$},
  author = {de Groot, R. A. and Buiting, J. J. M. and Weger, M. and Mueller, F. M.},
  journal = {Phys. Rev. B},
  volume = {31},
  issue = {5},
  pages = {2881--2885},
  numpages = {0},
  year = {1985},
  month = {Mar},
  publisher = {American Physical Society},
  doi = {10.1103/PhysRevB.31.2881},
  url = {https://link.aps.org/doi/10.1103/PhysRevB.31.2881}
}

@Article{drozdov_hydrogen_sulfide_experiment,
author={Drozdov, A. P.
and Eremets, M. I.
and Troyan, I. A.
and Ksenofontov, V.
and Shylin, S. I.},
title={Conventional superconductivity at 203 kelvin at high pressures in the sulfur hydride system},
journal={Nature},
year={2015},
month={Sep},
day={01},
volume={525},
number={7567},
pages={73-76},
issn={1476-4687},
doi={10.1038/nature14964},
url={https://doi.org/10.1038/nature14964}
}

@article{zakharov_and_cohen_sulfur,
  title = {Theory of structural, electronic, vibrational, and superconducting properties of high-pressure phases of sulfur},
  author = {Zakharov, Oleg and Cohen, Marvin L.},
  journal = {Phys. Rev. B},
  volume = {52},
  issue = {17},
  pages = {12572--12578},
  numpages = {0},
  year = {1995},
  month = {Nov},
  publisher = {American Physical Society},
  doi = {10.1103/PhysRevB.52.12572},
  url = {https://link.aps.org/doi/10.1103/PhysRevB.52.12572}
}

@article{Rudin,
  title = {Predicted Simple-Cubic Phase and Superconducting Properties for Compressed Sulfur},
  author = {Rudin, Sven P. and Liu, Amy Y.},
  journal = {Phys. Rev. Lett.},
  volume = {83},
  issue = {15},
  pages = {3049--3052},
  numpages = {0},
  year = {1999},
  month = {Oct},
  publisher = {American Physical Society},
  doi = {10.1103/PhysRevLett.83.3049},
  url = {https://link.aps.org/doi/10.1103/PhysRevLett.83.3049}
}

@article{sulfur_and_Se_Rudin,
  title = {Comparison of structural transformations and superconductivity in compressed sulfur and selenium},
  author = {Rudin, Sven P. and Liu, Amy Y. and Freericks, J. K. and Quandt, Alexander},
  journal = {Phys. Rev. B},
  volume = {63},
  issue = {22},
  pages = {224107},
  numpages = {9},
  year = {2001},
  month = {May},
  publisher = {American Physical Society},
  doi = {10.1103/PhysRevB.63.224107},
  url = {https://link.aps.org/doi/10.1103/PhysRevB.63.224107}
}

@Article{Haussermann_2002,
author={H{\"a}ussermann, Ulrich
and S{\"o}derberg, Karin
and Norrestam, Rolf},
title={Comparative Study of the High-Pressure Behavior of As, Sb, and Bi},
journal={Journal of the American Chemical Society},
year={2002},
month={Dec},
day={01},
publisher={American Chemical Society},
volume={124},
number={51},
pages={15359-15367},
issn={0002-7863},
doi={10.1021/ja020832s},
url={https://doi.org/10.1021/ja020832s}
}

@article{Morel_Anderson,
  title = {Calculation of the Superconducting State Parameters with Retarded Electron-Phonon Interaction},
  author = {Morel, P. and Anderson, P. W.},
  journal = {Phys. Rev.},
  volume = {125},
  issue = {4},
  pages = {1263--1271},
  numpages = {0},
  year = {1962},
  month = {Feb},
  publisher = {American Physical Society},
  doi = {10.1103/PhysRev.125.1263},
  url = {https://link.aps.org/doi/10.1103/PhysRev.125.1263}
}

@article{Rb-IV,
  title = {Rubidium-IV: A High Pressure Phase with Complex Crystal Structure},
  author = {Schwarz, U. and Grzechnik, A. and Syassen, K. and Loa, I. and Hanfland, M.},
  journal = {Phys. Rev. Lett.},
  volume = {83},
  issue = {20},
  pages = {4085--4088},
  numpages = {0},
  year = {1999},
  month = {Nov},
  publisher = {American Physical Society},
  doi = {10.1103/PhysRevLett.83.4085},
  url = {https://link.aps.org/doi/10.1103/PhysRevLett.83.4085}
}

@article{sulfur_melting_curve,
  title = {High-pressure melting curve of sulfur up to 65 GPa},
  author = {Arveson, Sarah M. and Meng, Yue and Lee, June and Lee, Kanani K. M.},
  journal = {Phys. Rev. B},
  volume = {100},
  issue = {5},
  pages = {054106},
  numpages = {8},
  year = {2019},
  month = {Aug},
  publisher = {American Physical Society},
  doi = {10.1103/PhysRevB.100.054106},
  url = {https://link.aps.org/doi/10.1103/PhysRevB.100.054106}
}

@article{simons-glatzel,
author = {Simon, Franz and Glatzel, Gunther},
title = {Bemerkungen zur Schmelzdruckkurve},
journal = {Zeitschrift f{\"u}r anorganische und allgemeine Chemie},
volume = {178},
number = {1},
pages = {309-316},
doi = {https://doi.org/10.1002/zaac.19291780123},
url = {https://onlinelibrary.wiley.com/doi/abs/10.1002/zaac.19291780123},
year = {1929}
}

%%%%%%%%%%%%%%%%%%%%%%%%%%%%%%%%%%%%%%%
% REFERENCES BELOW ONLY APPEAR IN SUPPLEMENTAL

@article{otfg_pdf,
  title = {On-the-fly pseudopotential generation in CASTEP},
  author = {C. J.	Pickard},
  year = {2006},
  month = {September},
  journal = {website publication},
  url = {http://www.tcm.phy.cam.ac.uk/castep/otfg.pdf},
}

@article{optados,
  title = {The	OptaDOS	code},
  author = {Andrew J.	Morris,	R. J.	Nicholls,	C. J.	Pickard	and	Jonathan R.	Yates},
  journal = {Comp. Phys. Comm. (2012)},
  year = {2012},
}

@article{niobium_Tc,
    title={Origins of low- and high-pressure discontinuities of $T_{c}$ in niobium},
    author={Malgorzata Wierzbowska and Stefano de Gironcoli and Paolo Giannozzi},
    eprint={cond-mat/0504077},
    archivePrefix={arXiv},
    primaryClass={cond-mat.supr-con},
}

@article{Matador,
  doi = {10.21105/joss.02563},
  url = {https://doi.org/10.21105/joss.02563},
  year = {2020},
  publisher = {The Open Journal},
  volume = {5},
  number = {54},
  pages = {2563},
  author = {Matthew L. Evans and Andrew J. Morris},
  title = {matador: a Python library for analysing, curating and performing high-throughput density-functional theory calculations},
  journal = {Journal of Open Source Software}
}

\end{document}